\documentclass{aa}  

\usepackage{graphicx}
\usepackage{xcolor}
\usepackage[colorlinks,citecolor={blue}]{hyperref}
\usepackage{ulem}

\usepackage{txfonts}

\newcommand{\JLM}[1]{ #1}

\begin{document}

   \title{3D RMHD simulations of jet-wind interactions in High Mass X-ray Binaries}

   \author{J. López-Miralles
          \inst{1}
          \and
          M. Perucho\inst{1,2}
          \and
          J. M. Martí\inst{1,2}          
          \and
          S. Migliari\inst{3,4}
          \and
          V. Bosch-Ramon\inst{4}
          }

   \institute{Departament d’Astronomia i Astrofísica, Universitat de València, Dr. Moliner 50, 46100, Burjassot, València, Spain,\\
              \email{jose.lopez-miralles@uv.es}
        \and      
        Observatori Astron\`omic, Universitat de Val\`encia, C/ Catedr\`atic Jos\'e Beltr\'an 2, 46980, Paterna, Val\`encia, Spain.      
         \and
             Aurora Technology for the European Space Agency, ESAC/ESA, Camino Bajo del Castillo s/n, Urb. Villafranca del Castillo, 28691 Villanueva de la Cañada, Madrid, Spain
         \and
             Institut de Ci\`encies del Cosmos (ICC), Universitat de Barcelona (IEEC-UB), Mart\'{i} i Franqu\`es 1, E08028 Barcelona, Spain.}

   \date{Received ??; accepted ??}

 
  \abstract
   {Relativistic jets are ubiquitous in the Universe. In microquasars, especially in High-Mass X-ray Binaries, the interaction of jets with the strong winds driven by the massive and hot companion star in the vicinity of the compact object is fundamental to understand the jet dynamics, non-thermal emission and long-term stability. However, the role of the jet magnetic field in this process is unclear. In particular, it is still under debate whether magnetic field favours jet collimation or triggers more instabilities that can jeopardize the jet evolution outside the binary.}
   {We study the dynamical role of weak and moderate-to-strong toroidal magnetic fields during the first hundreds of seconds of jet propagation through the stellar wind, focusing on the magnetized flow dynamics and the mechanisms of energy conversion.}
   {We have developed the code \JLM{Lóstrego v1.0}, a new 3D Relativistic Magnetohydrodynamics code to simulate astrophysical plasmas in \JLM{Cartesian} coordinates. Using this tool, we performed the first 3D RMHD numerical simulations of relativistic magnetized jets propagating through the clumpy stellar wind in a High-Mass X-ray Binary. To bring out the effect of the magnetic field in the jet dynamics, we compare the results of our analysis with the outcome of previous hydrodynamical simulations.}
   {The overall morphology and dynamics of weakly magnetized jet models is similar to previous hydrodynamical simulations, where the jet head generates a strong shock in the ambient medium and the initial over-pressure with respect to the stellar wind drives one -or more- recollimation shocks. In the time scales of our simulations (i.e., t$<200$~s), these jets are ballistic and seem to be more stable against internal instabilities than jets with the same power in the absence of fields. However, moderate-to-strong toroidal magnetic fields favour the development of current-driven instabilities and the disruption of the jet within the binary. A detailed analysis of the energy distribution in the relativistic outflow and the ambient medium reveals that both magnetic and internal energies can contribute to the effective acceleration of the jet. Moreover, 
  we certify that the jet feedback into the ambient medium is highly dependent on the jet energy distribution at injection, where hotter, more dilute and/or more magnetized jets are more efficient, as anticipated by feedback studies in the case of jets in active galaxies.
}
   {}


   \maketitle
%

\section{Introduction}


Jets are the most spectacular and powerful consequences of accretion onto compact objects; they have been observed in systems containing white dwarfs, neutron stars (NSs) and black holes (BHs) of all mass scales, from stellar-mass in X-ray binaries (XRBs) to super-massive in Active Galactic Nuclei (AGN). These outflows are relativistic and extract a large - possibly dominant- fraction of the total available accretion energy \citep{ghisellini14}, and may even tap the power of the black hole spin \citep{blandford77}. Studies of plasma dynamics and radiative processes in relativistic jets, and especially the comparison among different classes of compact objects \citep{migliari06}, allows us to investigate the physics of strong-gravity and curved space-time, the presence of a stellar surface in NSs or the existence of an event horizon in BHs, the role of magnetic fields in jet formation, collimation and evolution, and the shock acceleration mechanisms of plasma particles, just to name some representative examples \citep[see, e.g.,][]{EHT19,EHT20,EHT21}. Moreover, powerful relativistic jets are one of the main ways in which accreting black holes provide kinetic feedback to their surroundings \citep[see, e.g.,][in the case of microquasars and AGN, respectively]{bordas09, mcnamara07}. However, most of the fundamental questions regarding jet formation mechanisms, composition, acceleration, collimation and interaction with the  interstellar medium (ISM) are still under debate.


X-ray Binaries, or microquasars, are binary systems hosting a compact object (i.e., a stellar-mass BH or NS) and a companion non-collapsed star, which supplies matter to the compact object via Roche-lobe overflow (Low-Mass X-ray Binaries) or by the capture of stellar winds (High-Mass X-ray Binaries (HMXBs)). This matter accumulates around the central object in the form of an accretion disk. Most XRBs are transient objects \citep{belloni16}, but in some spectral states, these systems produce powerful bipolar relativistic jets launched by magnetocentrifugal forces \citep{blandford77,blandford82} that emit non-thermal synchroton radiation \citep{mirabel99}. These processes mimic, on smaller scales, most of the phenomena observed in Quasars and AGNs, but in XRBs accretion varies on much faster -humanly accessible- time scales than in supermassive BHs. This allows us to observe and follow the evolution of the system and to investigate the link between jet formation and the different accretion states. Therefore, numerical simulations that study the physical conditions that may reproduce existing observations are strongly needed.



In order to understand how XRBs produce relativistic ejections, and how these jets affect the binary system and the surrounding ISM, we may distinguish three different regions of interest: (1) the innermost scale close to the compact object, where the jet is launched by magnetocentrifugal forces from the inner accretion disk \citep{migliari03,migliari04,fender04,kylafis12,marino20}, (2) the binary scale, where the collimated outflow interacts with the wind of the companion star and may presumably be disrupted \citep{Perucho08,Perucho10,Perucho12,Bosch16,cita17,Barkov21} and (3) the interaction of the jet with the ISM, where non-thermal particles could produce significant amounts of radiation at different wavelengths \citep{bordas09,bosch11,yoon11,monceau14}.


For HMXBs, jet-wind interactions are particularly relevant due to the strong winds that the massive and hot companion star drives in the vicinity of the compact object. As mentioned above, this wind can have a strong impact on jet dynamics, jeopardizing its stability and eventually preventing jet detection \citep{Perucho08,Perucho10}. Moreover, the one-sided impact of the wind on the relativistic flow may produce strong collisionless shocks that lead to efficient particle acceleration \citep{rieger07}, resulting in significant non-thermal radiation from synchroton, Inverse Compton (IC) or even proton-proton collisions \citep[see, e.g.,][]{romero03,molina19}. These processes may be at the origin of the gamma-ray emission detected in some X-ray Binaries like LS~5039 \citep[e.g.,][]{paredes00,aharonian05}, LS~I+61~303 \citep[e.g.,][]{tavani98,albert06} (if they really host accreting BHs), Cygnus~X-1 \citep[e.g.,][]{albert07,zanin16} and Cygnus~X-3 \citep[e.g.,][]{albert07,zanin16}, among others.


There have been several attempts in the past to describe jet-wind interactions from the point of view of theory and numerical simulations. \cite{Perucho08} performed numerical two-dimensional simulations of jets crossing the stellar wind, reporting the formation of strong recollimation shocks that can accelerate particles efficiently and produce non-thermal radiation. Similar results, but with 3D hydrodynamical simulations, were found by \cite{Perucho10}. This work also confirmed that jets with total power $L_{\mathrm{j}}\lesssim 10^{36}\dot{M}_{\rm w,-6}$~erg~s$^{-1}$ \JLM{($\dot{M}_{\rm w,-6}\equiv\dot{M}_{\rm w}/10^{-6}$}~M$_\odot$~yr$^{-1}$: the wind mass-loss rate) may be disrupted by the impact of the stellar wind, conditioning the radio-band detection of this population of XRBs. \cite{Perucho12} analysed the effect of the wind clumpiness on the jet dynamics and found significant differences in the flow stability and disruption degree even with jets with luminosities $L_{\mathrm{j}}\sim 10^{37}\dot{M}_{\rm w,-6}$~erg~s$^{-1}$. Numerical calculations of the high-energy emission from one clump-jet interaction were performed by \cite{cita17}, who predicted that collectively such interactions may dominate the non-thermal radiation. \cite{Yoon15} developed numerical simulations to derive an analytic formula for the asymptotic jet bending and applied it to two well-known XRBs: Cygnus X-1 and Cygnus X-3. This work was then extended by \cite{yoon16} and \cite{Bosch16}. More recently, \cite{charlet21} analysed the dynamical and structural effects of radiative losses in the same two fiducial cases. They performed large-scale 3D RHD simulations of jet outbreak and early propagation, finding that radiative cooling effects are more relevant for Cygnus X-3 than Cygnus X-1. Furthermore, \cite{Barkov21} used 3D hydrodynamical simulations to study the combined effect of stellar winds and orbital motion in the scales of the binary and beyond, finding that jets with power $L_{\mathrm{j}}\sim 10^{37}\dot{M}_{\rm w,-6}$~erg~s$^{-1}$ can be disrupted on scales $\sim 1$ AU.

However, all of the aforementioned numerical works did not regard the dynamical effect of one fundamental ingredient: the existence of magnetic fields. As a matter of fact, toroidal magnetic fields may have a relevant role in jet evolution, either favouring collimation through magnetic hoop-stress or acting as a destabilizing agent via the growth of current-driven instabilities \citep[see, e.g.,][]{marti16,perucho192}.  

Jets are thought to be magnetically dominated close to the compact object, but as the jet propagates away from the binary center different MHD processes convert the magnetic energy into kinetic one, accelerating the flow, which may be also wind mass-loaded via mixing produced by small-scale instabilities. At the scale of the binary, which is the main focus of this paper, we expect the kinetic energy to dominate the total jet power, but it is still unclear to what extent even a relatively weak magnetic field can affect the jet dynamical evolution, favouring jet collimation or triggering more instabilities. Moreover, even moderately weak magnetic fields can be locally reinforced, and thus play an important role shaping the non-thermal emission of the jet by affecting particle acceleration, cooling and triggering synchrotron radiation. On the other hand, it is also possible that magnetic dissipation may not be as efficient as generally expected and thus jet magnetic and kinetic/thermal powers are still near equipartition on the scale of the binary.

\JLM{Previously, RMHD simulations have concentrated on the formation of jets from MHD mechanisms \citep{kinney09,tchek11,kinney12,porth13,osorio21} and the morphological characterization of kilo-parsec scale magnetized jets \citep[see e.g.,][and references therein]{marti19}. Different authors focused on the properties of stationary one-dimensional and two-dimensional -axisymmetric- relativistic magnetized jet models \citep{komissarov992,leismann05,mignone05,keppens08,komissarov15,marti153,moya21}. \cite{mizuno07} studied the long-term stability of 3D magnetized spine-sheath relativistic jets. The first high-resolution  numerical simulations of the propagation in three dimensions of a RMHD jet were performed by \cite{mignone10}. Using similar methods, \cite{guan14} studied the propagation of Poynting flux-driven jets in AGNs and provided a detailed analysis of energy conversion. \cite{porth14} also performed the first 3D RMHD simulations of pulsar wind nebulae with parameters most suitable for the Crab Nebula.}



In this work, we have performed the first 3D RMHD numerical simulations of magnetized microquasar jets interacting with a moderate-to-strong stellar wind in a HMXB. Our main goal is to analyse the role of a toroidal magnetic field configuration in the dynamics of a collection of relativistic jets with different parameters, comparing the results with those obtained in \cite{Perucho10} and \cite{Perucho12}. These simulations will be used as a numerical benchmark upon which to build new high-resolution simulations that can be directly compared with observations by radiative post-process calculations. We plan to address this work in the near future.


The paper is organized as follows: in Sec.~\ref{simulations}, we describe the physical scenario that we consider in this work and the numerical setup of the collection of simulations we have performed. We also present \JLM{Lóstrego v1.0}, our new 3D RMHD code, and provide the main technical information about the code configuration we use in this paper. In Sec.~\ref{results}, we present the results of the simulations. In Sec.~\ref{discussion}, we discuss the results comparing our outcome with previous studies of HMXBs jet-wind interactions and analyze different mechanisms of energy conversion. In Sec.~\ref{conclusions}, we summarize the main conclusions achieved.

\section{Simulations}
\label{simulations}


For our numerical simulations, we have developed a new proprietary code named \JLM{Lóstrego v1.0}, a program that solves the conservative equations of special relativistic magnetohydrodynamics in three dimensions using high-resolution shock capturing (HRSC) methods in \JLM{Cartesian} coordinates. This code is fully parallelized with an hybrid scheme with both parallel processes (Message Passing Interface, MPI) and parallel threads (OpenMP, OMP), based on the same configuration as the hydro code Ratpenat \citep{perucho101}. This hybrid MPI+OMP scheme exploits the architecture of modern supercomputers, keeping all the cores busy inside each node with OMP instructions. A description of the numerical methods employed to solve the relativistic magnetohydrodynamics equations and the performance of the new code solving classical tests in RMHD can be found in Appendix \ref{appA}. All simulations were performed in Tirant, the supercomputer at the Servei d'Inform\`atica de la Universitat de Val\`encia, using up to 1024 cores.  \\

\subsection{Physical scenario}

We study the scenario of a relativistic magnetized jet propagating through the clumpy wind of the companion star in a HMXB, similar to those of typical X-ray binaries like Cygnus X-1 \JLM{or LS 5039 (if it really hosts an accreting BH)}. The star is located for simplicity in the plane perpendicular to the direction of propagation of the jet, at $R_{\mathrm{orb}}=3\times 10^{12}$ cm from the compact object position, and no orbital motion is considered since the characteristic simulation time is much shorter than the typical orbital period of the system. The jet is injected with an initial radius $R_{\mathrm{j}}=6\times 10^9$ cm at a distance to the compact object $y_0=6\times 10^{10}$ cm, propagating along the y-axis with an initial mildly-relativistic velocity $v_{\mathrm{j}}=0.55c$. The distance between the injection point and the compact object is enough to guarantee that general relativistic effects (e.g., the curvature of space-time) can be disregarded from these simulations. For the ambient medium, we have considered an inhomogeneous stellar wind with radial velocity field $v_0=2\times 10^8$ \JLM{ $\mathrm{cm~s}^{-1}$}, mass-loss rate \JLM{$\sim 10^{-6}\,M_\odot~\mathrm{yr^{-1}}$}, temperature $T_{\mathrm{w}}\sim 10^4$~K, thermal pressure $p_{\mathrm{w}}\sim 1,6\times 10^{-3}$~\JLM{$\mathrm{erg~cm}^{-3}$} and mean density $ \rho_{\mathrm{w}}\sim 3\times 10^{-15} $ \JLM{$\mathrm{g~cm}^{-3}$} \JLM{(we made the assumption that the mean wind density is roughly constant at the scales of jet outbreak within the binary, i.e., $y<R_{\mathrm{orb}}$)}. These are typical values for a moderate-to-strong stellar wind from a primary OB-type star \JLM{\citep[see e.g.,][]{Perucho08,muijres12,krt14}}\footnote{\JLM{The stellar-wind parameters are based on a simplified model. For example, the wind may still be accelerating when it interacts with the jets, but on the other side, we do not take into account the beaming towards the compact object due to the ionization of the accelerating stellar wind close to the star \citep{molina19,vilhu21}. The characterization of these non-trivial effects, among other complexities of the wind structure, requires dedicated simulations that are beyond the scope of this paper.}}. The stellar wind fills the  computational box from the beginning and we do not account for radiation pressure from the star nor other related physical effects. \JLM{We also consider that the wind is dominated by the kinetic component at the scales of the binary, so we neglect the magnetization of the ambient medium}. We model the plasma as an ideal gas with $\Gamma=5/3$ and we neglect both thermal and non-thermal cooling.


\subsection{Numerical setup}\label{sec:setup}


We have simulated three different magnetized jets in order to study the effect of the magnetic field in the jet evolution inside the binary. The physical parameters of each simulation are listed in Table \ref{table1}. Total luminosities of jet A and jet B are respectively $\sim 10^{35}$ and $\sim 10^{37}$~erg~s$^{-1}$, similar to those studied in \cite{Perucho10}. Jet A and jet B are cold ($h\approx 1$) and kinetically-dominated, such that the internal and magnetic energy fluxes which are injected in the grid represent a small fraction of the total jet luminosity (i.e., $ L_{\mathrm{B}}\approx 0.004\times L_{\mathrm{h}}$, where $L_{\mathrm{h}}$ is the sum of the kinetic and internal energy powers and $L_{\mathrm{B}}$ is the magnetic power). Total luminosity of jet C is similar to jet B, but in this case the jet is set in equipartition between kinetic/thermal and magnetic energy fluxes. For all of our models, the $\beta$ ratio (i.e., the ratio between the average magnetic pressure, $\bar{p}_{\mathrm{m}}$, and the average gas pressure, $\bar{p}_{\mathrm{g}}$), remains close to unit, although magnetic pressure dominates over gas pressure by a factor of $1.5$ in jet C. The initial densities are $\rho_{\mathrm{A}}=0.088~\rho_{\mathrm{w}}$ for jet A, $\rho_{\mathrm{B}}=8.8~\rho_{\mathrm{w}}$ for jet B and $\rho_{\mathrm{C}}=0.88\,\rho_{\mathrm{w}}$ for jet C. Since jet C is more dilute and the total pressure (gas+magnetic) is higher than in jet B, the flow is initially hotter ($h\approx 1.66$).

\begin{table*}
\centering
\caption{Summary of the main parameters of the three jets simulated in this work. L$_{\mathrm{h}}$ is the sum of kinetic and internal power, L$_{\mathrm{B}}$ is the magnetic power, $\beta$ is the ratio between the average magnetic pressure, $\bar{p}_{\mathrm{m}}$, and the average gas pressure, $\bar{p}_{\mathrm{g}}$ and $B_{\mathrm{j,m}}^{\phi}$ is the maximum value of the toroidal magnetic field. }
\begin{tabular}[t]{lccccccc}
\hline
\hline
&L$_{\mathrm{h}}$ [erg~s$^{-1}$]&L$_{\mathrm{B}}$ [erg~s$^{-1}$]&$\beta$&Density [$\rho_{\mathrm{w}}$]&Specific enthalpy [$c^2$]&$B_{\mathrm{j,m}}^{\phi}\, [(4\pi\rho_0c^2)^{1/2}]$\\
\hline
jet A&$10^{35}$&$10^{33}$&1.03&0.088&1.00&0.011\\
jet B&$10^{37}$&$10^{35}$&1.03&8.8&1.00&0.11\\
jet C&$5\times 10^{36}$&$5\times 10^{36}$&1.56&0.88&1.66&1.24\\
\hline
\end{tabular}
\label{table1}
\end{table*}


We perform the simulations in a numerical grid box of physical dimensions $80\times 240\times 80$ $R_{\mathrm{j}}$ (in units of the jet radius at the injection plane) with an initial resolution of 6 cells/$R_{\mathrm{j}}$, so the box involves $480\times 1440\times 480$ cells. The resolution of the grid improves the resolution used in \cite{Perucho10} and \cite{Perucho12} by a factor of $1.5$, but we only consider $75\%$ of the longitudinal length in \cite{Perucho10} to keep the computational requirements under reasonable limits. No grid extensions nor mesh refinement were employed in our simulations. As the jets expand in the numerical box, the effective resolution of the simulation is increased as the jet transversal section involves more computational cells.


\JLM{Time resolved spectroscopic monitoring of hot-stars emission lines supports the idea that the stellar wind is formed by clumps, density inhomogenities created due to thermal instabilities during the wind acceleration process \citep{runacres02,cassinelli08,moffat08}. Moreover, the similarity of the scaling laws between these clumps and the Molecular Clouds favours the interpretation that the clumpy structure of the wind is also a result of turbulence, with scaling laws similar to the fractalized ISM \citep{moffat08}}. \JLM{Thus}, the initial inhomogenous density distribution of the stellar wind is constructed using the publicly available PyFC code\footnote{https://www2.ccs.tsukuba.ac.jp/Astro/Members/ayw/code/pyFC/

index.html} \citep{wagner11}, a Python package that is useful to represent a dense, inhomogenous component embedded in a smooth background. This approach is different from the one adopted in \cite{Perucho12}, where clouds were modelled as gaussian-shape spheres randomly distributed in the computational box. The distribution of inhomogeneities is established by an iterative process following the original work on terrestrial cloud models by \citet{Lewis02}. The three-dimensional density field follows a log-normal single-point statistics in real space, while the fractal structure of the system is achieved by first Fourier transforming and then multiplying the computational box by a power-law with Kolmogorov spectral index $\beta=-5/3$. The mean ($\mu$) of the log-normal parent distribution is $\mu=1.0$ and the variance $\sigma^2=0.6$, such that the minimum density of the wind is $\rho_{\mathrm{min}}\approx0.1\,\rho_{\mathrm{w}}$ and the maximum density is $\rho_{\mathrm{max}}\approx10\,\rho_{\mathrm{w}}$. Based on the computational box dimensions, we set a minimum sampling wave-number $k_\mathrm{min}=3$, so the radius of the largest fractal structure in the cube (i.e., the maximum cloud size) is approximately $R_{\mathrm{c,max}}=6.5$ $R_{\mathrm{j}}$ ($R_{\mathrm{c,max}}<2R_{\odot}$, $\sim 20$\% of the stellar radius).

\begin{figure}
  \centerline{\includegraphics[width=1.0\linewidth]{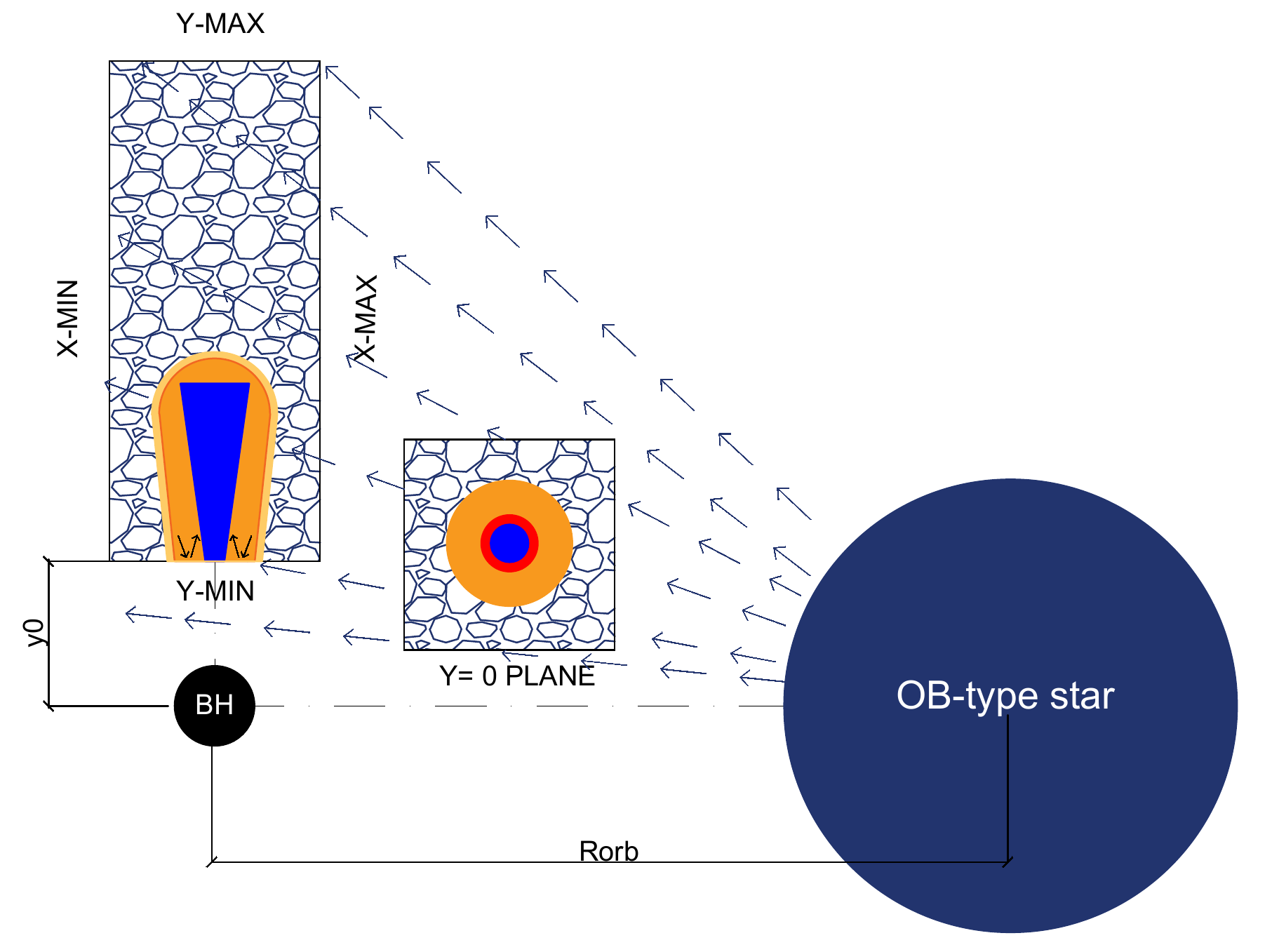}}
  \caption{Scheme of the simulation setup in the XY plane. The radial velocity field that represents the stellar wind fills the box from the beginning. We also show a top view of the injection plane at Y=0, where we distinguish the cylindrical nozzle (blue), the shear layer (red) and the jet cocoon (orange), where boundaries are reflecting.}
\label{scheme}
\end{figure}

In each simulation, the jet is injected in the $y=y_0$ plane (in the system of coordinates with origin in the compact object) as a cylindrical nozzle with $r = \sqrt{x^2 + z^2} < 1$ (where length units are normalized to the jet injection radius). In this plane, density ($\rho$) and velocity ($\boldsymbol{v}$) profiles are given by:
\begin{equation}
\rho(r) = \left\lbrace
\begin{array}{ll}
\rho_{\mathrm{jet}},& 0\leq r \leq 1,\\
\rho_{j=1}, & r>1,
\end{array}
\right.
\end{equation}
\begin{equation}
v^x(r) = \left\lbrace
\begin{array}{ll}
0,& 0\leq r \leq 1,\\
v^x_{j=1}, & r>1 \;\&\;  p_{\mathrm{g}}>p_{\mathrm{c}},\\
-v_{0}\cos{\theta}\cos{\phi}, & r>1 \;\&\; p_{\mathrm{g}}<p_{\mathrm{c}},
\end{array}
\right.
\end{equation}
\begin{equation}
v^y(r) = \left\lbrace
\begin{array}{ll}
v^y_{\mathrm{jet}},& 0\leq r \leq 1,\\
-v^y_{j=1}, & r>1\;\&\;  p_{\mathrm{g}}>p_{\mathrm{c}},\\
v_{0}\cos{\theta}\sin{\phi}, & r>1\;\&\;  p_{\mathrm{g}}<p_{\mathrm{c}},
\end{array}
\right.
\end{equation}
\begin{equation}
v^z(r) = \left\lbrace
\begin{array}{ll}
0,& 0\leq r \leq 1,\\
v^z_{j=1}, & r>1\;\&\; p_{\mathrm{g}}>p_{\mathrm{c}},\\
v_{0}\sin{\theta}, & r>1\;\&\; p_{\mathrm{g}}<p_{\mathrm{c}},
\end{array}
\right.
\end{equation}
where $\theta,\phi$ are the characteristic angles of spherical coordinates with respect to the star location:
\begin{equation}
    \theta=\arctan{\left(\frac{z}{\sqrt{(R_{\mathrm{orb}}-x)^2+(y+y_0)^2}}\right)},\hspace{0.5cm}\phi=\arctan{\left(\frac{y+y_0}{R_{\mathrm{orb}}-x}\right)},
\end{equation}
and the subscript \JLM{$\mathrm{j}=1$} indicates that the values of density and velocity in the y-boundary cells outside the nozzle $r<1$ are a copy of those values in the first plane inside the box. Beyond the cylindrical nozzle (i.e., $r>1$), we distinguish two different inlet boundaries: the jet cocoon, a dynamical region where boundaries are reflecting to simulate the presence of a counter-jet, and the stellar wind, following the radial velocity field with origin in the star location (see Fig.~\ref{scheme}). To discriminate between these two regions during the simulation, the jet cocoon is defined as the set of cells for which the gas pressure $p_{\mathrm{g}}>p_{\mathrm{c}}$, where we choose the threshold $p_{\mathrm{c}}=10^{-8}$ (in units of $\rho_0 c^2$). The azimuthal component of the jet axisymmetric magnetic field in the laboratory frame has the form \citep{lindt89,komissarov99,leismann05,marti153}:
\begin{equation}
B^\phi(r) = \left\lbrace
\begin{array}{ll}
2B^\phi_{\mathrm{j,m}}(r/R_{B^\phi,\mathrm{m}})/(1+(r/R_{B^\phi,\mathrm{m}})^2), & 0\leq r \leq 1,\\
0, & r>1,
\end{array}
\right.
\end{equation}
where $R_{B^\phi,\mathrm{m}}$ is the magnetization radius normalized to the jet radius and $B^\phi_{\mathrm{j,m}}$ is the maximum value of the magnetic field. Thus, the magnetic field grows linearly for $r<R_{B^\phi,\mathrm{m}}$, reaches a maximum at $r=R_{B^\phi,\mathrm{m}}$ and decreases as $1/r$ for $r>R_{B^\phi,\mathrm{m}}$. Hereinafter, we fix $R_{B^\phi,\mathrm{m}}=0.37$ and we choose $B^\phi_{\mathrm{j,m}}$ for each simulation according to the magnetic power of the jet (see Table \ref{table1}):
\begin{equation}
    L_{\mathrm{B}}=2\pi\int_0^{R_{\mathrm{j}}} (B_{\phi})^2v^yr dr.
    \label{Lb}
\end{equation}
In order to avoid the appearance of non-physical magnetic monopoles, boundaries are always free with respect to the magnetic field vector components.
For jets without rotation ($v^{\phi}=0$), the gas pressure profile can be derived from a single ordinary differential equation for the transversal equilibrium across the beam:
\begin{equation}
    \frac{dp(r)}{dr}=-\frac{B_{\phi}^2}{rW^2}-\frac{B_{\phi}}{W^2}\frac{dB^{\phi}}{dr},
    \label{partial}
\end{equation}
where $W$ is the Lorentz factor.
Integrating Eq. \ref{partial} by separation of variables, the gas pressure profile $p(r)$ yields:
\begin{equation}
p(r) = \left\lbrace
\begin{array}{ll}
2\left(\frac{B^{\phi}_{\mathrm{j,m}}}{W(1+(r/R_{B^{\phi},\mathrm{m}})^2)}\right)^2+C,& 0\leq r \leq 1\\
p_\mathrm{a}, & r>1
\end{array}
\right.
\end{equation}
where:
\begin{equation}
    C=p_{\mathrm{a}}-\frac{(B_1^{\phi})^2}{2W^2}(1+(R_{B^{\phi}_{j,m}})^2)\,.
    \label{C}
\end{equation}
In Eq. \ref{C}, $B_1^{\phi}$ represents the value of the toroidal magnetic field evaluated at the jet radius and we have assumed that $B^r=B^y=0$. \JLM{This assumption is supported by two main arguments: (1) an axial component contained in the boundary zone would lead to open field lines and thus to violate the divergence-free condition and (2) the scale of injection is likely beyond the jet acceleration region and so the magnetic field is expected to be mainly toroidal}. The total pressure in $r=R_\mathrm{j}$, $p_\mathrm{a}$, can be derived from the hydrodynamic (kinetic+internal) jet luminosity equation:
\begin{equation}
    L_\mathrm{h}=2\pi\int_0^{R_\mathrm{j}} \rho W(h(r) W-1)v^yr dr,
    \label{Lh}
\end{equation}
where $h(r)$ is the specific enthalpy. The stellar wind enters the grid from the X-MAX boundary, while in all the remaining boundaries of the computational box (i.e., X-MIN, Y-MAX, Z-MIN, Z-MAX), we consider free-flow conditions.

Simulations presented in Sec. 3 are performed with \JLM{Lóstrego v1.0}, a new conservative, finite-volume 3D-(S)RMHD code in \JLM{Cartesian} coordinates. The configuration of the code is based on our testing benchmark results (see Appendix A). We employ a second-order Godunov-type scheme with the HLLD Riemann solver \citep{Mignone09} and the piecewise linear method (PLM) for cell reconstruction, with the VAN LEER slope limiter \citep{vanLeer1974, Mignone06}. The limiter is degraded to MINMOD \citep{Roe86} when strong shocks are detected in order to avoid spurious numerical oscillations around shocks \citep{Mignone06}. When shock-flattening is applied to cell reconstruction, the HLLD Riemann solver is also degraded to the simpler and more diffusive HLL solver. The advance in time is performed using the third-order TVD-preserving Runge-Kutta \citep{shu89} with CFL = 0.2. The relativistic correction algorithm CA2 of \cite{marti152} has been used to correct the conserved variables after each time iteration. In our test simulations, this scheme has been probed to be robust even when the magnetic pressure dominates over the gas pressure by more than two orders of magnitude. The magnetic field divergence-free constraint is preserved with the Constrained Transport (CT) method \citep{evans88,Ryu98,balsara99}, where electromotive forces at cell corners are interpolated following \cite{gardiner05}. An additional equation that describes the advection of a tracer function $f$ is included in the system of equations to indicate the composition of the fluid in every cell of the box as a function of time. This tracer is used to discriminate between jet material ($f=1$), wind material ($f=0$) and those regions where jet has been mixed with the environment ($0<f<1$).



In RMHD jet simulations, the plane of injection is one of the most challenging regions of the grid and it is critical for the long-term survival of the whole simulation. This might become specially relevant when jets are injected as axisymmetric top-hat distributions with low numerical resolution in \JLM{Cartesian} coordinates. For example, we have found that analyzing the first seconds of jet propagation, the axial velocity is slightly perturbed along the jet surface. This perturbation is translated to the toroidal magnetic field, which starts to develop an axial component that grows with time and may lead to local non-physical solutions. \cite{porth13} also described the introduction of a significant amount of noise originated due to the \JLM{Cartesian} discretization and quadrantal symmetry of the grid, that led to the pump of multiples of the $m=4$ mode in all flow quantities \citep[see Sec. 3.2 in][]{porth13}. These pathologies are controlled in our simulations by first, using inhomogenous stellar winds that naturally break the original \JLM{Cartesian} symmetry, and second, smoothing the initial top-hat field distribution. In order to smooth the initial profile to avoid the growth of random perturbations at the jet base, we replace the discontinuous functions that we have described previously in this section by smooth functions of the form \citep{bodo94}:
\begin{equation}
    v^y(r)=\frac{v_\mathrm{j}}{\cosh{(r^4)}},\hspace{1cm}B^{\phi}(r)=\frac{B^{\phi}(r)}{\cosh{(r^4})}\,.
    \label{shear}
\end{equation}
 To guarantee that the injected toroidal field has zero divergence up to machine accuracy, we fix one of the field components according to its analytical expression (smoothed with the shear layer of Eq. \ref{shear}) and calculate the other component numerically using the solenoidal condition (i.e., $\nabla\cdot B=0$), where spatial derivatives are approximated with finite differences. For the sake of completeness, a schematic representation of the whole setup is shown in Fig. \ref{scheme}.


Units are used in which the light speed (\textit{c}), the mean density of the stellar-wind ($\rho_{\mathrm{w}}$) and the jet radius ($R_{\mathrm{j}}$) are set to unity. A factor of $1/\sqrt{4\pi}$ is absorbed in the definition of the magnetic field, so the actual field is smaller by $1/\sqrt{4\pi}$.

\section{Results}
\label{results}

\begin{figure}
  \centerline{\includegraphics[width=\linewidth]{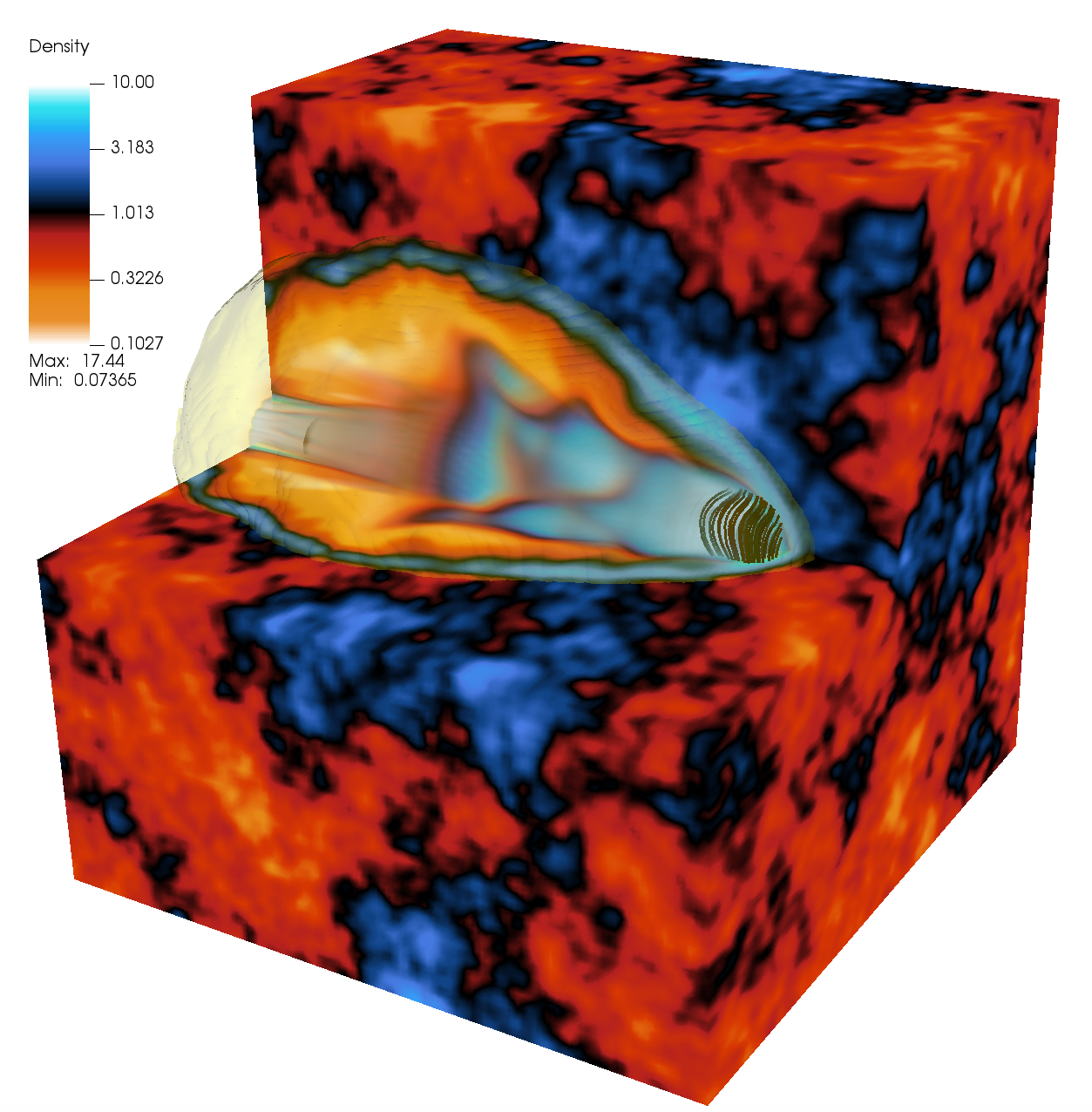}}
  \caption{Propagation of a fiducial jet (jet B) during the first seconds after injection. The dimensions of the box are $20\,R_{\mathrm{j}}^3$, with an effective resolution of 6 cells/$R_{\mathrm{j}}$. The density distribution of the ambient medium is represented in logarithmic scale, where we limited the maximum density to $\rho_{\mathrm{max}}=10~\rho_{\mathrm{w}}$ to highlight the wind clumps. Magnetic field lines are represented in the head of the jet, where density is maximum. The 3D jet render is constructed using the jet tracer function and coloured with the same color scale than for density distribution (although we do not show the tracer legend to avoid redundancy), where jet particles ($f=1$) are represented in light blue/white and the ambient medium ($f=0$) in yellow/white. A gas pressure contour (faint yellow) is also included to show the position of the jet bow shock.}
\label{20rj}
\end{figure}
\begin{figure}
    \centerline{\includegraphics[width=\linewidth]{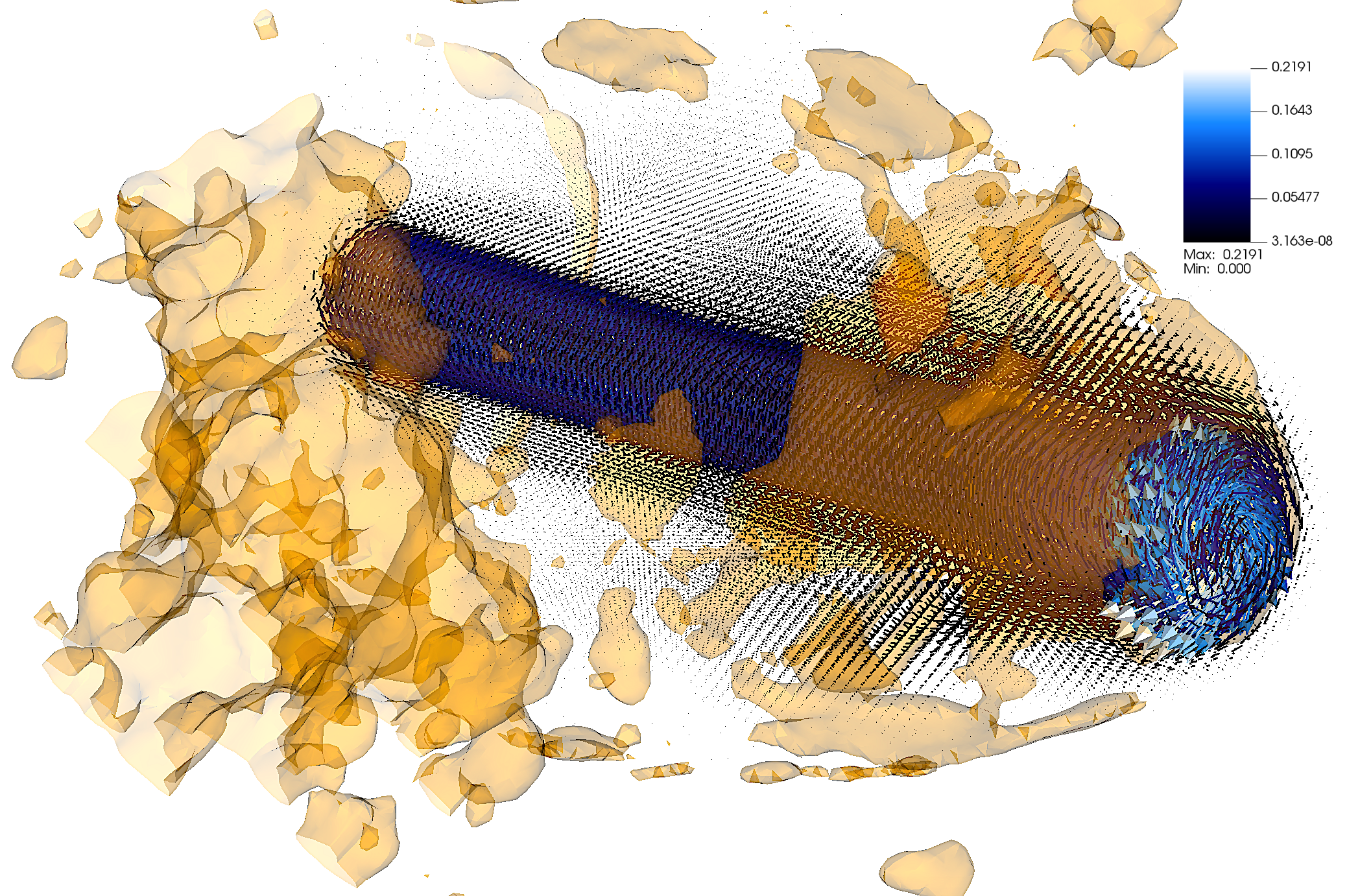}}
  \caption{Toroidal magnetic field vectors in the same time frame as in Fig. \ref{20rj}. Stellar wind clumps are represented with gold isosurfaces at $\rho=3~\rho_{\mathrm{w}}$.}
\label{20rjfield}
\end{figure}

To illustrate the early evolution of a fiducial jet model of our three simulations, Fig.~\ref{20rj} shows the rest-mass density distribution of jet B within a small computational box soon after injection (after the first $20\,R_{\mathrm{j}}$ of propagation through the ambient medium). In the plot, we show a 3D cut of the logarithmic density of the ambient medium, a volume render of the tracer function and a gas pressure contour to show the position of the jet bow shock. When the jet starts to propagate, the highly-supersonic gas generates a strong forward shock that pushes the ambient medium and a reverse shock that decelerates the jet gas at the terminal region (i.e., the hot spot), deflecting off the jet flow backwards forming a hot and light cavity surrounding the jet called cocoon. In the following, we shall talk indistinctly about cocoon, shocked cavity or jet cavity. Surrounding the cocoon, there is a dense shell of shocked gas, which is separated from the cocoon by a contact discontinuity and isolated from the unshocked ambient medium by the bow shock. In these simulations, both the cocoon and the shear layer are subject to the development of irregular structures due to the interaction of the bow shock with the clumps of the stellar wind. In Fig.~\ref{20rjfield}, we show that the toroidal magnetic field is advected following the jet gas through the clumps of the stellar wind, represented with gold isosurfaces at $\rho=3~\rho_{\mathrm{w}}$. Field vectors of Fig.~\ref{20rjfield} have been integrated to represent a collection of magnetic field lines in the hot spot of Fig. \ref{20rj} (black solid lines), where both density of lines and rest-mass density reach their maxima.

\subsection{Jet A}

Jet A propagates up to $y\le 220\,R_{\mathrm{j}}$ at $t=1020\,(R_{\mathrm{j}}/c)$, which corresponds to $y\le 1.3\times 10^{12}$~cm at $t\approx200$~s in physical units. This means that the propagation velocity of the jet head through the stellar wind is \JLM{$v_{\mathrm{j}}\approx 0.21~c$}, slightly higher than the one dimensional relativistic approximation of \cite{marti97}\footnote{The relativistic velocity of the jet head can be estimated based on a one-dimensional hydrodynamical momentum balance across the working surface.}, \JLM{$v_{\mathrm{1D}}\approx 0.15~c$}. This small difference in the velocity of propagation can be explained considering the ballistic nature of jet A, as described later on this section.

\begin{figure*}
\centering
  \includegraphics[width=1.02\linewidth]{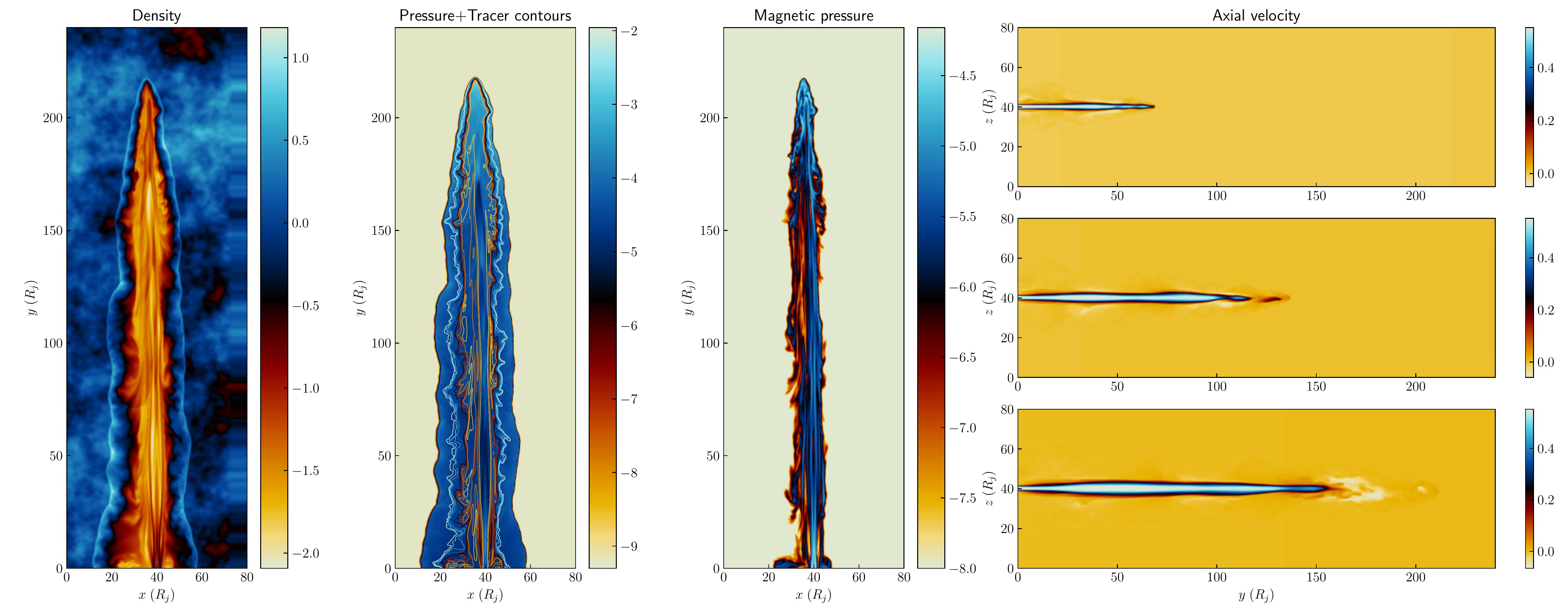}
  \caption{Vertical panels: logarithmic rest-mass density, logarithmic gas pressure and logarithmic magnetic pressure of jet A at t=1020 $(R_{\mathrm{j}}/c)$. Ten tracer contours are over-plotted together with the gas pressure, from $f\approx 0$ (white-to-blue) to $f=1$ (yellow-to-white). Horizontal panels: evolution of the jet velocity at t=310 (top), t=610 (middle) and t=1020 (bottom) $(R_{\mathrm{j}}/c)$.}
\label{panel35}
\end{figure*}

\begin{figure*}
\centering
  \includegraphics[width=\textwidth]{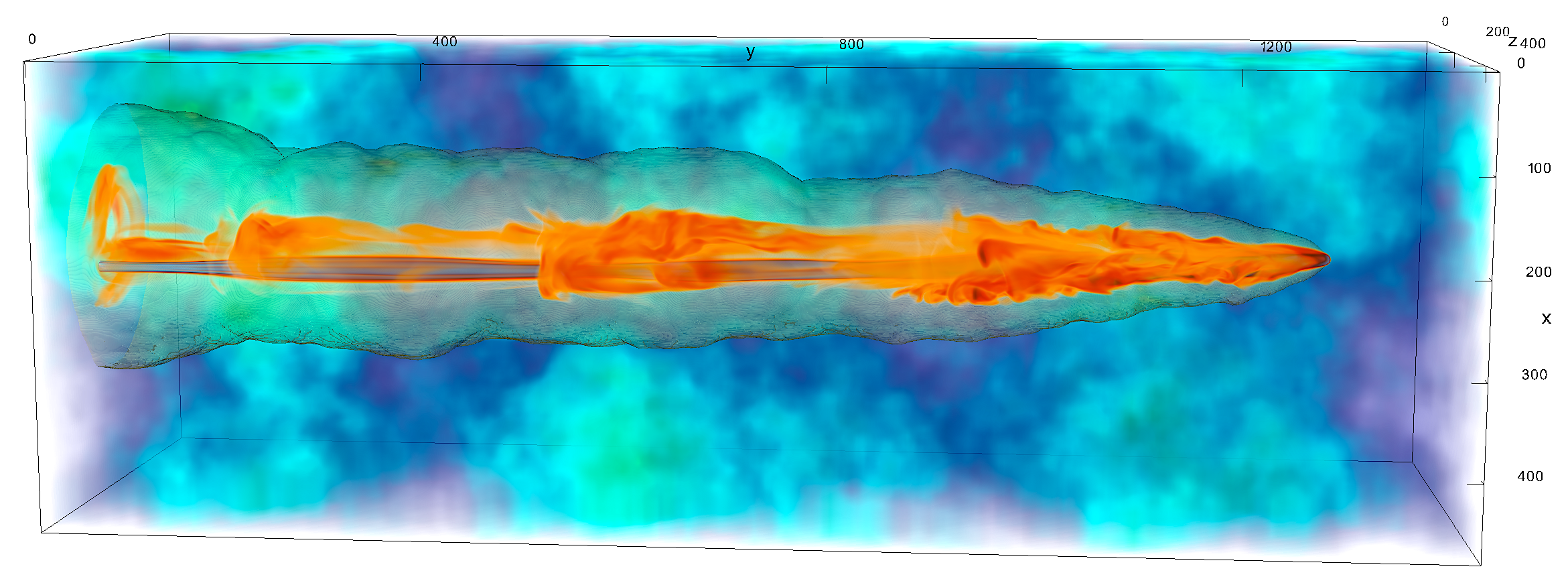}
    \includegraphics[width=0.95\textwidth]{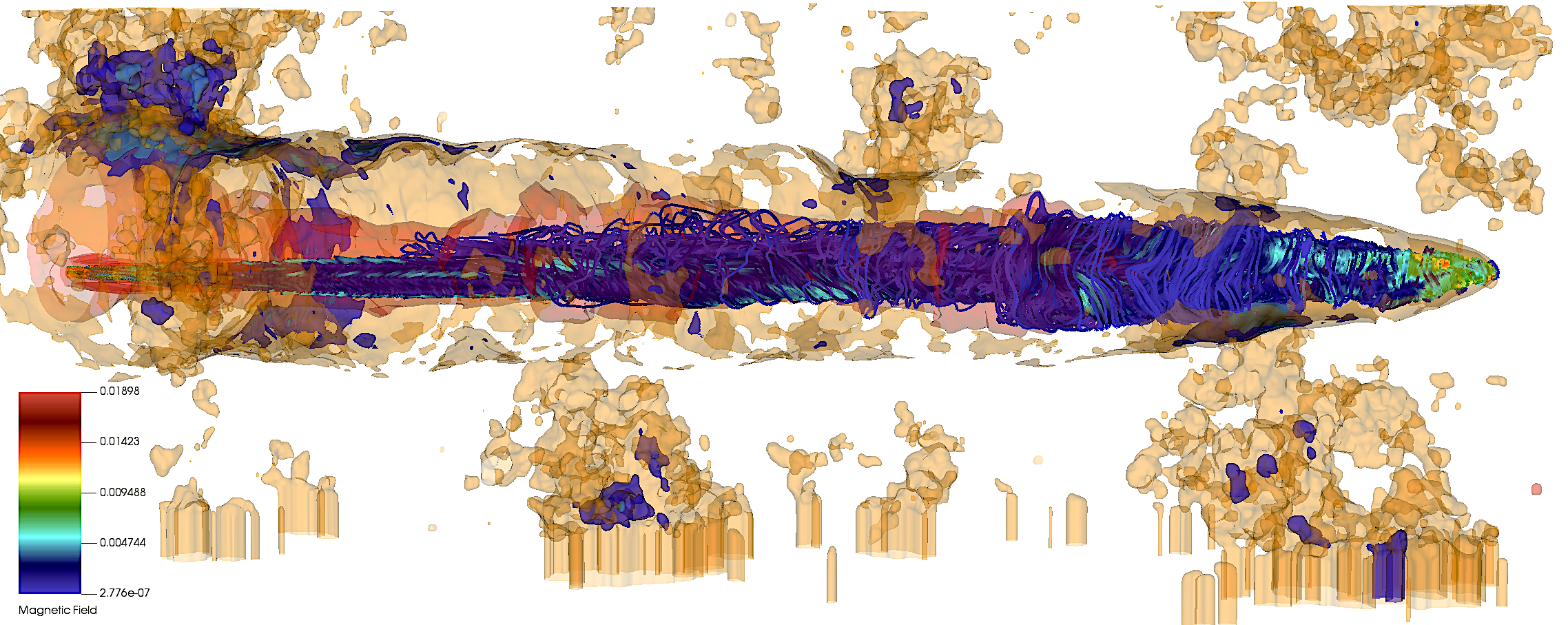}
  \caption{Top panel: 3D render of jet A tracer at t=1010 $(R_{\mathrm{j}}/c)$ and stellar wind clumps. A gas pressure contour (faint yellow) is included to show the position of the jet bow shock. Bottom panel: density isocontours at $\rho=0.1~\rho_{\mathrm{w}}$ (red), $\rho=2.5~\rho_{\mathrm{w}}$ (gold), $\rho=5.0~\rho_{\mathrm{w}}$ (dark blue), $\rho=7.5~\rho_{\mathrm{w}}$ (light blue) together with a collection of magnetic field lines.}
\label{panel353D}
\end{figure*}

On the vertical panels of Fig. \ref{panel35}, we show the logarithms of rest-mass density, gas pressure and magnetic pressure at $t=1020\,(R_{\mathrm{j}}/c)$. A collection of 10 tracer contours are over-plotted together with the logarithm of gas pressure to show the position of the jet core and the composition of the jet cavity. On the horizontal panels of this figure, we show the distribution of axial velocity at three different time frames: $t=310$, $t=610$ and $t=1020\,(R_{\mathrm{j}}/c)$. On the top panel of Fig. \ref{panel353D}, we show a 3D render of the jet morphology and the clumpy wind distribution after jet propagation through the numerical domain, using the jet tracer and the rest-mass density, respectively. We also include a gas pressure contour (in faint yellow) to show the position of the jet bow shock at the last time frame. The overall structure of the jet is highly ballistic along the time scales of the simulation, but the interaction with the denser clumps of the stellar wind deforms the bow shock and produces an irregular shell. Near the base of the jet, the cavity seems to be inflated because of the effect of reflecting boundary conditions inside the shocked gas, producing a small equatorial bulge. The jet central spine shows several pinches due to weak reconfinement shocks, but jet collimation is preserved during the whole evolution. The jet bends slightly to the left (in the XY plane of Fig.~\ref{panel35}) due to the lateral impact of the stellar wind and the subsequent difference in the total pressure on both sides of the jet. Instabilities caused by this asymmetry trigger some degree of mixing between the jet material and the ambient gas at the jet boundaries close to the head. In this simulation, the jet spine mantains the velocity of injection until the terminal region (i.e., \JLM{$v_{\mathrm{j}}\simeq0.55~c$}). On the bottom panel of Fig. \ref{panel353D}, we show a collection of density isocontours that represent the jet cocoon ($\rho=0.1\,\rho_{\mathrm{w}}$, red), the unperturbed low-density clumps of the stellar wind ($\rho=2.5\,\rho_{\mathrm{w}}$, gold) and the clumps compressed by the bow shock at two density levels: $\rho=5\,\rho_{\mathrm{w}}$ (dark blue) and $\rho=7.5\,\rho_{\mathrm{w}}$ (light blue). Toroidal magnetic field lines remain anchored to the jet central spine, whereas the gas that fills the cocoon drives a large scale mildly entangled toroidal-to-helical field structure. At $y\le 80\,R_{\mathrm{j}}$, magnetic field lines near the jet surface are subject to shear effects, which lead to the development of a poloidal component. This effect, which explains the melted appearance of the toroidal field in this model, might be produced due to the combined effect of numerical -possibly non physical- instabilities and the irregular structure of the poloidal velocity profile.

\subsection{Jet B}

\begin{figure*}
\centering
  \includegraphics[width=\linewidth]{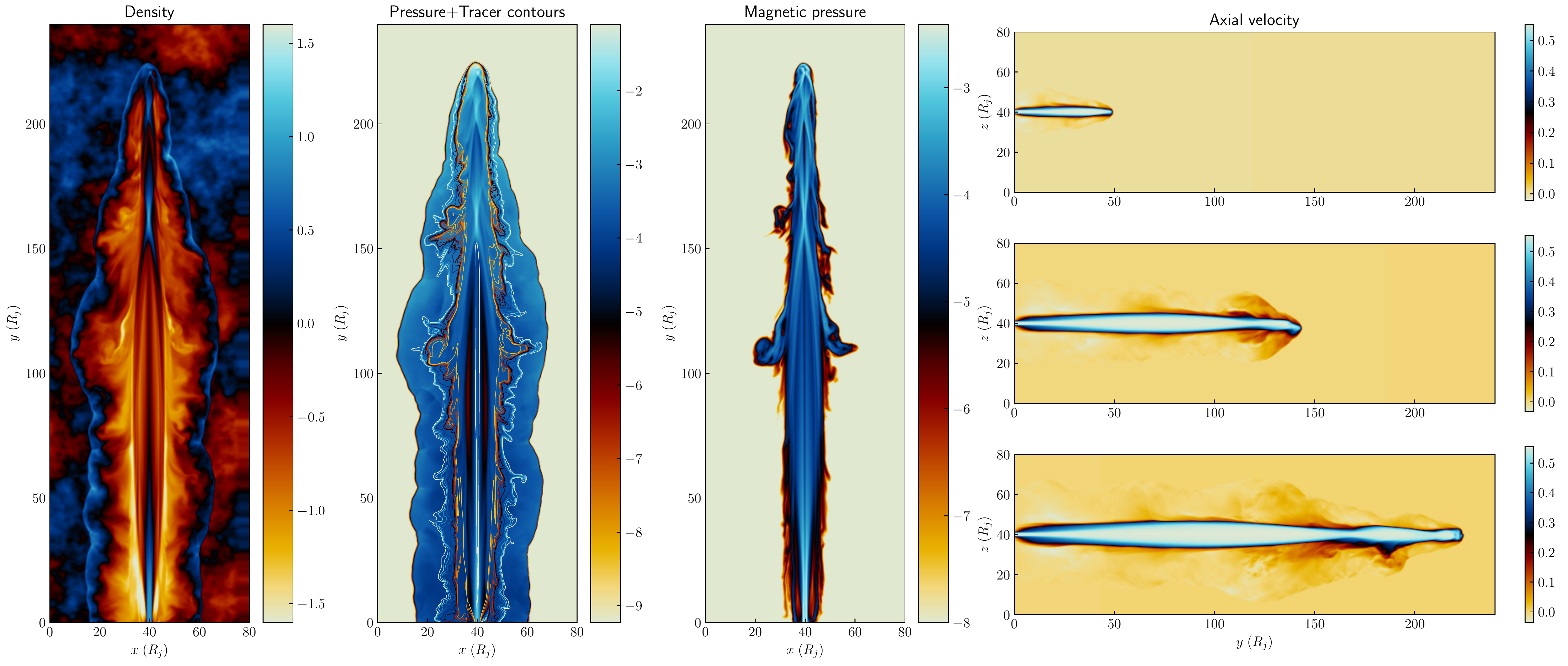}
  \caption{Vertical panels: logarithmic rest-mass density, logarithmic gas pressure and logarithmic magnetic pressure of jet B at t=560 $(R_{\mathrm{j}}/c)$. Ten tracer contours are over-plotted together with the gas pressure, from $f\approx 0$ (white-to-blue) to $f=1$ (yellow-to-white). Horizontal panels: evolution of the jet velocity at t=110 (top), t=360 (middle) and t=560 (bottom) $(R_{\mathrm{j}}/c)$.}
\label{panel37}
\end{figure*}

Propagation of jet A and jet B is similar from a qualitative point of view, although there are important dynamical differences triggered by the asymmetric impact of the stellar wind and the reconfinement shocks produced within the jet. Fig. \ref{panel37} is similar to Fig. \ref{panel35}, but it shows the solution of jet B at $t=560\,(R_{\mathrm{j}}/c)$ and three frames of the jet axial velocity  along its evolution, at $t=110$, $t=360$ and $t=560\,(R_{\mathrm{j}}/c)$. This jet propagates rather ballistically up to $y\le 220\,R_{\mathrm{j}}$ at $t=560\,(R_{\mathrm{j}}/c)$, which corresponds to $y\le1.3\times 10^{12}$~cm at $t\approx110$~s, in physical units. This means that the propagation velocity of the jet head through the stellar wind is \JLM{$v_{\mathrm{j}}\approx 0.39~c$}, in good agreement with the one dimensional relativistic estimate, \JLM{$v_{\mathrm{1D}}\approx 0.43~c$}.

Since the total power of jet B is two orders of magnitude larger than for jet A, the effect of the impact of the lateral wind on the long-term evolution of the jet is clearly smaller and the symmetry of the cocoon is almost preserved. The interaction of the backflow and the shocked ambient medium in the shear layer leads to the development of Kelvin-Helmholtz instabilities and turbulence, but the jet core is unmixed ($f\approx1$) and maintains the velocity of the injection point with good accuracy until the end of the simulation (i.e., \JLM{$v_{\mathrm{j}}\simeq0.55~c$}). The cocoon is more elongated than in jet A because of the faster jet head advance. In this case, the jet is initially denser than the average ambient medium (i.e, $\rho_{\mathrm{j}}=8.8~\rho_{\mathrm{w}}$) and the total pressure is also higher. This produces an initial lateral expansion near the base, with the corresponding drop in total pressure, which finally leads to underpressure with respect to the cocoon and the generation of a recollimation shock far from the injection plane. The position of the shock changes from $y\approx 120\,R_{\mathrm{j}}$ at $t=360\,(R_{\mathrm{j}}/c)$ to $y\approx160\,R_{\mathrm{j}}$ at $t=560\,(R_{\mathrm{j}}/c)$, meaning that it moves with \JLM{$v_{\mathrm{shock}}\approx 0.2~c$}. At the terminal shock, the magnetic pressure also increases due to compression of the field lines. Near the reconfinement shock region, jet material cannot flow freely and accumulates in the head of the jet, leading to backflow blobs of jet gas which drag the magnetic field filling the cocoon (Fig. \ref{panel37}, bottom panel). As the shock moves downstream, the stretched nozzle deposits more plasma in the cocoon and creates a filamentous structure, where $f<1$ because of mixing with the ambient medium gas.

\begin{figure*}
\centering
  \includegraphics[width=\linewidth]{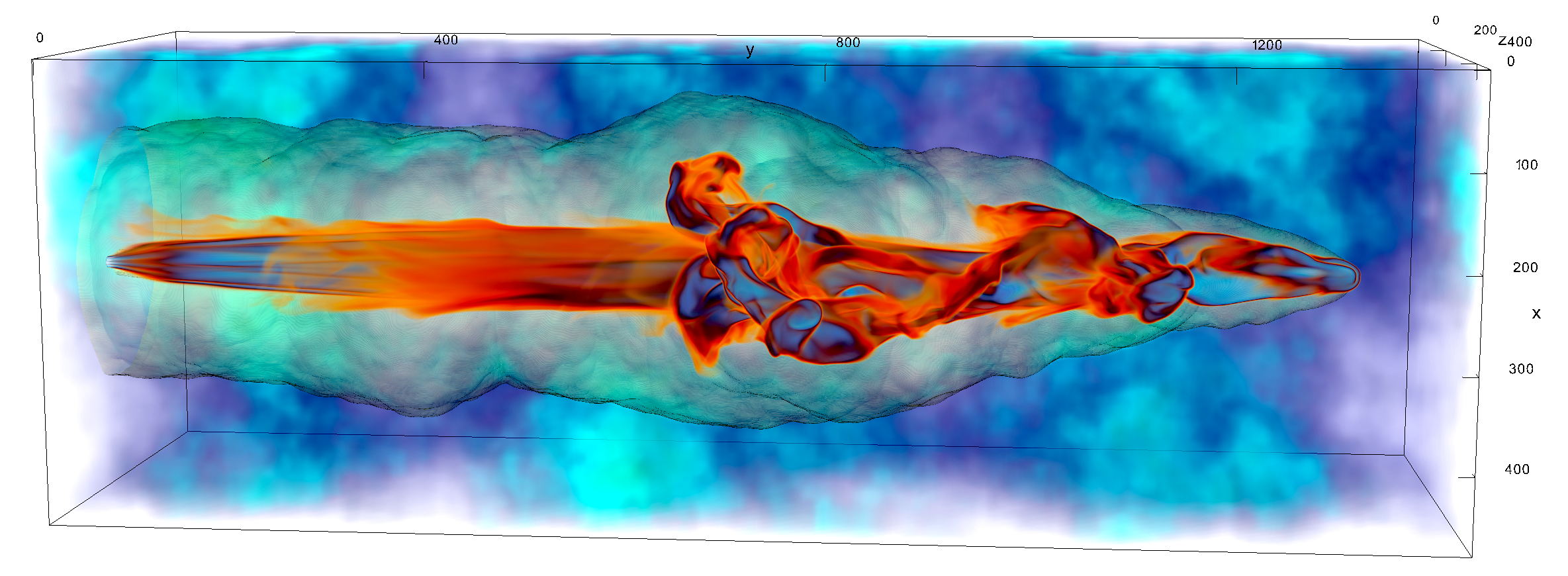}
   \includegraphics[width=0.95\textwidth]{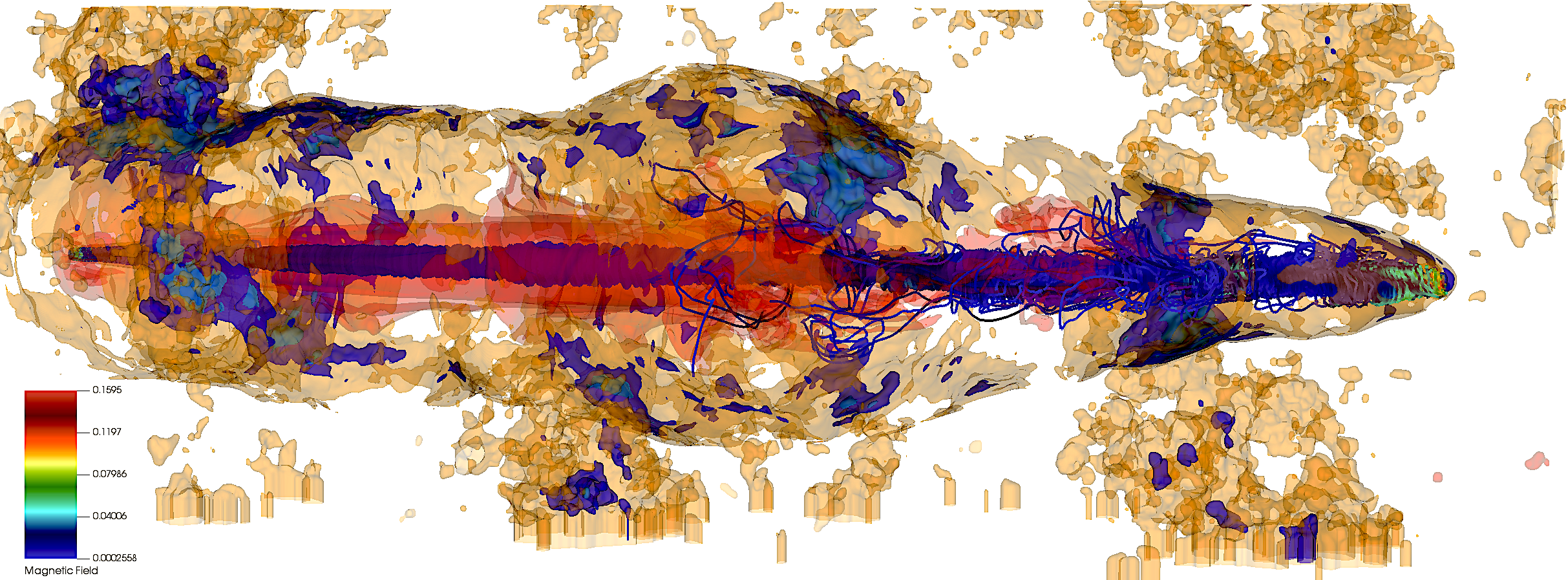}
  \caption{Top panel: 3D render of jet B tracer at t=560 $(R_{\mathrm{j}}/c)$ and stellar wind clumps. A gas pressure contour (faint yellow) is included to show the position of the jet bow-shock. Bottom panel: density isosurfaces at $\rho=0.1~\rho_{\mathrm{w}}$ (red), $\rho=2.5~\rho_{\mathrm{w}}$ (gold), $\rho=5.0~\rho_{\mathrm{w}}$ (dark blue), $\rho=7.5~\rho_{\mathrm{w}}$ (light blue) together with a collection of magnetic field lines.}
\label{panel373D}
\end{figure*}


\subsection{Jet C}

\begin{figure*}
\centering
  \includegraphics[width=1.02\linewidth]{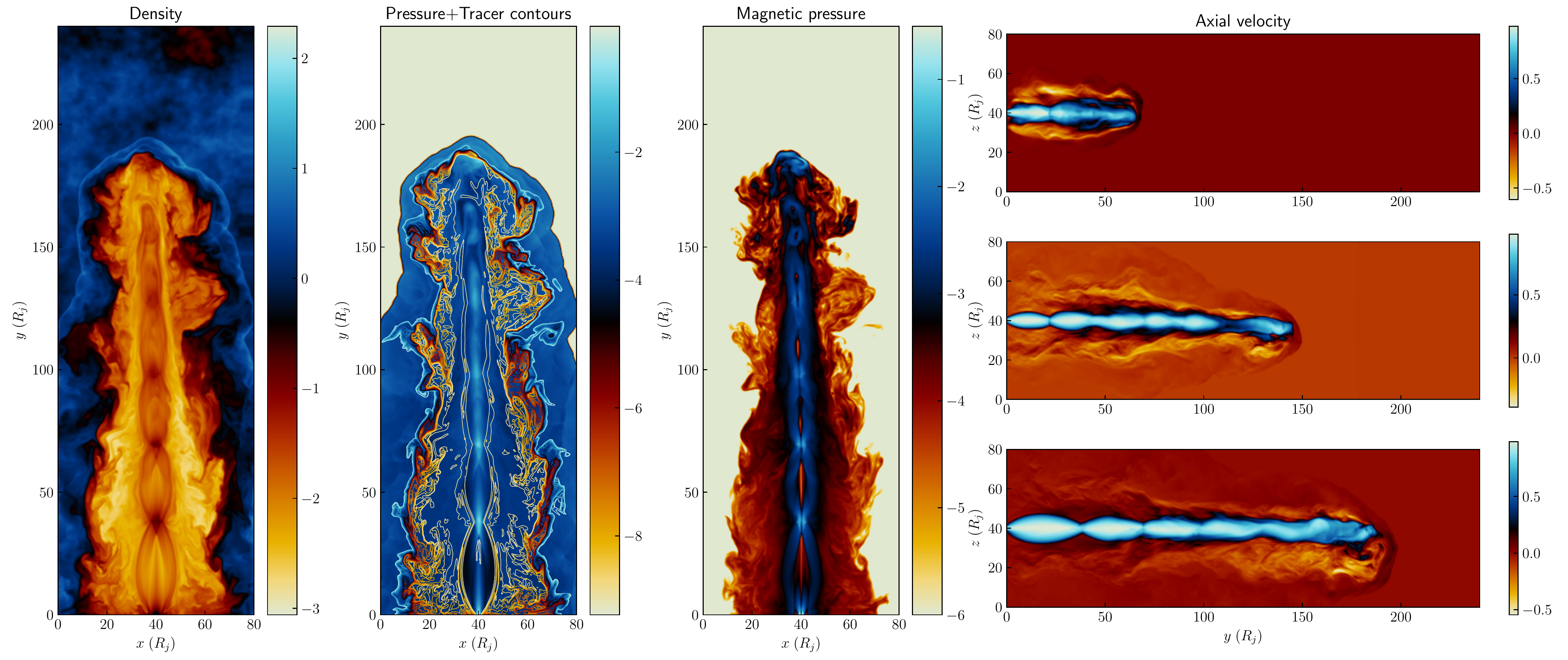}
  \caption{Vertical panels: logarithmic rest-mass density, logarithmic gas pressure and logarithmic magnetic pressure of jet C at t=720 $(R_{\mathrm{j}}/c)$. Ten tracer contours are over-plotted together with the gas pressure, from $f\approx 0$ (white-to-blue) to $f=1$ (yellow-to-white). Horizontal panels: evolution of the jet velocity at t=210 (top), t=490 (middle) and t=720 (bottom) $(R_{\mathrm{j}}/c)$. }
\label{panel37e}
\end{figure*}

Despite total luminosity of jet C is the same as in jet B, there are vast differences in the jet dynamics, evolution and overall morphology triggered by the presence of moderate-to-strong magnetic fields within the jet. Fig. \ref{panel37e} is similar to Figs. \ref{panel35} and \ref{panel37}, but it shows the solution of jet C at $t=720\,(R_{\mathrm{j}}/c)$, and three panels of the axial velocity distribution at times $t=210$, $t=490$ and $t=720\,(R_{\mathrm{j}}/c)$ to illustrate the jet evolution. In this simulation, the jet propagates up to $y\le 200\,R_{\mathrm{j}}$ at $t=720\,(R_{\mathrm{j}}/c)$, which corresponds to $y\le 1.2\times 10^{12}$~cm at $t\approx 142$~s, in physical units. This means that the propagation velocity of the jet head through the stellar wind is \JLM{$v_{\mathrm{j}}\approx 0.26~c$}, which is lower than the one dimensional relativistic estimate of \cite{marti97}, \JLM{$v_{\mathrm{1D}}\approx 0.33~c$}. It is important to note, however, that this simulation cannot be prolonged in time since the bow shock has touched the lateral walls of the grid at $t<720\,(R_{\mathrm{j}}/c)$. Since the magnetic field is initially larger than in the two previous models (as it is required to achieve power equipartition with the energy flux associated to the particles), the initial over-pressure of the jet nozzle with respect to the medium is even larger than in jet B, leading to a pronounced lateral expansion and a quick strong recollimation. During this quick process, the jet flow increases its bulk velocity from the mildly-relativistic value at injection (i.e., \JLM{$v_{\mathrm{j}}\simeq0.55~c$}) to a maximum of \JLM{$v_{\mathrm{j}}\simeq0.95~c$} at the jet core.  The difference with respect to the one dimensional estimate is even more evident if we calculate the velocity of propagation considering the maximum velocity after the first quick expansion (i.e., $v_{\mathrm{j}}\simeq0.95$c), yielding \JLM{$v_{\mathrm{1D}}\approx 0.76~c$} (a factor $\sim 3$ larger than the actual velocity of propagation).

\begin{figure*}
\centering
  \includegraphics[width=\textwidth]{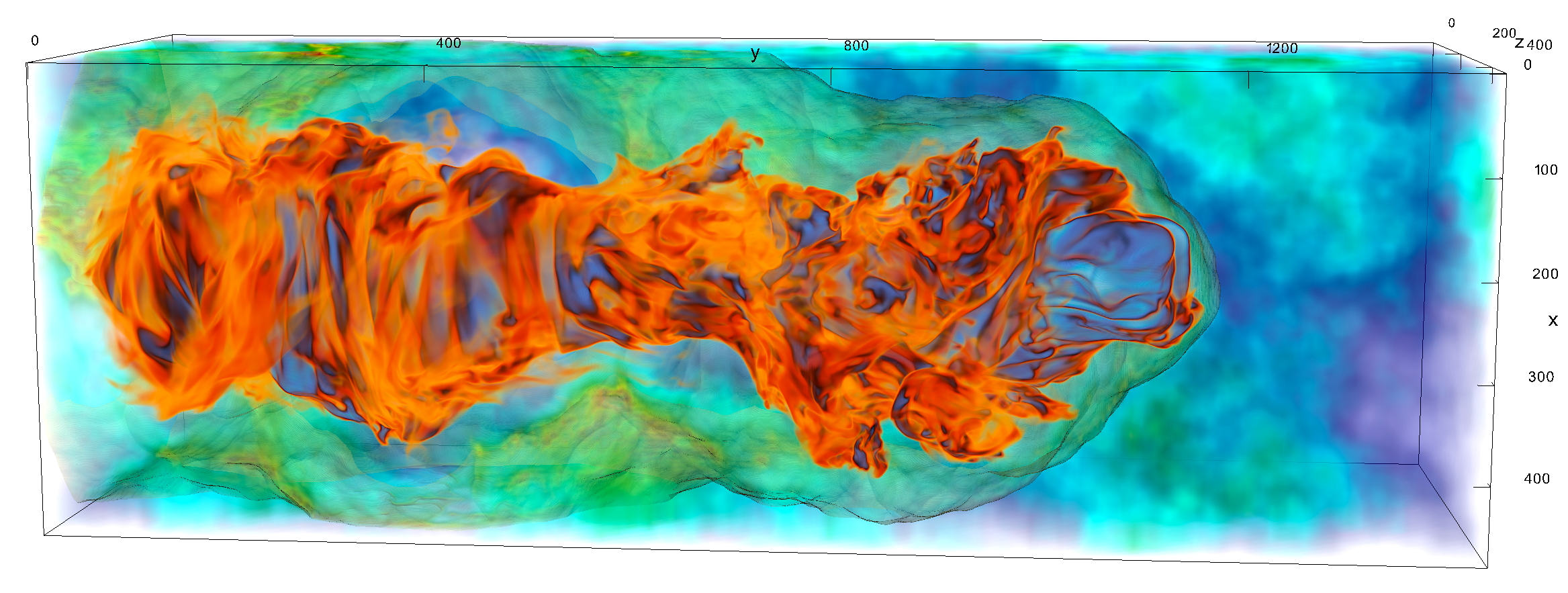}
  \includegraphics[width=0.95\textwidth]{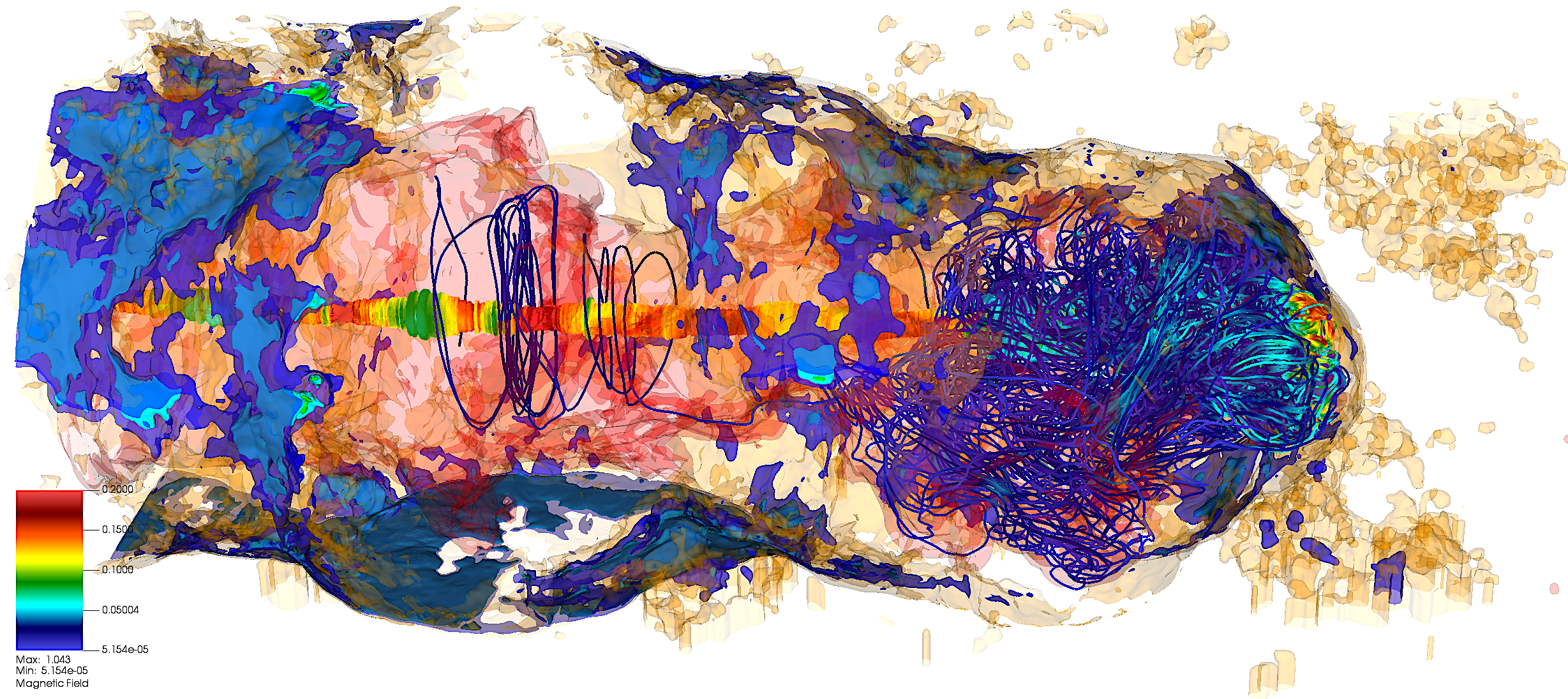}
  \caption{Top panel: 3D render of jet C tracer at t=720 $(R_{\mathrm{j}}/c)$ and stellar wind clumps. A gas pressure contour (faint yellow) is included to show the position of the jet bow shock. Bottom panel: density isosurfaces at $\rho=0.1~\rho_{\mathrm{w}}$ (red), $\rho=2.5~\rho_{\mathrm{w}}$ (gold), $\rho=5.0~\rho_{\mathrm{w}}$ (dark blue), $\rho=7.5~\rho_{\mathrm{w}}$ (light blue) together with a collection of magnetic field lines.}
\label{panel37e3D}
\end{figure*}

The effect of the toroidal field on the jet dynamics manifests from the early jet evolution by the development of current-driven instabilities that destabilise the jet beyond the first reconfinement shock. The consequent increase of the cross-section in which the head deposits its momentum makes the velocity of propagation of the head slower with respect to jet B. The morphology and dynamics of the cocoon are also different than in the other two simulations (Fig. \ref{panel37e3D}, top panel) for this same reason. The first reconfinement shock drives a chain of shocks which are evident all along the jet spine, where magnetic pressure reaches the maximum within the jet due to the compression of toroidal field lines. At $t=720\,(R_{\mathrm{j}}/c)$, at least four reconfinement shocks have appeared after jet evolution, at $y\approx 35\,R_{\mathrm{j}}$, $y\approx 70\,R_{\mathrm{j}}$, $y\approx 100\,R_{\mathrm{j}}$ and $y\approx 130\,R_{\mathrm{j}}$. This means that at the end of the simulation there is a shock every $y\approx30\,R_{\mathrm{j}}$, approximately. The backflow interacts strongly with the shocked ambient medium, developing Kelvin-Helmholtz instabilities that lead to turbulence and mixing with the jet gas within the cocoon ($f<1$). The morphology of the magnetic field lines is also different in the lower body of the cocoon and in the jet head (Fig. \ref{panel37e3D}, bottom panel): at $y\le 100\,R_{\mathrm{j}}$, the magnetic field remains roughly ordered and field lines preserve the toroidal morphology. However, at  $y> 100\,R_{\mathrm{j}}$ the field lines are highly tousled forming an entangled lobe structure around the head of the central axis. This structure of the magnetic field is translated to the morphology of the cocoon material, where gas exhibits toroidal filaments at $y\le 100\,R_{\mathrm{j}}$ but it appears heavily disordered downstream from the chain of shocks (Fig. \ref{panel37e3D}, top panel).

\section{Discussion}
\label{discussion}


\subsection{Jet power and jet morphology}


In the RMHD simulations presented in this work, total jet power has been chosen to be comparable to those of the RHD simulations performed in \citet{Perucho10}. Jet A and  jet B have similar power to Jet 1 and Jet 2, respectively. Focusing first in the case of Jet A compared to Jet 1, we see that, although the magnetized jet propagates up to almost the same distance ($y\simeq 1.3\times 10^{12}$ \JLM{vs.} $y\simeq 1.5\times 10^{12}\,{\rm cm}$), there is a remarkable difference in the amount of time that these two models take to achieve such distance: whereas Jet A needs approximately $200$~s, Jet 1 takes $1250$~s (i.e., a difference of a factor $\sim 6$). The reason for this difference is related with the destabilization and disruption of Jet 1 beyond a moderate-to-strong recollimation shock at $y\simeq 5\times 10^{11}\,{\rm cm}$. The jet flow is strongly decelerated following that shock and thus prone to the fast development of helical instabilities triggered by the impact of the stellar wind. In Jet A, magnetic tension of the toroidal field acts against expansion (furthermore, the gas pressure profile is set up initially to ensure transversal equilibrium with the magnetic pressure), whereas in Jet 1, the jet expands only as determined by the jet-to-cocoon total pressure ratio. 

Although Jet B is also slightly faster than Jet 2 in \citet{Perucho10} (the former reaches $y\simeq1.3\times10^{12}\,{\rm cm}$ in $t\simeq 110$~s, and the latter takes almost twice as much in reaching $y\simeq2\times 10^{12}\,{\rm cm}$), the difference is reduced precisely because Jet 2 is more collimated and develops a very oblique recollimation shock, which does not decelerate the flow enough to favour its disruption, in contrast to Jet 1. Both jets are fairly similar in their evolution, with the minor difference in both advance velocity and location of the first recollimation shock (which is slightly farther downstream in Jet 2), probably due to the longer simulation time in the RHD simulation. Nevertheless, the hydrodynamical jet shows a lower degree of collimation than the RMHD one (see Fig.~\ref{panel373D}). Again, although in this case the magnetic tension does not seem to play such an obvious role as in the case of Jet A, its effect can be crucial for jet evolution and long-term stability. Whether this difference is related to the development of a poloidal sheared component around the jet flow (see bottom panel in Fig.~\ref{panel373D}) should be a matter of study, but it is out of the scope of this paper.

Overall, we find that the presence of a toroidal field, as long as it is non dominant from an energetic point of view, might contribute to stabilize the jet evolution in microquasars. Nevertheless, a stronger magnetic field (as it is the case for Jet C) makes the jet prone to the development of fast recollimation shocks and current driven instabilities, which do not develop in the case of Jet A and Jet B. Therefore, future work should address the threshold from which the magnetic field can become a destabilizing factor for the jet structure at the scales of the binary. This comment is of remarkable relevance in terms of the long-term stability of relativistic jets not only for microquasars, but also for AGNs: a non dominant toroidal field, as expected to be the case beyond collimation and acceleration scales in both microquasar and AGN jets could have a stabilizing role with respect to purely hydrodynamical jets. The reasons being magnetic tension to avoid rapid expansion and large angle recollimation shocks with the consequent flow deceleration, on the one hand, and the possible generation of a sheared poloidal component in the backflow that shields the jet against the development of instabilities, on the other.\footnote{Note that Kelvin-Helmholtz instabilities, mainly helical, can still develop.} 

Interestingly, Jet C is more similar to Jet 1 (which is two orders of magnitude less luminous) than to Jet 2 in \citet{Perucho10} or Jet B in this paper, which stresses the relevance of  the development -or not- of instabilities in outflows in terms of energy dissipation and the subsequent kinematics and morphology. This jet resembles the structures observed in axisymmetric simulations \citep[see e.g.,][]{marti16,moya21} for the case of jets in which the Poynting flux is relevant or dominates the total energy flux. In this kind of models, the presence of a chain of recollimation shocks forced by the toroidal field favours the dissipation of kinetic energy. However, it is important to note that this process takes place after the jet has reached larger velocities than expected in microquasar jets, which is precisely a consequence of the acceleration provided by the presence of strong fields and high internal energies at injection.\footnote{Jets with larger kinetic energies -or inertia- are expected to be less prone to develop disrupting instabilities, at least in the non-magnetized case \citep{perucho05}.} Despite the large inertia achieved by the jet after acceleration, kinetic energy is efficiently dissipated at the series of recollimation shocks. These shocks therefore destabilize the flow not only by themselves but also by making the jet sensitive to the effect of the lateral wind, as reported in \cite{Perucho10} for Jet 1. Whether microquasar jets eventually reach such velocities and are later decelerated at larger distances (beyond the binary system) should be tested by comparison between synthetic radiative output from simulations and observational features from known sources.

Finally, we want to stress that Jet 1 and Jet 2 in \cite{Perucho10} were facing a homogeneous medium, but this is not the case of Jet A and Jet B in this paper. However, our results show that the differences due to the presence of clumps are likely minor at the time scales of our simulations. In contrast, in \cite{Perucho12} the authors adopted a setup in which the jet was established in equilibrium with the ambient medium, a simplified version of the clumpy stellar wind. Indeed, the stage of direct wind-jet interaction is only plausible once the forward bow shock has propagated far enough from the simulated binary region, i.e., at larger time-scales than those studied here. 

\subsection{Energy channels}

The analysis of the energy distribution and the evolution of the energy fluxes along the jet can shed light on the interplay among the different energy channels (i.e., kinetic, internal and magnetic) and, in particular, the possibility of jet acceleration via extraction of magnetic and/or internal energies. Furthermore, it is well understood that the relativistic nature of jets favour that a good fraction of the kinetic and internal energy fluxes can be exchanged very efficiently with the ambient medium, mediated by the presence of strong shocks. This effect was previously described based on 2D axisymmetric models \citep{perucho11,perucho14} and extended to long-term three-dimensional hydrodynamical simulations in \cite{perucho191} for AGN jets evolution. However, this effect has not been explored yet neither with magnetized jet models nor in the particular context of microquasars. 

The accumulated energies in each of the three aforementioned channels can be computed numerically considering the relativistic version of the energy equations and summation over all cells in the 3D numerical box:
\begin{equation}
\begin{aligned}
\tau_k(t)\approx\sum_{\mathrm{cells}}\xi \rho W(W-1)\Delta x\,\Delta y\,\Delta z,\\
\tau_{\epsilon}(t)\approx\sum_{\mathrm{cells}}\xi\left[ \rho\left(\epsilon+\frac{p_{\mathrm{g}}}{\rho}\right)W^2-p_{\mathrm{g}}\right]\Delta x\,\Delta y\,\Delta z,\\
\tau_B(t)\approx\sum_{\mathrm{cells}}\xi\left[ B^2-\frac{B^2}{2W^2}-\frac{1}{2}(\vec{v} \cdot \vec{B})^2\right]\Delta x\,\Delta y\,\Delta z,
\end{aligned}
\label{energies}
\end{equation}
where $\tau_{\mathrm{k}}(t)$ is the kinetic energy, $\tau_{\epsilon}(t)$ is the internal energy, $\tau_{\mathrm{B}}(t)$ is the magnetic energy (all expressed in the observer's reference frame), and $\xi$ represents the jet tracer, $\xi=f$, to calculate the energy of jet particles, or $\xi=1-f$, to calculate the energy transferred to the ambient medium. Finally, $\Delta x, \Delta y, \Delta z$ are the cell-sizes for each spatial dimension. Considering that the wind is stationary and the bow shock has not crossed the outer boundaries of the numerical grid, the sum of the different energies should be equal to the total energy injected by the jet nozzle at a given time, $t$ (plus the internal and kinetic energy of the stellar wind). 





\begin{figure*}
\centering
  \includegraphics[width=0.5\textwidth]{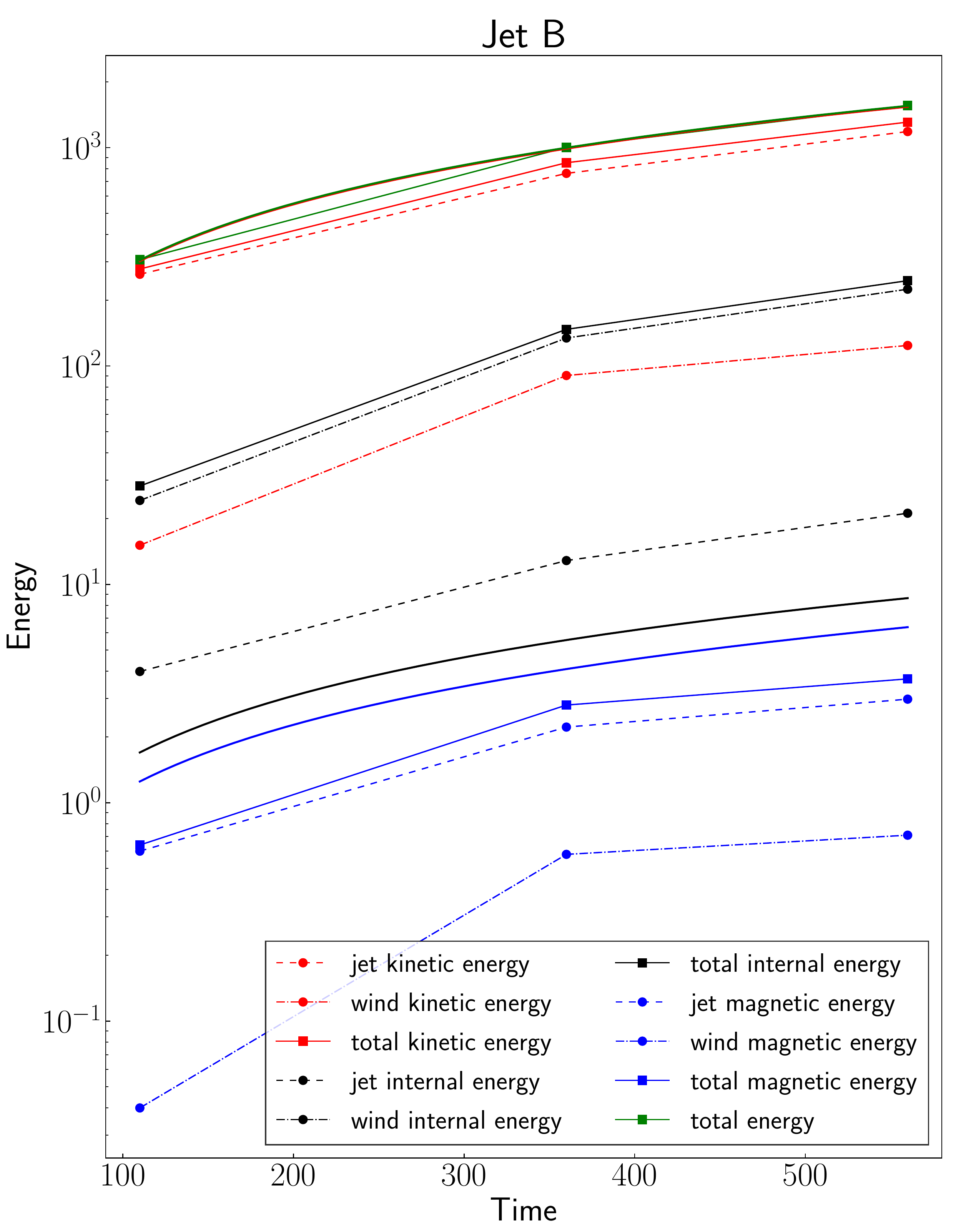}~
  \includegraphics[width=0.5\textwidth]{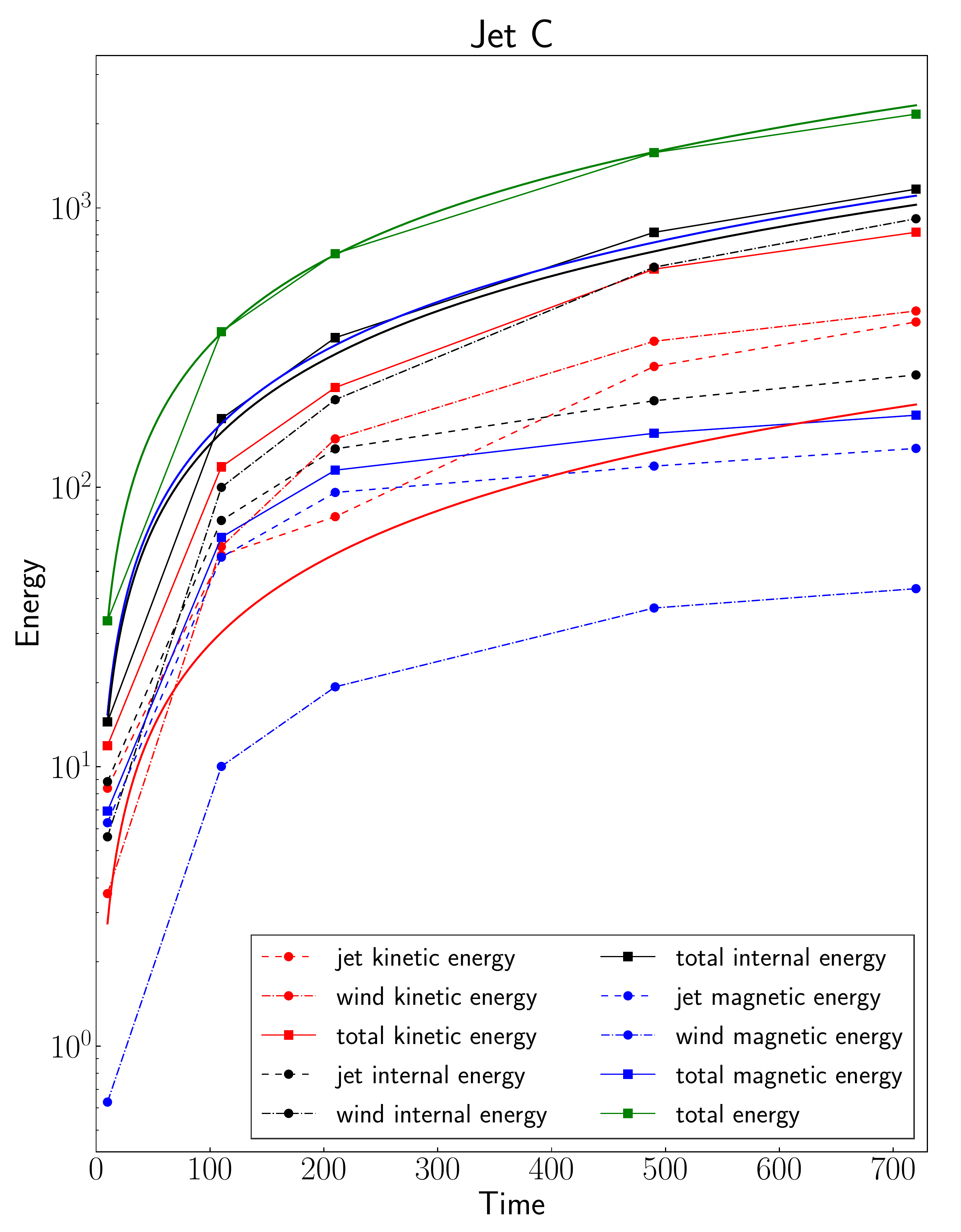}

  \caption{Time evolution of the logarithm of the energy for jet B (left) and jet C (right) simulation at the three time frames of Fig. \ref{panel37} and Fig. \ref{panel37e}, respectively. In the right panel (jet C), we have included two additional time frames, at $t=10~(R_{\mathrm{j}}/c)$ and $t=110~(R_{\mathrm{j}}/c)$. Energy stored by the jet plasma ($f\ne 0$) is represented with dot marks and dashed lines, ambient medium energies ($f=0$) with dot marks and dashed-dot lines and total energies (jet+ambient medium) with square marks and solid lines. Continuous solid lines (with no marks) are the theoretical curves (i.e., the injected values) for each type of energy: kinetic (red), internal (black), magnetic (blue) and total (green). \JLM{Energy units are $\rho_{\mathrm{w}}c^2R_{\mathrm{j}}^3$}.}
\label{energies-t}
\end{figure*}

For the purposes of this section, we will restrict our analysis to jet B and jet C simulations, since these two models have an equivalent total jet power, but the energy flux is distributed in a different way among the three energy channels; while in jet B the kinetic energy flux dominates over the internal energy flux and the magnetic energy flux by almost two orders of magnitude, in jet C the magnetic flux is initially in equipartition with the sum of internal and kinetic energy fluxes (whereas the internal energy flux is also a factor $\sim 5$ larger than the kinetic energy flux). Fig.~\ref{energies-t} shows the time evolution of the logarithm of the different types of energies for jet B (left panel) and jet C (right panel) at $t=110,360,560\, (R_{\mathrm{j}}/c)$ and $t=10,110,210,490,720\, (R_{\mathrm{j}}/c)$, respectively. In the case of jet C, we have included two additional time frames to those shown in Fig.~\ref{panel37e} in order to analyse energy conversion during the early moments of jet propagation, because some of these processes happen very quickly after the injection of the jet. In Fig.~\ref{energies-t}, solid lines (without marks) show the logarithm of the energy injected in the computational box as a function of time. We remind the reader that the inclusion of a shear layer to smooth the initial top-hat axial velocity and magnetic field profiles as described in Sec.~\ref{sec:setup} implies that the eventual injected power is lower than the theoretical value given at Table~\ref{table1}. This is translated in a power reduction of a $\sim 20\%$ for Jet B and $\sim 10\%$ for Jet C simulation. The difference in the percentages between both models can be explained because kinetic energy flux is more sensitive to our shear layer than the magnetic and internal energy fluxes.

As said before, the sum of the jet and ambient medium energies accumulated in the numerical grid must be equal to the injected energy plus the original internal and kinetic energies of the ambient medium. The {small difference (a few percent; see next paragraphs)} comes from the use of a correction algorithm for the conserved variables (see the Appendix) and, more importantly, the low resolution of the numerical representation of the circular injection nozzle by a Cartesian grid. On the other hand, assuming that no energy conversion between the different channels (internal, magnetic and kinetic) takes place within the jet, and that the only transfer of energy within each of those channels is that between the jet and the ambient medium, the sum of the jet and the ambient energies (solid lines with square marks) should be close to the theoretical injected energy (solid lines) for each energy channel. Thus, the difference between the numerical values for the total energy in each channel and the theoretical curves implies that different physical processes of energy conversion are operating during jet evolution. We describe these processes for the two jet simulations in the remaining paragraphs of the section.

\paragraph{Jet B:}

In the left panel of Fig.~\ref{energies-t}, we show that total energy (green solid line with square marks) is roughly conserved during the evolution from $t=110$ to $t=560\, (R_{\mathrm{j}}/c)$, since the three points we consider for this analysis overlap (with an error less than 1\%) with the total injected energy (green solid line with no marks). Although the total kinetic energy (red solid line with square marks), total internal energy (black solid line with square marks) and total magnetic energy (blue solid line with square marks) do not coincide with their respective injected energies (because of effective energy conversion), the initial global energy distribution is preserved: kinetic energy dominates the whole evolution and magnetic energy remains residual (below 1\% of the total energy) even when the jet is fully developed. Jet magnetic energy (blue dashed line with dot marks) is lower than the injected value even at early $t=110 \, (R_{\mathrm{j}}/c)$ (we have also checked that this is also true for earlier times).
Discarding a conversion of magnetic energy into kinetic energy (since the jet is kinetically dominated), the origin of the difference between the total magnetic energy and the injected one can be attributed to the numerical resistivity of the code, which can produce continuous conversion of magnetic into internal energy. 

On the other hand, blue points connected with dash-dot lines shown in the left panel of Fig.~\ref{energies-t} represent the total magnetic energy accumulated in the ambient medium. Optimally, this energy would have to be zero since in ideal MHD there is no physical mechanism to transfer magnetic energy from the jet fluid to the non-magnetized ambient medium. However, it is important to note that the magnetic energy assigned to the ambient medium comes mainly from numerical cells with a mixture of jet and ambient fluids, where according to Eq.~\ref{energies} the magnetic energy is distributed -arbitrarily- proportional to the corresponding mass fraction. Then, the amount of magnetic energy accumulated in the ambient medium ($\sim 10$\% of the injected magnetic energy at the end of the simulation) may be interpreted as an ample upper bound of the magnetic diffusion of our numerical code.


In this simulation, the main manifestation of energy transfer concerns the conversion of jet kinetic energy (red dashed line with dot marks) into kinetic and internal energy of the ambient medium (red and black dashdot lines with dot marks, respectively), mediated by the propagation of the bow shock, and into internal energy of jet plasma in the shocked cavity (black dashed line with dot marks). At the end of the simulation, the jet plasma preserves more than $70\%$ of the injected kinetic energy. By the same time, the transferred portion ($\sim 30\%$) has been used to accelerate ($\sim 10\%$) and heat ($\sim 20\%$) the shocked stellar wind, and to increase the total internal energy of the jet plasma by a factor $\sim 2$.


\paragraph{Jet C:}

In the right panel of Fig. \ref{energies-t}, we show the energy distribution for Jet C simulation between $t=10$ and $t=720 ~(R_{\mathrm{j}}/c)$, where we have included two additional time frames at $t=10$ and $t=110 ~(R_{\mathrm{j}}/c)$ with respect to those shown in the horizontal panels of Fig.~\ref{panel37e}. The total energy is conserved during the evolution with an error smaller than $1\%$,\footnote{This error increases up to $8\%$ in the last time frame because some fraction of the shocked gas has left the box through the lateral boundaries.} whereas the magnetic energy assigned to the ambient medium by the end of the simulation is less than 5\% of the total injected magnetic energy.

The evolution of the energy distribution in this simulation is strongly conditioned by the physical processes operating at very early times. At $t=10~(R_{\mathrm{j}}/c)$, magnetic/internal and kinetic energies are already lower and higher than their injected values, respectively, with $\tau_{\mathrm{B,jet}}/\tau_{\mathrm{B,inj}}\approx 0.40$, $\tau_{\epsilon,\mathrm{jet}}/\tau_{\epsilon,\mathrm{inj}}\approx 0.60$ and $\tau_{\mathrm{k,jet}}/\tau_{\mathrm{k,inj}}\approx 3$, meaning that about $60\%$ of the jet magnetic energy and $40\%$ of the internal energy have been converted into jet kinetic energy by that time.  Thermal acceleration occurs in hot, expanding flows by the Bernoulli process. Magnetic acceleration, on the contrary, needs a particular expanding flow geometry \citep[differential collimation; see e.g.,][and references therein]{vlahakis04,vlahakis042,komissarov11}. According to the results of \cite{komissarov07} for axisymmetric models, magnetic acceleration is very efficient with as much as $\sim 75 \%$ of the Poynting flux converted into kinetic energy at long enough spatial scales. The fast expansion suffered by Jet C close to the injection nozzle (see Fig.~\ref{panel37e}) could provide the correct dynamical and geometric conditions to trigger the magnetic acceleration process, which would have been then observed in microquasar jets simulations for the first time and at much shorter spatial scales.

\begin{figure}
\centering
  \includegraphics[width=1.0\linewidth]{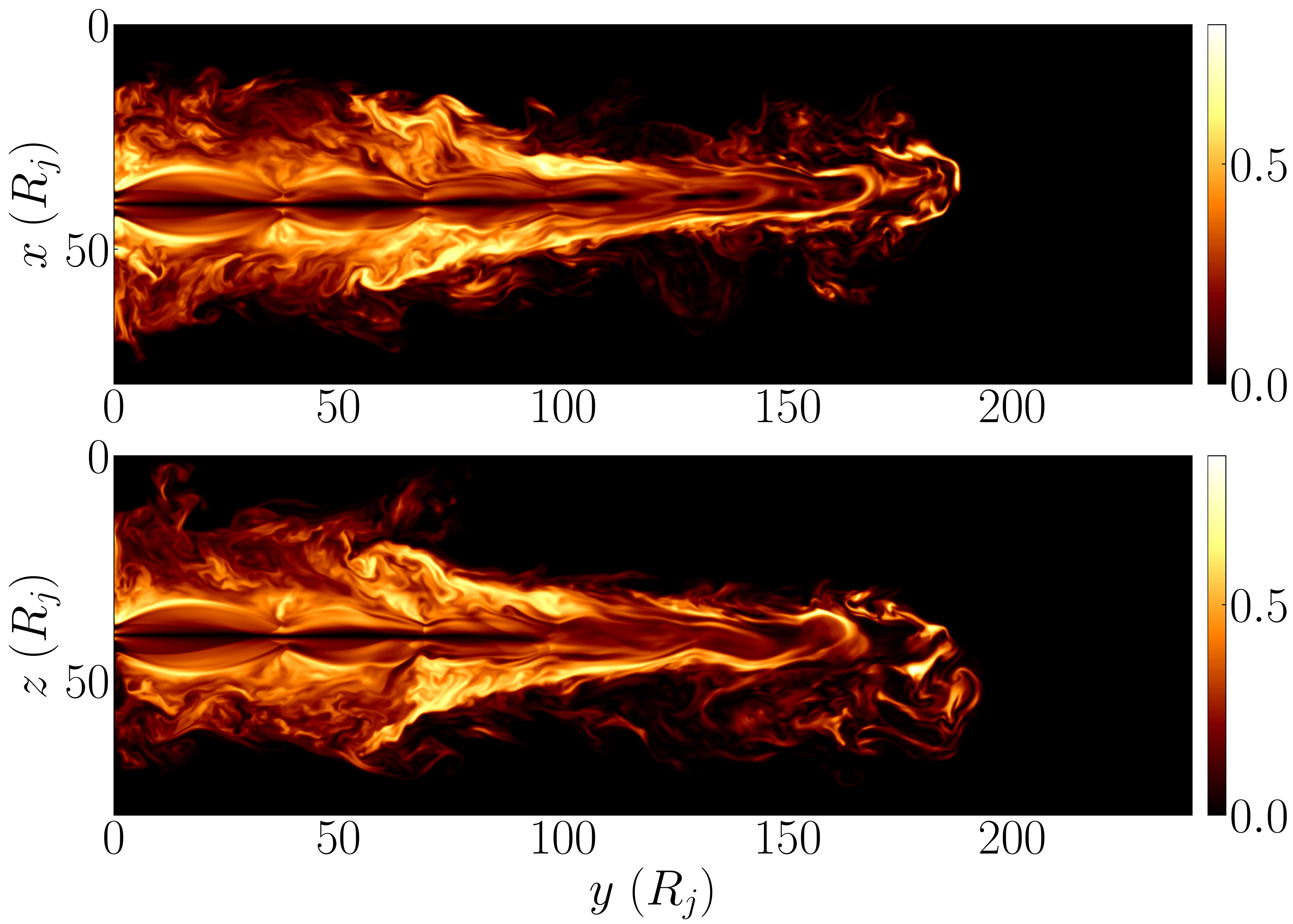}
  \caption{\JLM{Fraction of magnetic energy density with respect to total energy density for jet C in the XY (top) and ZY (bottom) plane.}}
\label{ubc}
\end{figure}

The plot of the evolution of the energy accumulated in the different channels, with the curves overtaking each other several times, reveals the complex nature of jet propagation in model C. As an example, let us note that the jet internal energy dominates over the magnetic and kinetic ones between $t=10~(R_{\mathrm{j}}/c)$ and $t=210~(R_{\mathrm{j}}/c)$,
but at $t=490~(R_{\mathrm{j}}/c)$ it has exchanged its position with the kinetic energy. Additionally, the figure shows that the magnetic channel loses energy steadily with time. The decrease of magnetic energy can take place within the jet at the magnetic acceleration sites after the reconfinement shocks (where the flow expansion could admit the differential collimation), but also at the magnetically dominated regions of the turbulent shocked cavity (see Fig. \ref{ubc}).

Finally, we observe a relevant difference between models B and C in terms of kinetic and internal energy distribution: in the case of jet B, most of the energy in the grid is hosted by the jet matter, whereas the opposite seems to be the case for jet C. It has been shown in a series of papers \citep{perucho11,perucho14,perucho191} that relativistic jets are very efficient in transferring energy to the ambient medium. In \cite{perucho17}, the authors attributed this efficiency to the relativistic nature of jets and showed that massive, cold jets are less efficient in the process of energetic transfer because a large fraction of the energy flux is in the form of kinetic energy dominated by the mass of the particles. This is in agreement with the results derived in the present paper, where jet B is significantly denser and colder than jet C, which makes it highly inefficient with respect to jet C in terms of energy transfer to the ambient medium. This shows the relevance of the relativistic nature of outflows in the feedback process.


\subsection{\JLM{Radiative} processes in the magnetized jet}

As discussed, for instance, in \cite{rieger07} and \cite{bosch12b}, particle acceleration is expected to be more efficient in strong shocks in microquasar jets, with electrons and positrons typically cooling much more efficiently than protons and nuclei \citep[see, e.g.,][]{bosch09}. In these systems, strong magnetic field dissipation through magnetic reconnection could also be relevant for particle acceleration  \citep{bosch12a,bosch12b}.

With respect to hydrodynamical shocks, the strong recollimation shock(s) in relativistic jets can be a perfect candidate for generating a significant non-thermal population of particles. The non-thermal emission related to the jet-wind interaction has been discussed in previous works \citep[see, e.g.,][and references therein]{molina19}, although in all but one case \citep[i.e.,][where the one clump-jet interaction was simulated using relativistic hydrodynamics]{cita17} jet dynamics was semi-analytical. Nevertheless, in these works the magnetic field was not informed using devoted RMHD simulations, but assuming either a fixed magnetic-to-internal energy density ratio, or establishing frozen-in conditions in the hydrodynamical flow. Therefore, in this context, the simulations presented here provide with new insights into the non-thermal processes of HMXB jets. 

The most straightforward conclusion that one can derive from the results of this work is that synchrotron emission and inverse Compton (IC) may be anti-correlated space-wise in the jet. Non-thermal electrons in highly magnetized regions may emit more synchrotron and less IC radiation, whereas in non-strongly magnetized regions it may be the opposite  \citep[see, e.g.,][for subtle effects of synchrotron and IC losses on the radiation spectra]{khangulyan05}. All this applies in particular to electrons emitting synchrotron X-rays or very high-energy IC photons, as they are more sensitive to radiative cooling versus advection in the jet. 

On the other hand, our results for jet C show more shocks than in jet B. In these shocks, kinetic energy turns into internal energy, favouring plausible particle acceleration within the scales of the binary. Magnetic energy is also transferred to matter rather efficiently through magnetic acceleration at favourable expansions. Magnetic acceleration introduces farther complexity to radiation coming from the region where this occurs through a complex pattern of Doppler boosting. Another non-thermal process linked to the magnetic field, and difficult to capture here but which could be relevant as well in highly magnetized jets, is magnetic reconnection. Magnetic energy can be injected into particles by reconnecting magnetic lines, which can trigger (multiple) shocks in the region where this occurs. These non-thermal processes and others also possibly relevant (e.g., acceleration driven by turbulence or shear motion, gamma-ray reprocessing in the complex jet-wind environment, etc.), are very difficult to quantify unless specific detailed studies are carried out, which is beyond the scope of this work. 

As shown, jets might be more stable if moderately magnetized. If so, they could reduce the amount of energy transferred to non-thermal particles within the binary system. However, as suggested in \cite{Bosch16} and shown by \cite{Barkov21}, the impact of orbital motion in longer time-scales than simulated here can further destabilize the jet, and so most of the energy dissipation may not take place within the system, but farther out, where orbital motion strongly shapes the jet evolution \citep[see][for an example of such  situation]{molina18}.

If the jet kinetic and magnetic powers are roughly similar, or the latter dominates over the former, the jet may get severely disrupted on the scales of the binary. This may seem incompatible with milliarcsecond observations of the jet of Cygnus~X-1 \citep{stirling01}, which show that the jet is still rather collimated on scales of tens of orbital separation distances. However, the combined effects of the stellar wind and orbital motion may lead to the binary system acting as a nozzle for the highly magnetized, disordered jet, allowing the jet to reaccelerate and potentially recollimate in relatively short time scales given the fast drop of density and pressure outside that region. In addition, the magnetic field of the jet may act as a protective shield against mixing with the stellar wind, which would allow the jet reaching again relativistic speeds if the environment and/or the magnetic field were appropriate \citep[e.g.,][]{komissarov11}. Such an effect is not expected from a hydrodynamical jet, which could hardly become relativistic again due to heavy wind-entrainment \citep{Bosch16,Barkov21}. 

Depending on the magnetic field strength, the jet will present different emission properties. Specifying these differences for comparison with observations requires specific simulations covering regions larger than the orbital separation distance. Simulations of quasi-steady state jets would be also needed, as they are complementary to the present ones because the timescales involved in observations are often long enough to capture that jet regime. Moreover, the magnetic flux in the jet may be different depending on the particular accretion-ejection state of HMXBs. This different magnetization in the ejections, which would be continuous or discrete depending on the state of the source, would be a major factor (in addition to the flow energetics) determining the properties of the non-thermal emission. Such a possibility requires a detailed analysis of the observed outflow properties that is left for future work.

\JLM{Nevertheless, recent large-scale 3D hydrodynamical simulations \citep{charlet21} show that, for two fiducial HMXBs (namely, Cygnus X-1 and Cygnus X-3), thermal cooling (i.e., free-free losses) could dominate the emission over non-thermal mechanisms in the beam and inner cocoon. In particular, the authors analysed the jet outbreak and early propagation beyond the scale of the binary, and conclude that these radiative losses do not play a dynamical or structural role for the space of parameters in Cygnus X-1, although they can be relevant for the evolution of Cygnus X-3 due to the denser stellar winds of the Wolf-Rayet primary. Thus, according to our models and the time scales of our simulations, we do not expect that (non)-thermal cooling produces a significant effect on the relativistic jet dynamics, but the characterization of the system emission would require a dedicated analysis which is beyond the scope of this paper.}

\section{Conclusions}
\label{conclusions}


We have performed three different numerical simulations of microquasar jets propagating through an inhomogeneous stellar wind, using the new code \JLM{Lóstrego v1.0} for computations in relativistic astrophysics. Our numerical setup is based on the space of parameters described in previous 3D hydrodynamical jet-wind simulations, including a toroidal magnetic field in transversal equilibrium with the gas to analyze its role for jet dynamics and long-term stability. We present the jet evolution throughout the initial minutes after the trigger of jet formation, which allows us to keep a steady injection point in the grid, neglecting the orbital motion of the binary.

Our simulations show that the magnetic field could play a stabilizing role in jet evolution, as long as the flux of magnetic energy is low compared to the kinetic energy flux. This stabilizing effect provides the simulated, magnetized jets, with extra collimation along the grids, as compared to similar, non-magnetized jets. On the contrary, a non negligible magnetic energy flux translates into a destabilization of the flow, with the consequence of increased energy dissipation and a slower jet head propagation velocity. The study of the threshold at which the magnetic field changes its role should be addressed in future publications. 

The evaluation of the probability of jet propagation beyond the binary region has thus to be revisited in the presence of magnetic fields. In \cite{Perucho08,Perucho10} it was concluded that this depended basically on the jet power, whereas in this work we see that low power jets threaded by relatively weak fields could propagate longer distances without being destabilized. However, more powerful jets could, in contrast, be disrupted by current-driven instabilities if the magnetic field is still dynamically relevant at the scales of the binary. Therefore, we conclude that a more thorough characterization of the magnetic field role in microquasar jets at these scales is needed in order to set a proper limit on the minimum energy flux to allow for propagation beyond the progenitor system. 

In RMHD jet simulations, the presence of recollimation shocks within the binary region is more probable than in the case of RHD jets due to the role of magnetic tension which acts against expansion and contributes to jet collimation \citep{marti16,fuentes18}, independently from drops in the ambient pressure \citep[in contrast to RHD flows,][]{Perucho08,Perucho10},
but the pressure jumps at these shocks and its consequent radiative properties will still depend on the obliqueness with respect to the jet stream lines. 

A detailed analysis of the energy distribution in the jet shows that both magnetic and internal energies can contribute very efficiently to jet acceleration along expansion regions. At shock waves (both recollimation, standing, and terminal shocks), both magnetic and internal energies may grow at the expense of the kinetic energy budget. In the turbulent backflow, those magnetically dominated regions can convert magnetic energy into kinetic and internal energy of the flow (the latter by magnetic diffusion), while if the kinetic energy is dominant, this can be used as a reservoir to reinforced locally the magnetic field and to increase the internal energy. Moreover, we have also contributed further evidence to the relevant role of the relativistic nature of outflows in the study of jet energy transfer to the ambient medium, as predicted before by \cite{perucho17} for AGNs, and analyzed here in the context of microquasar jet simulations.


\begin{acknowledgements}

The project that gave rise to these results received the support of a fellowship from ”la Caixa” Foundation (ID 100010434). The fellowship code is LCF/BQ/DR19/11740030. J.L.M acknowledges additional support from the Spanish Ministerio de Ciencia through grant PID2019-105510GB-C31.

M.P. and J.M.-M acknowledges support by the Spanish Ministerio de Ciencia through grants PID2019-107427GB-C33 and PGC2018-095984-B-I00, and from the Generalitat Valenciana through grant PROMETEU/2019/071. M.P acknowledges additional support from the Spanish Ministerio de Ciencia through grant PID2019-105510GB-C31.

V.B-R. acknowledges financial support from the State Agency for Research of the Spanish Ministry of Science and Innovation under grant PID2019-105510GB-C31 and through the ''Unit of Excellence Mar\'ia de Maeztu 2020-2023'' award to the Institute of Cosmos Sciences (CEX2019-000918-M), and by the Catalan DEC grant 2017 SGR 643. V.B-R. is Correspondent Researcher of CONICET, Argentina, at the IAR. Computer simulations have been carried out  in the Servei d'Inform\`atica de la Universitat de Val\`encia (Tirant supercomputer).

We thank the referee for all the constructive comments and suggestions that really helped to improve the quality of the manuscript.

\end{acknowledgements}

\bibliographystyle{aa} 
\bibliography{bibfile} 

\begin{thebibliography}{125}
\expandafter\ifx\csname natexlab\endcsname\relax\def\natexlab#1{#1}\fi

\bibitem[{{Aharonian} {et~al.}(2005){Aharonian}, {Akhperjanian}, {Aye},
  {Bazer-Bachi}, {Beilicke}, {Benbow}, {Berge}, {Berghaus}, {Bernl{\"o}hr},
  {Boisson}, {Bolz}, {Borrel}, {Braun}, {Breitling}, {Brown}, {Gordo},
  {Chadwick}, {Chounet}, {Cornils}, {Costamante}, {Degrange}, {Dickinson},
  {Djannati-Ata{\"\i}}, {Drury}, {Dubus}, {Emmanoulopoulos}, {Espigat},
  {Feinstein}, {Fleury}, {Fontaine}, {Fuchs}, {Funk}, {Gallant}, {Giebels},
  {Gillessen}, {Glicenstein}, {Goret}, {Hadjichristidis}, {Hauser},
  {Heinzelmann}, {Henri}, {Hermann}, {Hinton}, {Hofmann}, {Holleran}, {Horns},
  {Jacholkowska}, {de Jager}, {Kh{\'e}lifi}, {Komin}, {Konopelko}, {Latham},
  {Le Gallou}, {Lemi{\`e}re}, {Lemoine-Goumard}, {Leroy}, {Lohse}, {Marcowith},
  {Martin}, {Martineau-Huynh}, {Masterson}, {McComb}, {de Naurois}, {Nolan},
  {Noutsos}, {Orford}, {Osborne}, {Ouchrif}, {Panter}, {Pelletier}, {Pita},
  {P{\"u}hlhofer}, {Punch}, {Raubenheimer}, {Raue}, {Raux}, {Rayner}, {Reimer},
  {Reimer}, {Ripken}, {Rob}, {Rolland}, {Rowell}, {Sahakian}, {Saug{\'e}},
  {Schlenker}, {Schlickeiser}, {Schuster}, {Schwanke}, {Siewert}, {Sol},
  {Spangler}, {Steenkamp}, {Stegmann}, {Tavernet}, {Terrier}, {Th{\'e}oret},
  {Tluczykont}, {Vasileiadis}, {Venter}, {Vincent}, {V{\"o}lk}, \&
  {Wagner}}]{aharonian05}
{Aharonian}, F., {Akhperjanian}, A.~G., {Aye}, K.~M., {et~al.} 2005, Science,
  309, 746

\bibitem[{{Albert} {et~al.}(2006){Albert}, {Aliu}, {Anderhub}, {Antoranz},
  {Armada}, {Asensio}, {Baixeras}, {Barrio}, {Bartelt}, {Bartko}, {Bastieri},
  {Bavikadi}, {Bednarek}, {Berger}, {Bigongiari}, {Biland}, {Bisesi}, {Bock},
  {Bordas}, {Bosch-Ramon}, {Bretz}, {Britvitch}, {Camara}, {Carmona},
  {Chilingarian}, {Ciprini}, {Coarasa}, {Commichau}, {Contreras}, {Cortina},
  {Curtef}, {Danielyan}, {Dazzi}, {De Angelis}, {de los Reyes}, {De Lotto},
  {Domingo-Santamar{\'\i}a}, {Dorner}, {Doro}, {Errando}, {Fagiolini},
  {Ferenc}, {Fern{\'a}ndez}, {Firpo}, {Flix}, {Fonseca}, {Font}, {Fuchs},
  {Galante}, {Garczarczyk}, {Gaug}, {Giller}, {Goebel}, {Hakobyan},
  {Hayashida}, {Hengstebeck}, {H{\"o}hne}, {Hose}, {Hsu}, {Isar}, {Jacon},
  {Kalekin}, {Kosyra}, {Kranich}, {Laatiaoui}, {Laille}, {Lenisa}, {Liebing},
  {Lindfors}, {Lombardi}, {Longo}, {L{\'o}pez}, {L{\'o}pez}, {Lorenz},
  {Lucarelli}, {Majumdar}, {Maneva}, {Mannheim}, {Mansutti}, {Mariotti},
  {Mart{\'\i}nez}, {Mase}, {Mazin}, {Merck}, {Meucci}, {Meyer}, {Miranda},
  {Mirzoyan}, {Mizobuchi}, {Moralejo}, {Nilsson}, {O{\~n}a-Wilhelmi},
  {Ordu{\~n}a}, {Otte}, {Oya}, {Paneque}, {Paoletti}, {Paredes}, {Pasanen},
  {Pascoli}, {Pauss}, {Pavel}, {Pegna}, {Persic}, {Peruzzo}, {Piccioli},
  {Poller}, {Pooley}, {Prandini}, {Raymers}, {Rhode}, {Rib{\'o}}, {Rico},
  {Riegel}, {Rissi}, {Robert}, {Romero}, {R{\"u}gamer}, {Saggion},
  {S{\'a}nchez}, {Sartori}, {Scalzotto}, {Scapin}, {Schmitt}, {Schweizer},
  {Shayduk}, {Shinozaki}, {Shore}, {Sidro}, {Sillanp{\"a}{\"a}}, {Sobczynska},
  {Stamerra}, {Stark}, {Takalo}, {Temnikov}, {Tescaro}, {Teshima}, {Tonello},
  {Torres}, {Torres}, {Turini}, {Vankov}, {Vitale}, {Wagner}, {Wibig},
  {Wittek}, {Zanin}, \& {Zapatero}}]{albert06}
{Albert}, J., {Aliu}, E., {Anderhub}, H., {et~al.} 2006, Science, 312, 1771

\bibitem[{{Albert} {et~al.}(2007){Albert}, {Aliu}, {Anderhub}, {Antoranz},
  {Armada}, {Baixeras}, {Barrio}, {Bartko}, {Bastieri}, {Becker}, {Bednarek},
  {Berger}, {Bigongiari}, {Biland}, {Bock}, {Bordas}, {Bosch-Ramon}, {Bretz},
  {Britvitch}, {Camara}, {Carmona}, {Chilingarian}, {Coarasa}, {Commichau},
  {Contreras}, {Cortina}, {Costado}, {Curtef}, {Danielyan}, {Dazzi}, {De
  Angelis}, {Delgado}, {de los Reyes}, {De Lotto}, {Domingo-Santamar{\'\i}a},
  {Dorner}, {Doro}, {Errando}, {Fagiolini}, {Ferenc}, {Fern{\'a}ndez}, {Firpo},
  {Flix}, {Fonseca}, {Font}, {Fuchs}, {Galante}, {Garc{\'\i}a-L{\'o}pez},
  {Garczarczyk}, {Gaug}, {Giller}, {Goebel}, {Hakobyan}, {Hayashida},
  {Hengstebeck}, {Herrero}, {H{\"o}hne}, {Hose}, {Hsu}, {Jacon}, {Jogler},
  {Kosyra}, {Kranich}, {Kritzer}, {Laille}, {Lindfors}, {Lombardi}, {Longo},
  {L{\'o}pez}, {L{\'o}pez}, {Lorenz}, {Majumdar}, {Maneva}, {Mannheim},
  {Mansutti}, {Mariotti}, {Mart{\'\i}nez}, {Mazin}, {Merck}, {Meucci}, {Meyer},
  {Miranda}, {Mirzoyan}, {Mizobuchi}, {Moralejo}, {Nieto}, {Nilsson},
  {Ninkovic}, {O{\~n}a-Wilhelmi}, {Otte}, {Oya}, {Panniello}, {Paoletti},
  {Paredes}, {Pasanen}, {Pascoli}, {Pauss}, {Pegna}, {Persic}, {Peruzzo},
  {Piccioli}, {Prandini}, {Puchades}, {Raymers}, {Rhode}, {Rib{\'o}}, {Rico},
  {Rissi}, {Robert}, {R{\"u}gamer}, {Saggion}, {Saito}, {S{\'a}nchez},
  {Sartori}, {Scalzotto}, {Scapin}, {Schmitt}, {Schweizer}, {Shayduk},
  {Shinozaki}, {Shore}, {Sidro}, {Sillanp{\"a}{\"a}}, {Sobczynska}, {Stamerra},
  {Stark}, {Takalo}, {Temnikov}, {Tescaro}, {Teshima}, {Torres}, {Turini},
  {Vankov}, {Vitale}, {Wagner}, {Wibig}, {Wittek}, {Zandanel}, {Zanin}, \&
  {Zapatero}}]{albert07}
{Albert}, J., {Aliu}, E., {Anderhub}, H., {et~al.} 2007, \apjl, 665, L51

\bibitem[{{Anile}(1989)}]{anile89}
{Anile}, A.~M. 1989, {Relativistic fluids and magneto-fluids : with
  applications in astrophysics and plasma physics}

\bibitem[{{Ant{\'o}n} {et~al.}(2010){Ant{\'o}n}, {Miralles}, {Mart{\'\i}},
  {Ib{\'a}{\~n}ez}, {Aloy}, \& {Mimica}}]{anton10}
{Ant{\'o}n}, L., {Miralles}, J.~A., {Mart{\'\i}}, J.~M., {et~al.} 2010, \apjs,
  188, 1

\bibitem[{{Balsara}(2001)}]{balsara01}
{Balsara}, D. 2001, \apjs, 132, 83

\bibitem[{{Balsara} \& {Spicer}(1999)}]{balsara99}
{Balsara}, D.~S. \& {Spicer}, D.~S. 1999, Journal of Computational Physics,
  149, 270

\bibitem[{{Barkov} \& {Bosch-Ramon}(2022)}]{Barkov21}
{Barkov}, M.~V. \& {Bosch-Ramon}, V. 2022, \mnras, 510, 3479

\bibitem[{{Beckwith} \& {Stone}(2011)}]{beckwith11}
{Beckwith}, K. \& {Stone}, J.~M. 2011, \apjs, 193, 6

\bibitem[{{Belloni} \& {Motta}(2016)}]{belloni16}
{Belloni}, T.~M. \& {Motta}, S.~E. 2016, {Transient Black Hole Binaries}, ed.
  C.~{Bambi}, Vol. 440, 61

\bibitem[{{Blandford} \& {Payne}(1982)}]{blandford82}
{Blandford}, R.~D. \& {Payne}, D.~G. 1982, \mnras, 199, 883

\bibitem[{{Blandford} \& {Znajek}(1977)}]{blandford77}
{Blandford}, R.~D. \& {Znajek}, R.~L. 1977, \mnras, 179, 433

\bibitem[{{Bodo} {et~al.}(1994){Bodo}, {Massaglia}, {Ferrari}, \&
  {Trussoni}}]{bodo94}
{Bodo}, G., {Massaglia}, S., {Ferrari}, A., \& {Trussoni}, E. 1994, \aap, 283,
  655

\bibitem[{{Bordas} {et~al.}(2009){Bordas}, {Bosch-Ramon}, {Paredes}, \&
  {Perucho}}]{bordas09}
{Bordas}, P., {Bosch-Ramon}, V., {Paredes}, J.~M., \& {Perucho}, M. 2009, \aap,
  497, 325

\bibitem[{{Bosch-Ramon}(2012)}]{bosch12a}
{Bosch-Ramon}, V. 2012, \aap, 542, A125

\bibitem[{{Bosch-Ramon} \& {Barkov}(2016)}]{Bosch16}
{Bosch-Ramon}, V. \& {Barkov}, M.~V. 2016, \aap, 590, A119

\bibitem[{{Bosch-Ramon} \& {Khangulyan}(2009)}]{bosch09}
{Bosch-Ramon}, V. \& {Khangulyan}, D. 2009, International Journal of Modern
  Physics D, 18, 347

\bibitem[{{Bosch-Ramon} {et~al.}(2011){Bosch-Ramon}, {Perucho}, \&
  {Bordas}}]{bosch11}
{Bosch-Ramon}, V., {Perucho}, M., \& {Bordas}, P. 2011, \aap, 528, A89

\bibitem[{{Bosch-Ramon} \& {Rieger}(2012)}]{bosch12b}
{Bosch-Ramon}, V. \& {Rieger}, F.~M. 2012, in Astroparticle, 219--225

\bibitem[{{Brackbill} \& {Barnes}(1980)}]{brackbill80}
{Brackbill}, J.~U. \& {Barnes}, D.~C. 1980, Journal of Computational Physics,
  35, 426

\bibitem[{{Brio} \& {Wu}(1988)}]{brio88}
{Brio}, M. \& {Wu}, C.~C. 1988, Journal of Computational Physics, 75, 400

\bibitem[{{Cassinelli} {et~al.}(2008){Cassinelli}, {Ignace}, {Waldron}, {Cho},
  {Murphy}, \& {Lazarian}}]{cassinelli08}
{Cassinelli}, J.~P., {Ignace}, R., {Waldron}, W.~L., {et~al.} 2008, \apj, 683,
  1052

\bibitem[{{Castro} {et~al.}(2017){Castro}, {Gallardo}, \&
  {Marquina}}]{castro17}
{Castro}, M.~J., {Gallardo}, J.~M., \& {Marquina}, A. 2017, Computer Physics
  Communications, 219, 108

\bibitem[{{Charlet} {et~al.}(2022){Charlet}, {Walder}, {Marcowith}, {Folini},
  {Favre}, \& {Dieckmann}}]{charlet21}
{Charlet}, A., {Walder}, R., {Marcowith}, A., {et~al.} 2022, \aap, 658, A100

\bibitem[{{Cruz-Osorio} {et~al.}(2021){Cruz-Osorio}, {Fromm}, {Mizuno},
  {Nathanail}, {Younsi}, {Porth}, {Davelaar}, {Falcke}, {Kramer}, \&
  {Rezzolla}}]{osorio21}
{Cruz-Osorio}, A., {Fromm}, C.~M., {Mizuno}, Y., {et~al.} 2021, Nature
  Astronomy [\eprint[arXiv]{2111.02517}]

\bibitem[{{de la Cita} {et~al.}(2017){de la Cita}, {del Palacio},
  {Bosch-Ramon}, {Paredes-Fortuny}, {Romero}, \& {Khangulyan}}]{cita17}
{de la Cita}, V.~M., {del Palacio}, S., {Bosch-Ramon}, V., {et~al.} 2017, \aap,
  604, A39

\bibitem[{{Dedner} {et~al.}(2002){Dedner}, {Kemm}, {Kr{\"o}ner}, {Munz},
  {Schnitzer}, \& {Wesenberg}}]{dedner02}
{Dedner}, A., {Kemm}, F., {Kr{\"o}ner}, D., {et~al.} 2002, Journal of
  Computational Physics, 175, 645

\bibitem[{{Del Zanna} \& {Bucciantini}(2002)}]{Zanna02}
{Del Zanna}, L. \& {Bucciantini}, N. 2002, \aap, 390, 1177

\bibitem[{{Del Zanna} {et~al.}(2003){Del Zanna}, {Bucciantini}, \&
  {Londrillo}}]{zanna03}
{Del Zanna}, L., {Bucciantini}, N., \& {Londrillo}, P. 2003, \aap, 400, 397

\bibitem[{{Dolezal} \& {Wong}(1995)}]{Dolezal95}
{Dolezal}, A. \& {Wong}, S.~S.~M. 1995, Journal of Computational Physics, 120,
  266

\bibitem[{{Dubal}(1991)}]{dubal91}
{Dubal}, M.~R. 1991, Computer Physics Communications, 64, 221

\bibitem[{{Evans} \& {Hawley}(1988)}]{evans88}
{Evans}, C.~R. \& {Hawley}, J.~F. 1988, \apj, 332, 659

\bibitem[{{Event Horizon Telescope Collaboration} {et~al.}(2019){Event Horizon
  Telescope Collaboration}, {Akiyama}, {Alberdi}, {Alef}, {Asada}, {Azulay},
  {Baczko}, {Ball}, {Balokovi{\'c}}, {Barrett}, {Bintley}, {Blackburn},
  {Boland}, {Bouman}, {Bower}, {Bremer}, {Brinkerink}, {Brissenden}, {Britzen},
  {Broderick}, {Broguiere}, {Bronzwaer}, {Byun}, {Carlstrom}, {Chael}, {Chan},
  {Chatterjee}, {Chatterjee}, {Chen}, {Chen}, {Cho}, {Christian}, {Conway},
  {Cordes}, {Crew}, {Cui}, {Davelaar}, {De Laurentis}, {Deane}, {Dempsey},
  {Desvignes}, {Dexter}, {Doeleman}, {Eatough}, {Falcke}, {Fish}, {Fomalont},
  {Fraga-Encinas}, {Freeman}, {Friberg}, {Fromm}, {G{\'o}mez}, {Galison},
  {Gammie}, {Garc{\'\i}a}, {Gentaz}, {Georgiev}, {Goddi}, {Gold}, {Gu},
  {Gurwell}, {Hada}, {Hecht}, {Hesper}, {Ho}, {Ho}, {Honma}, {Huang}, {Huang},
  {Hughes}, {Ikeda}, {Inoue}, {Issaoun}, {James}, {Jannuzi}, {Janssen},
  {Jeter}, {Jiang}, {Johnson}, {Jorstad}, {Jung}, {Karami}, {Karuppusamy},
  {Kawashima}, {Keating}, {Kettenis}, {Kim}, {Kim}, {Kim}, {Kino}, {Koay},
  {Koch}, {Koyama}, {Kramer}, {Kramer}, {Krichbaum}, {Kuo}, {Lauer}, {Lee},
  {Li}, {Li}, {Lindqvist}, {Liu}, {Liuzzo}, {Lo}, {Lobanov}, {Loinard},
  {Lonsdale}, {Lu}, {MacDonald}, {Mao}, {Markoff}, {Marrone}, {Marscher},
  {Mart{\'\i}-Vidal}, {Matsushita}, {Matthews}, {Medeiros}, {Menten}, {Mizuno},
  {Mizuno}, {Moran}, {Moriyama}, {Moscibrodzka}, {M{\"u}ller}, {Nagai},
  {Nagar}, {Nakamura}, {Narayan}, {Narayanan}, {Natarajan}, {Neri}, {Ni},
  {Noutsos}, {Okino}, {Olivares}, {Ortiz-Le{\'o}n}, {Oyama}, {{\"O}zel},
  {Palumbo}, {Patel}, {Pen}, {Pesce}, {Pi{\'e}tu}, {Plambeck}, {PopStefanija},
  {Porth}, {Prather}, {Preciado-L{\'o}pez}, {Psaltis}, {Pu}, {Ramakrishnan},
  {Rao}, {Rawlings}, {Raymond}, {Rezzolla}, {Ripperda}, {Roelofs}, {Rogers},
  {Ros}, {Rose}, {Roshanineshat}, {Rottmann}, {Roy}, {Ruszczyk}, {Ryan},
  {Rygl}, {S{\'a}nchez}, {S{\'a}nchez-Arguelles}, {Sasada}, {Savolainen},
  {Schloerb}, {Schuster}, {Shao}, {Shen}, {Small}, {Sohn}, {SooHoo}, {Tazaki},
  {Tiede}, {Tilanus}, {Titus}, {Toma}, {Torne}, {Trent}, {Trippe}, {Tsuda},
  {van Bemmel}, {van Langevelde}, {van Rossum}, {Wagner}, {Wardle},
  {Weintroub}, {Wex}, {Wharton}, {Wielgus}, {Wong}, {Wu}, {Young}, {Young},
  {Younsi}, {Yuan}, {Yuan}, {Zensus}, {Zhao}, {Zhao}, {Zhu}, {Algaba},
  {Allardi}, {Amestica}, {Anczarski}, {Bach}, {Baganoff}, {Beaudoin}, {Benson},
  {Berthold}, {Blanchard}, {Blundell}, {Bustamente}, {Cappallo},
  {Castillo-Dom{\'\i}nguez}, {Chang}, {Chang}, {Chang}, {Chen}, {Chilson},
  {Chuter}, {C{\'o}rdova Rosado}, {Coulson}, {Crawford}, {Crowley}, {David},
  {Derome}, {Dexter}, {Dornbusch}, {Dudevoir}, {Dzib}, {Eckart}, {Eckert},
  {Erickson}, {Everett}, {Faber}, {Farah}, {Fath}, {Folkers}, {Forbes},
  {Freund}, {G{\'o}mez-Ruiz}, {Gale}, {Gao}, {Geertsema}, {Graham}, {Greer},
  {Grosslein}, {Gueth}, {Haggard}, {Halverson}, {Han}, {Han}, {Hao},
  {Hasegawa}, {Henning}, {Hern{\'a}ndez-G{\'o}mez}, {Herrero-Illana},
  {Heyminck}, {Hirota}, {Hoge}, {Huang}, {Impellizzeri}, {Jiang}, {Kamble},
  {Keisler}, {Kimura}, {Kono}, {Kubo}, {Kuroda}, {Lacasse}, {Laing}, {Leitch},
  {Li}, {Lin}, {Liu}, {Liu}, {Lu}, {Marson}, {Martin-Cocher}, {Massingill},
  {Matulonis}, {McColl}, {McWhirter}, {Messias}, {Meyer-Zhao}, {Michalik},
  {Monta{\~n}a}, {Montgomerie}, {Mora-Klein}, {Muders}, {Nadolski}, {Navarro},
  {Neilsen}, {Nguyen}, {Nishioka}, {Norton}, {Nowak}, {Nystrom}, {Ogawa},
  {Oshiro}, {Oyama}, {Parsons}, {Paine}, {Pe{\~n}alver}, {Phillips}, {Poirier},
  {Pradel}, {Primiani}, {Raffin}, {Rahlin}, {Reiland}, {Risacher}, {Ruiz},
  {S{\'a}ez-Mada{\'\i}n}, {Sassella}, {Schellart}, {Shaw}, {Silva}, {Shiokawa},
  {Smith}, {Snow}, {Souccar}, {Sousa}, {Sridharan}, {Srinivasan}, {Stahm},
  {Stark}, {Story}, {Timmer}, {Vertatschitsch}, {Walther}, {Wei}, {Whitehorn},
  {Whitney}, {Woody}, {Wouterloot}, {Wright}, {Yamaguchi}, {Yu}, {Zeballos},
  {Zhang}, \& {Ziurys}}]{EHT19}
{Event Horizon Telescope Collaboration}, {Akiyama}, K., {Alberdi}, A., {et~al.}
  2019, \apjl, 875, L1

\bibitem[{{Fender} {et~al.}(2004){Fender}, {Belloni}, \& {Gallo}}]{fender04}
{Fender}, R.~P., {Belloni}, T.~M., \& {Gallo}, E. 2004, \mnras, 355, 1105

\bibitem[{{Fuentes} {et~al.}(2018){Fuentes}, {G{\'o}mez}, {Mart{\'\i}}, \&
  {Perucho}}]{fuentes18}
{Fuentes}, A., {G{\'o}mez}, J.~L., {Mart{\'\i}}, J.~M., \& {Perucho}, M. 2018,
  \apj, 860, 121

\bibitem[{{Gardiner} \& {Stone}(2005)}]{gardiner05}
{Gardiner}, T.~A. \& {Stone}, J.~M. 2005, Journal of Computational Physics,
  205, 509

\bibitem[{{Ghisellini} {et~al.}(2014){Ghisellini}, {Tavecchio}, {Maraschi},
  {Celotti}, \& {Sbarrato}}]{ghisellini14}
{Ghisellini}, G., {Tavecchio}, F., {Maraschi}, L., {Celotti}, A., \&
  {Sbarrato}, T. 2014, \nat, 515, 376

\bibitem[{{Giacomazzo} \& {Rezzolla}(2006)}]{giacomazzo06}
{Giacomazzo}, B. \& {Rezzolla}, L. 2006, Journal of Fluid Mechanics, 562, 223

\bibitem[{{Giacomazzo} \& {Rezzolla}(2007)}]{giacomazzo07}
{Giacomazzo}, B. \& {Rezzolla}, L. 2007, Classical and Quantum Gravity, 24,
  S235

\bibitem[{{Guan} {et~al.}(2014){Guan}, {Li}, \& {Li}}]{guan14}
{Guan}, X., {Li}, H., \& {Li}, S. 2014, \apj, 781, 48

\bibitem[{Harten {et~al.}(1983)Harten, Lax, \& van Leer}]{harten83}
Harten, A., Lax, P., \& van Leer, B. 1983, SIAM Rev, 25, 35

\bibitem[{{Janssen et al.}(2021)}]{EHT21}
{Janssen et al.}, M. 2021, Nature Astronomy

\bibitem[{{Keppens} {et~al.}(2008){Keppens}, {Meliani}, {van der Holst}, \&
  {Casse}}]{keppens08}
{Keppens}, R., {Meliani}, Z., {van der Holst}, B., \& {Casse}, F. 2008, \aap,
  486, 663

\bibitem[{{Khangulyan} \& {Aharonian}(2005)}]{khangulyan05}
{Khangulyan}, D. \& {Aharonian}, F. 2005, in American Institute of Physics
  Conference Series, Vol. 745, High Energy Gamma-Ray Astronomy, ed. F.~A.
  {Aharonian}, H.~J. {V{\"o}lk}, \& D.~{Horns}, 359--364

\bibitem[{{Kim et al.}(2020)}]{EHT20}
{Kim et al.}, J.-Y. 2020, \aap, 640, A69

\bibitem[{{Komissarov}(1999{\natexlab{a}})}]{komissarov99}
{Komissarov}, S.~S. 1999{\natexlab{a}}, \mnras, 303, 343

\bibitem[{{Komissarov}(1999{\natexlab{b}})}]{komissarov992}
{Komissarov}, S.~S. 1999{\natexlab{b}}, \mnras, 308, 1069

\bibitem[{{Komissarov}(2011)}]{komissarov11}
{Komissarov}, S.~S. 2011, \memsai, 82, 95

\bibitem[{{Komissarov} {et~al.}(2007){Komissarov}, {Barkov}, {Vlahakis}, \&
  {K{\"o}nigl}}]{komissarov07}
{Komissarov}, S.~S., {Barkov}, M.~V., {Vlahakis}, N., \& {K{\"o}nigl}, A. 2007,
  \mnras, 380, 51

\bibitem[{{Komissarov} {et~al.}(2015){Komissarov}, {Porth}, \&
  {Lyutikov}}]{komissarov15}
{Komissarov}, S.~S., {Porth}, O., \& {Lyutikov}, M. 2015, Computational
  Astrophysics and Cosmology, 2, 9

\bibitem[{{Krti{\v{c}}ka}(2014)}]{krt14}
{Krti{\v{c}}ka}, J. 2014, \aap, 564, A70

\bibitem[{{Kylafis} {et~al.}(2012){Kylafis}, {Contopoulos}, {Kazanas}, \&
  {Christodoulou}}]{kylafis12}
{Kylafis}, N.~D., {Contopoulos}, I., {Kazanas}, D., \& {Christodoulou}, D.~M.
  2012, \aap, 538, A5

\bibitem[{{Leismann} {et~al.}(2005){Leismann}, {Ant{\'o}n}, {Aloy},
  {M{\"u}ller}, {Mart{\'\i}}, {Miralles}, \& {Ib{\'a}{\~n}ez}}]{leismann05}
{Leismann}, T., {Ant{\'o}n}, L., {Aloy}, M.~A., {et~al.} 2005, \aap, 436, 503

\bibitem[{{Lewis} \& {Austin}(2002)}]{Lewis02}
{Lewis}, G.~M. \& {Austin}, P.~H. 2002, in 11th Conference on Atmospheric
  Radiation, American Meteorological Society Conference Series, ed. G.~{Smith}
  \& J.~{Brodie}, 123--126

\bibitem[{{Lind} {et~al.}(1989){Lind}, {Payne}, {Meier}, \&
  {Blandford}}]{lindt89}
{Lind}, K.~R., {Payne}, D.~G., {Meier}, D.~L., \& {Blandford}, R.~D. 1989,
  \apj, 344, 89

\bibitem[{{Londrillo} \& {Del Zanna}(2000)}]{Londrillo00}
{Londrillo}, P. \& {Del Zanna}, L. 2000, \apj, 530, 508

\bibitem[{{Marino} {et~al.}(2020){Marino}, {Malzac}, {Del Santo}, {Migliari},
  {Belmont}, {Di Salvo}, {Russell}, {Lopez Miralles}, {Perucho}, {D'A{\`\i}},
  {Iaria}, \& {Burderi}}]{marino20}
{Marino}, A., {Malzac}, J., {Del Santo}, M., {et~al.} 2020, \mnras, 498, 3351

\bibitem[{{Mart{\'\i}}(2015{\natexlab{a}})}]{marti152}
{Mart{\'\i}}, J.-M. 2015{\natexlab{a}}, Computer Physics Communications, 191,
  100

\bibitem[{{Mart{\'\i}}(2015{\natexlab{b}})}]{marti153}
{Mart{\'\i}}, J.-M. 2015{\natexlab{b}}, \mnras, 452, 3106

\bibitem[{{Mart{\'\i}}(2019)}]{marti19}
{Mart{\'\i}}, J.-M. 2019, Galaxies, 7, 24

\bibitem[{{Mart{\'\i}} \& {M{\"u}ller}(2015)}]{marti15}
{Mart{\'\i}}, J.~M. \& {M{\"u}ller}, E. 2015, Living Reviews in Computational
  Astrophysics, 1, 3

\bibitem[{{Mart{\'\i}} {et~al.}(1997){Mart{\'\i}}, {M{\"u}ller}, {Font},
  {Ib{\'a}{\~n}ez}, \& {Marquina}}]{marti97}
{Mart{\'\i}}, J.~M., {M{\"u}ller}, E., {Font}, J.~A., {Ib{\'a}{\~n}ez},
  J.~M.~Z., \& {Marquina}, A. 1997, \apj, 479, 151

\bibitem[{{Mart{\'\i}} {et~al.}(2016){Mart{\'\i}}, {Perucho}, \&
  {G{\'o}mez}}]{marti16}
{Mart{\'\i}}, J.~M., {Perucho}, M., \& {G{\'o}mez}, J.~L. 2016, \apj, 831, 163

\bibitem[{{McKinney} \& {Blandford}(2009)}]{kinney09}
{McKinney}, J.~C. \& {Blandford}, R.~D. 2009, \mnras, 394, L126

\bibitem[{{McKinney} {et~al.}(2012){McKinney}, {Tchekhovskoy}, \&
  {Blandford}}]{kinney12}
{McKinney}, J.~C., {Tchekhovskoy}, A., \& {Blandford}, R.~D. 2012, \mnras, 423,
  3083

\bibitem[{{McNamara} \& {Nulsen}(2007)}]{mcnamara07}
{McNamara}, B.~R. \& {Nulsen}, P.~E.~J. 2007, \araa, 45, 117

\bibitem[{{Migliari} \& {Fender}(2006)}]{migliari06}
{Migliari}, S. \& {Fender}, R.~P. 2006, \mnras, 366, 79

\bibitem[{{Migliari} {et~al.}(2003){Migliari}, {Fender}, {Rupen}, {Jonker},
  {Klein-Wolt}, {Hjellming}, \& {van der Klis}}]{migliari03}
{Migliari}, S., {Fender}, R.~P., {Rupen}, M., {et~al.} 2003, \mnras, 342, L67

\bibitem[{{Migliari} {et~al.}(2004){Migliari}, {Fender}, {Rupen}, {Jonker},
  {Klein-Wolt}, {Hjellming}, {Wachter}, {Homan}, \& {van der
  Klis}}]{migliari04}
{Migliari}, S., {Fender}, R.~P., {Rupen}, M., {et~al.} 2004, Nuclear Physics B
  Proceedings Supplements, 132, 628

\bibitem[{Mignone \& Bodo(2006)}]{Mignone06}
Mignone, A. \& Bodo, G. 2006, Monthly Notices of the Royal Astronomical
  Society, 368, 1040

\bibitem[{{Mignone} \& {Del Zanna}(2021)}]{mignone21}
{Mignone}, A. \& {Del Zanna}, L. 2021, Journal of Computational Physics, 424,
  109748

\bibitem[{{Mignone} {et~al.}(2005){Mignone}, {Massaglia}, \&
  {Bodo}}]{mignone05}
{Mignone}, A., {Massaglia}, S., \& {Bodo}, G. 2005, \ssr, 121, 21

\bibitem[{{Mignone} \& {McKinney}(2007)}]{mignone07}
{Mignone}, A. \& {McKinney}, J.~C. 2007, \mnras, 378, 1118

\bibitem[{{Mignone} {et~al.}(2010){Mignone}, {Rossi}, {Bodo}, {Ferrari}, \&
  {Massaglia}}]{mignone10}
{Mignone}, A., {Rossi}, P., {Bodo}, G., {Ferrari}, A., \& {Massaglia}, S. 2010,
  \mnras, 402, 7

\bibitem[{Mignone {et~al.}(2009)Mignone, Ugliano, \& Bodo}]{Mignone09}
Mignone, A., Ugliano, M., \& Bodo, G. 2009, Monthly Notices of the Royal
  Astronomical Society, 393, 1141

\bibitem[{{Mignone} {et~al.}(2012){Mignone}, {Zanni}, {Tzeferacos}, {van
  Straalen}, {Colella}, \& {Bodo}}]{mignone12}
{Mignone}, A., {Zanni}, C., {Tzeferacos}, P., {et~al.} 2012, \apjs, 198, 7

\bibitem[{{Mirabel} \& {Rodr{\'\i}guez}(1999)}]{mirabel99}
{Mirabel}, I.~F. \& {Rodr{\'\i}guez}, L.~F. 1999, \araa, 37, 409

\bibitem[{{Mizuno} {et~al.}(2007){Mizuno}, {Hardee}, \& {Nishikawa}}]{mizuno07}
{Mizuno}, Y., {Hardee}, P., \& {Nishikawa}, K.-I. 2007, \apj, 662, 835

\bibitem[{{Moffat}(2008)}]{moffat08}
{Moffat}, A. F.~J. 2008, in Clumping in Hot-Star Winds, ed. W.-R. {Hamann},
  A.~{Feldmeier}, \& L.~M. {Oskinova}, 17

\bibitem[{{Molina} \& {Bosch-Ramon}(2018)}]{molina18}
{Molina}, E. \& {Bosch-Ramon}, V. 2018, \aap, 618, A146

\bibitem[{{Molina} {et~al.}(2019){Molina}, {del Palacio}, \&
  {Bosch-Ramon}}]{molina19}
{Molina}, E., {del Palacio}, S., \& {Bosch-Ramon}, V. 2019, \aap, 629, A129

\bibitem[{{Monceau-Baroux} {et~al.}(2014){Monceau-Baroux}, {Porth}, {Meliani},
  \& {Keppens}}]{monceau14}
{Monceau-Baroux}, R., {Porth}, O., {Meliani}, Z., \& {Keppens}, R. 2014, \aap,
  561, A30

\bibitem[{{Moya-Torregrosa} {et~al.}(2021){Moya-Torregrosa}, {Fuentes},
  {Mart{\'\i}}, {G{\'o}mez}, \& {Perucho}}]{moya21}
{Moya-Torregrosa}, I., {Fuentes}, A., {Mart{\'\i}}, J.~M., {G{\'o}mez}, J.~L.,
  \& {Perucho}, M. 2021, \aap, 650, A60

\bibitem[{{Muijres} {et~al.}(2012){Muijres}, {Vink}, {de Koter}, {M{\"u}ller},
  \& {Langer}}]{muijres12}
{Muijres}, L.~E., {Vink}, J.~S., {de Koter}, A., {M{\"u}ller}, P.~E., \&
  {Langer}, N. 2012, \aap, 537, A37

\bibitem[{Noh(1987)}]{noh87}
Noh, W. 1987, Journal of Computational Physics, 72, 78

\bibitem[{{Orszag} \& {Tang}(1979)}]{Orszag79}
{Orszag}, S.~A. \& {Tang}, C.~M. 1979, Journal of Fluid Mechanics, 90, 129

\bibitem[{{Paredes} {et~al.}(2000){Paredes}, {Mart{\'\i}}, {Rib{\'o}}, \&
  {Massi}}]{paredes00}
{Paredes}, J.~M., {Mart{\'\i}}, J., {Rib{\'o}}, M., \& {Massi}, M. 2000,
  Science, 288, 2340

\bibitem[{{Perucho}(2019)}]{perucho192}
{Perucho}, M. 2019, Galaxies, 7, 70

\bibitem[{{Perucho} \& {Bosch-Ramon}(2008)}]{Perucho08}
{Perucho}, M. \& {Bosch-Ramon}, V. 2008, \aap, 482, 917

\bibitem[{{Perucho} \& {Bosch-Ramon}(2012)}]{Perucho12}
{Perucho}, M. \& {Bosch-Ramon}, V. 2012, \aap, 539, A57

\bibitem[{{Perucho} {et~al.}(2010{\natexlab{a}}){Perucho}, {Bosch-Ramon}, \&
  {Khangulyan}}]{Perucho10}
{Perucho}, M., {Bosch-Ramon}, V., \& {Khangulyan}, D. 2010{\natexlab{a}}, \aap,
  512, L4

\bibitem[{{Perucho} {et~al.}(2010{\natexlab{b}}){Perucho}, {Mart{\'\i}},
  {Cela}, {Hanasz}, {de La Cruz}, \& {Rubio}}]{perucho101}
{Perucho}, M., {Mart{\'\i}}, J.~M., {Cela}, J.~M., {et~al.} 2010{\natexlab{b}},
  \aap, 519, A41

\bibitem[{{Perucho} {et~al.}(2005){Perucho}, {Mart{\'\i}}, \&
  {Hanasz}}]{perucho05}
{Perucho}, M., {Mart{\'\i}}, J.~M., \& {Hanasz}, M. 2005, \aap, 443, 863

\bibitem[{{Perucho} {et~al.}(2019){Perucho}, {Mart{\'\i}}, \&
  {Quilis}}]{perucho191}
{Perucho}, M., {Mart{\'\i}}, J.-M., \& {Quilis}, V. 2019, \mnras, 482, 3718

\bibitem[{{Perucho} {et~al.}(2017){Perucho}, {Mart{\'\i}}, {Quilis}, \&
  {Borja-Lloret}}]{perucho17}
{Perucho}, M., {Mart{\'\i}}, J.-M., {Quilis}, V., \& {Borja-Lloret}, M. 2017,
  \mnras, 471, L120

\bibitem[{{Perucho} {et~al.}(2014){Perucho}, {Mart{\'\i}}, {Quilis}, \&
  {Ricciardelli}}]{perucho14}
{Perucho}, M., {Mart{\'\i}}, J.-M., {Quilis}, V., \& {Ricciardelli}, E. 2014,
  \mnras, 445, 1462

\bibitem[{{Perucho} {et~al.}(2011){Perucho}, {Quilis}, \&
  {Mart{\'\i}}}]{perucho11}
{Perucho}, M., {Quilis}, V., \& {Mart{\'\i}}, J.-M. 2011, \apj, 743, 42

\bibitem[{{Porth}(2013)}]{porth13}
{Porth}, O. 2013, \mnras, 429, 2482

\bibitem[{{Porth} {et~al.}(2014){Porth}, {Komissarov}, \& {Keppens}}]{porth14}
{Porth}, O., {Komissarov}, S.~S., \& {Keppens}, R. 2014, \mnras, 438, 278

\bibitem[{{Powell}(1994)}]{powell94}
{Powell}, K.~G. 1994, {Approximate Riemann solver for magnetohydrodynamics
  (that works in more than one dimension)}, Unknown

\bibitem[{{Quirk}(1994)}]{quirk94}
{Quirk}, J.~J. 1994, International Journal for Numerical Methods in Fluids, 18,
  555

\bibitem[{{Rieger} {et~al.}(2007){Rieger}, {Bosch-Ramon}, \&
  {Duffy}}]{rieger07}
{Rieger}, F.~M., {Bosch-Ramon}, V., \& {Duffy}, P. 2007, \apss, 309, 119

\bibitem[{{Roe}(1986)}]{Roe86}
{Roe}, P.~L. 1986, Annual Review of Fluid Mechanics, 18, 337

\bibitem[{{Romero} {et~al.}(2003){Romero}, {Torres}, {Kaufman Bernad{\'o}}, \&
  {Mirabel}}]{romero03}
{Romero}, G.~E., {Torres}, D.~F., {Kaufman Bernad{\'o}}, M.~M., \& {Mirabel},
  I.~F. 2003, \aap, 410, L1

\bibitem[{{Runacres} \& {Owocki}(2002)}]{runacres02}
{Runacres}, M.~C. \& {Owocki}, S.~P. 2002, \aap, 381, 1015

\bibitem[{{Ryu} {et~al.}(1998){Ryu}, {Miniati}, {Jones}, \& {Frank}}]{Ryu98}
{Ryu}, D., {Miniati}, F., {Jones}, T.~W., \& {Frank}, A. 1998, \apj, 509, 244

\bibitem[{{Shu} \& {Osher}(1989)}]{shu89}
{Shu}, C.-W. \& {Osher}, S. 1989, Journal of Computational Physics, 83, 32

\bibitem[{{Stirling} {et~al.}(2001){Stirling}, {Spencer}, {de la Force},
  {Garrett}, {Fender}, \& {Ogley}}]{stirling01}
{Stirling}, A.~M., {Spencer}, R.~E., {de la Force}, C.~J., {et~al.} 2001,
  \mnras, 327, 1273

\bibitem[{{Suresh} \& {Huynh}(1997)}]{suresh97}
{Suresh}, A. \& {Huynh}, H.~T. 1997, Journal of Computational Physics, 136, 83

\bibitem[{{Tavani} {et~al.}(1998){Tavani}, {Kniffen}, {Mattox}, {Paredes}, \&
  {Foster}}]{tavani98}
{Tavani}, M., {Kniffen}, D., {Mattox}, J.~R., {Paredes}, J.~M., \& {Foster},
  R.~S. 1998, \apjl, 497, L89

\bibitem[{{Tchekhovskoy} {et~al.}(2011){Tchekhovskoy}, {Narayan}, \&
  {McKinney}}]{tchek11}
{Tchekhovskoy}, A., {Narayan}, R., \& {McKinney}, J.~C. 2011, \mnras, 418, L79

\bibitem[{{T{\'o}th}(2000)}]{toth00}
{T{\'o}th}, G. 2000, Journal of Computational Physics, 161, 605

\bibitem[{van Leer(1974)}]{vanLeer1974}
van Leer, B. 1974, Journal of Computational Physics, 14, 361

\bibitem[{{van Leer}(1977)}]{vanLeer1977}
{van Leer}, B. 1977, Journal of Computational Physics, 23, 276

\bibitem[{{van Putten}(1993)}]{putten93}
{van Putten}, M. H.~P.~M. 1993, Journal of Computational Physics, 105, 339

\bibitem[{{Vilhu} {et~al.}(2021){Vilhu}, {Kallman}, {Koljonen}, \&
  {Hannikainen}}]{vilhu21}
{Vilhu}, O., {Kallman}, T.~R., {Koljonen}, K.~I.~I., \& {Hannikainen}, D.~C.
  2021, \aap, 649, A176

\bibitem[{{Vlahakis}(2004)}]{vlahakis04}
{Vlahakis}, N. 2004, \apss, 293, 67

\bibitem[{{Vlahakis} \& {K{\"o}nigl}(2004)}]{vlahakis042}
{Vlahakis}, N. \& {K{\"o}nigl}, A. 2004, \apj, 605, 656

\bibitem[{{Wagner} \& {Bicknell}(2011)}]{wagner11}
{Wagner}, A.~Y. \& {Bicknell}, G.~V. 2011, \apj, 728, 29

\bibitem[{{Wang} {et~al.}(2008){Wang}, {Abel}, \& {Zhang}}]{wang08}
{Wang}, P., {Abel}, T., \& {Zhang}, W. 2008, \apjs, 176, 467

\bibitem[{{Wright} \& {Hawke}(2019)}]{wright19}
{Wright}, A.~J. \& {Hawke}, I. 2019, \apjs, 240, 8

\bibitem[{{Yoon} \& {Heinz}(2015)}]{Yoon15}
{Yoon}, D. \& {Heinz}, S. 2015, \apj, 801, 55

\bibitem[{{Yoon} {et~al.}(2011){Yoon}, {Morsony}, {Heinz}, {Wiersema},
  {Fender}, {Russell}, \& {Sunyaev}}]{yoon11}
{Yoon}, D., {Morsony}, B., {Heinz}, S., {et~al.} 2011, \apj, 742, 25

\bibitem[{{Yoon} {et~al.}(2016){Yoon}, {Zdziarski}, \& {Heinz}}]{yoon16}
{Yoon}, D., {Zdziarski}, A.~A., \& {Heinz}, S. 2016, \mnras, 456, 3638

\bibitem[{{Zanin} {et~al.}(2016){Zanin}, {Fern{\'a}ndez-Barral}, {de O{\~n}a
  Wilhelmi}, {Aharonian}, {Blanch}, {Bosch-Ramon}, \& {Galindo}}]{zanin16}
{Zanin}, R., {Fern{\'a}ndez-Barral}, A., {de O{\~n}a Wilhelmi}, E., {et~al.}
  2016, \aap, 596, A55

\end{thebibliography}

\begin{appendix}
\section{\JLM{Lóstrego v1.0}. A new tool for Computational Relativistic Astrophysics}
\label{appA}

In this appendix, we present \JLM{Lóstrego v1.0}, a new computational tool to simulate astrophysical relativistic plasmas in 1D, 2D and 3D \JLM{Cartesian} coordinates. The code, which is entirely written in the FORTRAN programming language, solves the conservative equations of Special Relativistic Magnetohydrodynamics (SRMHD) in finite volumes, and it is especially designed to run in multiple cores to exploit the capacity of modern parallel architectures. \JLM{Lóstrego v1.0} is based on the High Resolution Shock Capturing (HRSC) methods that have been probed to be robust and accurate in other similar codes \citep[see e.g.][and references therein]{marti15}, extending to three dimensions the techniques implemented in \cite{marti153}. Essentially, the algorithm follows the reconstruct-solve-average (RSA) strategy: first, cell-average primitive variables are reconstructed to the cell interfaces, where a Riemann problem is then solved numerically by any approximate Riemann solver. Once fluxes are known at each cell face, the conserved variables are evolved explicitly in time using Total Variation Diminishing (TVD) Runge-Kutta algorithms. Finally, an inversion scheme is applied to recover the primitive variables from the time-evolved solution. 

Several one-dimensional and multi-dimensional tests are presented in this appendix to probe the ability of the new code to solve classical problems in the field of relativistic magnetohydrodynamics.

\subsection{Basic equations}


The Relativistic Magnetohydrodynamics (RMHD) system of partial differential equations in the Minkowski metric\footnote{We assume a metric signature (-,+,+,+). Greek subscripts in 4-vectors run from 0 to 3. Latin indices run from 1 to 3. In the following, we use a system of units where $c=1$ and a factor of $1/\sqrt{4\pi}$ is absorbed in the definition of the magnetic field.} and Cartesian coordinates can be written as a system of conservation laws \citep{komissarov99}:
\begin{equation}
    \partial_t\boldsymbol{U}+\partial_i\boldsymbol{F}^i=0,
    \label{eqa0}
\end{equation}
where $\boldsymbol{U}=\{D,S^j,\tau,B^j\}$ is a vector of conserved variables and $\boldsymbol{F^i}$ are the vector of fluxes for each spatial direction. Both the vector of conserved variables $\boldsymbol{U}$ and the vector of fluxes $\boldsymbol{F^i}$ can be expressed in terms of a set of primitive variables $\boldsymbol{V}=\{\rho,v^j,p,B^j\}$ using the following relations:
\begin{equation}
\label{eqa1}
\boldsymbol{U}=
    \begin{pmatrix}
D\\ 
S^j\\ 
\tau\\ 
B^j\\

\end{pmatrix}=
    \begin{pmatrix}
\rho W \\ 
\rho h^* W^2v^j-b^0b^j\\ 
\rho h^* W^2-p^*-b^0b^0-\rho W\\ 
B^j\\ 
\end{pmatrix},
\end{equation}
\begin{equation}
\label{eqa2}
\boldsymbol{F^i}=
        \begin{pmatrix}
\rho W v^i \\ 
\rho h^* W^2v^iv^j+p^*\delta^{ij}-b^ib^j\\ 
\rho h^* W^2v^i-b^0b^i-\rho W v^i\\ 
v^iB^j-B^iv^j\\ 
\end{pmatrix},
\end{equation}
where $D$ is the relativistic rest mass density, $S^j$ is the momentum density of the magnetized fluid, $\tau$ the energy density measured in the Eulerian frame and $\delta^{ij}$ is the Kronecker delta. The set of primitive variables are the fluid rest-mass density $\rho$, fluid 3-velocity $v_j$ and gas pressure $p$. The magnetic field vector $B^j$ is at the same time primitive and conserved variable, so the conversion from one set to the other is trivial. In Eqs. \ref{eqa1} and \ref{eqa2} we introduced
the hydromagnetic specific enthalpy $h^*$:
\begin{equation}
    h^*=h+\frac{b^{\alpha}b_{\alpha}}{\rho}=1+\epsilon+\frac{p}{\rho}+\frac{|b|^2}{\rho},
\end{equation}
where $\epsilon$ is the specific internal energy. The total pressure $p^*$ of the magnetic fluid is given by:
\begin{equation}
    p^*=p_{\mathrm{g}}+p_{\mathrm{mag}}=p+\frac{|b|^2}{2}.
\end{equation}
In terms of the vectors  $\boldsymbol{v}$ and $\boldsymbol{B}$ in the laboratory frame, the 4-velocity $u^{\alpha}$ and magnetic field $b^{\alpha}$ covariant vectors are:
\begin{equation}
    u^{\alpha} = W(1,\boldsymbol{v}),
\end{equation}
\begin{equation}
    b^0=W(\boldsymbol{v}\cdot\boldsymbol{B}),
\end{equation}
\begin{equation}
    b^i=\frac{B^i}{W}+v^ib^0,
\end{equation}
with the normalizations:
\begin{equation}
    u^{\alpha}u_{\alpha}=-1, \hspace{0.3cm} u^{\alpha}b_{\alpha}=0,
\end{equation}
\begin{equation}
    |b|^2=b_{\alpha}b^{\alpha}=\frac{B^2}{W^2}+(\boldsymbol{v}\cdot\boldsymbol{B})^2,
\end{equation}
where $W=1/\sqrt{1-\boldsymbol{v}^2}$ is the Lorentz factor and summation over repeated indices is assumed. The time evolution of the system of equations of classical and relativistic MHD must also preserve
the magnetic field divergence-free constraint:
\begin{equation}
    \boldsymbol{\nabla}\cdot\boldsymbol{B}=0.
    \label{solenoidal}
\end{equation}
Finally, the system of RMHD equations is closed with an equation of state (EoS) of the form:
\begin{equation}
    p=p (\rho,\epsilon),
\end{equation}
which relates the thermodynamic primitive variables with each other. For the particular case of ideal gases, this relation becomes:
\begin{equation}
    p=(\Gamma - 1)\rho\epsilon,
\end{equation}
where $\Gamma$ is the adiabatic index. This index should be set as $\Gamma=5/3$ for non-relativistic or mildly-relativistic flows and as $\Gamma=4/3$ in the ultrarelativistic limit.

\subsection{Numerical methods}
\label{methods}

\subsubsection{Spatial and temporal discretization}
We have implemented multi-dimensional HRSC methods following the Finite Volumes (FV) strategy and dimensional splitting, taking the integral form of the conservation laws (Eq. \ref{eqa0}) and cell averaged values. This means that operators that involve spatial derivatives are applied dimension by dimension, permuting cyclically the spatial coordinates. Eq. \ref{eqa0} admits a conservative semi-discretization in a rectangular grid of size $L_x\times L_y\times L_z$ and resolution $N_x\times N_y\times N_z$:

\begin{equation}
\begin{aligned}
    \frac{d\boldsymbol{U}_{i,j,k}}{dt}=-\frac{1}{\Delta x}\left(\hat{\boldsymbol{F}}^x_{i+1/2,j,k}-\hat{\boldsymbol{F}}^x_{i-1/2,j,k}\right)\\
    -\frac{1}{\Delta y}\left(\hat{\boldsymbol{F}}^y_{i,j+1/2,k}-\hat{\boldsymbol{F}}^y_{i,j-1/2,k}\right)\\
    -\frac{1}{\Delta z}\left(\hat{\boldsymbol{F}}^z_{i,j,k+1/2}-\hat{\boldsymbol{F}}^z_{i,j,k-1/2}\right)
\end{aligned}
\label{timeadv}
\end{equation}
where $\{i,j,k\}$ is the position of each cell center and $\{i\pm 1/2 ,j\pm 1/2, k \pm 1/2\}$ is the position of the right/left cell interface, respectively. The elements $\Delta x=L_x/N_x, \Delta y=L_y/N_y, \Delta z=L_z/N_z$ are the cell sizes for each spatial dimension. As mentioned before, in the FV formalism the conserved variables represent an approximation to the average over the cell volume:
\begin{equation}
     \boldsymbol{U}_{i,j,k}^{n}=\frac{1}{\Delta x\Delta y\Delta z}\int_{z_{k-1/2}}^{z_{k+1/2}}\int_{y_{j-1/2}}^{y_{j+1/2}}\int_{x_{i-1/2}}^{x_{i+1/2}}\boldsymbol{U}(x,y,z,t)\; dxdydz,
\end{equation}
while the numerical fluxes represent an approximation to the average in the surface of the cell:
\begin{equation}
     \boldsymbol{\hat{F}}_{i+1/2,j,k}^{x}=\frac{1}{\Delta y\Delta z}\int_{z_{k-1/2}}^{z_{k+1/2}}\int_{y_{j-1/2}}^{y_{j+1/2}}\int_{t^n}^{t^{n+1}}\boldsymbol{F}^x(x_{i+1/2},y,z,t)\; dydz,
\end{equation}
\begin{equation}
     \boldsymbol{\hat{F}}_{i,j+1/2,k}^{y}=\frac{1}{\Delta x\Delta z}\int_{z_{k-1/2}}^{z_{k+1/2}}\int_{x_{i-1/2}}^{x_{i+1/2}}\int_{t^n}^{t^{n+1}}\boldsymbol{F}^y(x,y_{j+1/2},z,t)\; dxdz,
\end{equation}
\begin{equation}
     \boldsymbol{\hat{F}}_{i,j,k+1/2}^{z}=\frac{1}{\Delta y\Delta x}\int_{x_{i-1/2}}^{x_{i+1/2}}\int_{y_{j-1/2}}^{y_{j+1/2}}\int_{t^n}^{t^{n+1}}\boldsymbol{F}^z(x,y,z_{k+1/2},t)\; dydx.
\end{equation}

\subsubsection{Spatial cell reconstruction}

To advance in time the set of conserved variables, numerical fluxes must be known at cell interfaces, the latter being functions of the primitive variables. These are reconstructed from cell-centered values by means of interpolation routines that are designed to reconstruct a piece-wise polynomial approximation inside each cell preserving the monotonicity and TV-stability properties of the algorithm. Consistently with our dimensional splitting, reconstruction is performed in one dimension following each spatial coordinate. The current version of \JLM{Lóstrego} admits either Godunov (first order) or linear (second order) spatial reconstruction. The piece-wise linear method (PLM) in the $x$-direction follows:
\begin{equation}
    V(x,t^n)=V_i^n+s_i^n(x-x_i),
\end{equation}
where $s_i^n$ is the linear slope at cell $i$. Similar expressions hold for the $y,z$ spatial directions. The most popular slope limiters are MINMOD \citep{Roe86}, MC \citep{vanLeer1977} and VAN LEER \citep{vanLeer1974}. These limiters avoid the generation of spurious extrema at cell interfaces and reduce the accuracy of the method to first order at extrema with a vanishing slope. Although conserved variables can be directly reconstructed to cell interfaces, we interpolate the primitive variables since experience has probed it to be more robust than the former approach. If the PLM reconstruction leads to non-physical situations where $\boldsymbol{v}^2>1$, we reduce the method to first order (i.e., Godunov reconstruction). High order schemes like the MP reconstruction \citep{suresh97} are already available in the code and we plan to test them in future publications.\\

\subsubsection{Numerical fluxes and the Riemann problem}

Once the cell-centered primitive variables are spatially reconstructed  to obtain $V^L_{i+1/2},V^R_{i+1/2}$ (i.e., the values of $V$ at the left and right sides of the interface $i+1/2$), computation of numerical fluxes requires solving a Riemann problem at each zone edge of the type:
\begin{equation}
V(x,0)=\left\{\begin{matrix}
V^L_{i+1/2} & \text{if}\hspace{0.3cm} x<x_{i+1/2} \\ 
V^R_{i+1/2} & \text{if}\hspace{0.3cm} x>x_{i+1/2}. 
\end{matrix}\right.
\end{equation}

\begin{figure}
  \centerline{\includegraphics[width=\columnwidth]{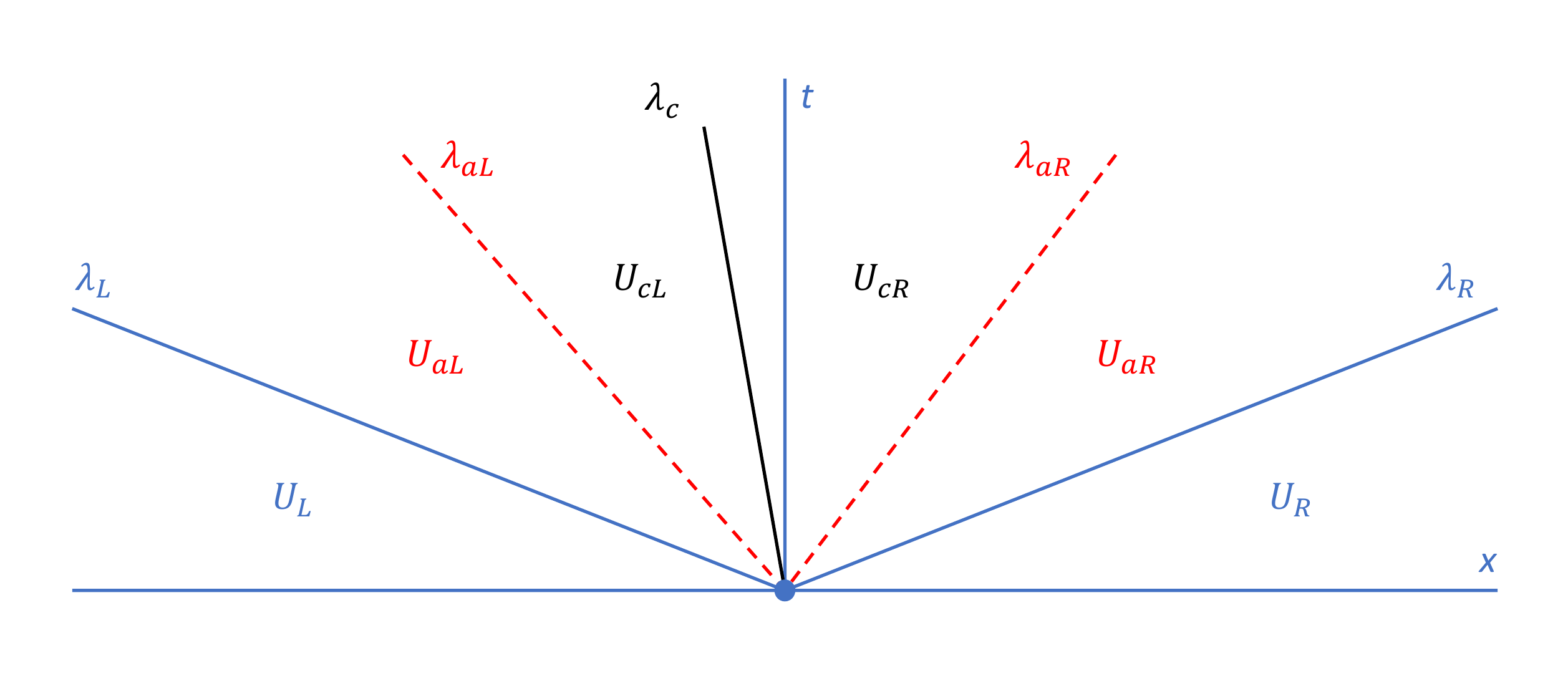}}
  \caption{Riemann fan for the HLL-family of Riemann solvers: HLL (blue), HLLC (blue+black), HLLD (blue+black+red).}
\label{fan}
\end{figure}

Although exact Riemann solvers are available in the literature for RMHD \citep{giacomazzo06}, computation of numerical fluxes usually involves upwind strategies based on approximate solutions. In the current version of \JLM{Lóstrego}, there are available three different approximate solvers, all pertained to the Harten-Lax-van Leer (HLL) family: HLL \citep{harten83}, HLLC \citep{Mignone06} and HLLD \citep{Mignone09}. In this collection of non-linear solvers, the solution is approximated by $N<7$ waves that travel with characteristic speeds $\lambda_{k+1}>\lambda_k, k=1,...,N-1$, separated by $N+1$ intermediate states. The characteristic speed of the outermost waves in the Riemann fan (i.e, fast magnetosonic waves), which are the maximum and minimum eigenvalues of the jacobian matrix of the left and right states, require finding the roots of a quartic equation \citep[see e.g.,][]{anile89,anton10}, which must be solved by any root-finding numerical method. As the number of waves and intermediate states increases, the diffusion of the method is reduced and thus we obtain more accurate representations of the real solution. Figure \ref{fan} shows a typical Riemann fan representing all different states in the three approximate solvers contained in \JLM{Lóstrego}. In HLL ($N=2$), which is the simplest method of the family, the initial discontinuity is decomposed in two fast magnetosonic waves with characteristic speeds $\lambda_L, \lambda_R$, such that the internal flux in the fan can be derived from the Rankine-Hugoniot jump conditions accross the magnetosonic waves. This method is robust and attractive for RMHD applications since it preserves positivity of pressure and $|v|<1$, but it may suffer from high levels of numerical diffusion compared to more sophisticated solvers like HLLC or HLLD. HLLC ($N=3$) involves two internal states separated by a contact wave. This solver increases significantly the accuracy of HLL, but it requires a separate treatment of the hydrodynamical limit ($B=0$) and may suffer from different pathologies when $B\rightarrow 0$, specially in three dimensional applications. For this reason, we have also included the five-wave Riemann solver HLLD ($N=5$), where besides the contact mode and the two fast magnetosonic waves, it also considers two rotational discontinuities. In this case, the solution of the system requires applying the Rankine-Hugoniot jump coditions accros all the waves in the fan of Fig. \ref{fan}. Although this solver is more complex and thus more expensive from the point of view of computational efficiency, we find it very robust and accurate for multidimensional applications. However, less dissipative solvers like HLLC and HLLD might develop undesirable artifacts \citep[see e.g.,]{quirk94,wang08} which require a careful treatment, for example, degrading the solver to HLL in the proximity of strong shocks.\\

\subsubsection{Time integration}

Once numerical fluxes are resolved at cell interfaces, conserved variables can be advance one time step following Eq. \ref{timeadv} and predictor-corrector methods. Numerical time integration is based on second (RK2) and third-order (RK3) total variation dimishing (TVD) Runge-Kutta time integrators described by \cite{shu89}. The second-order method (RK2) follows: 
\begin{equation}
\begin{aligned}
U^{(1)}=U^{(0)}+\Delta t L(U^{(0)})\\
U^{(2)}=U^{(0)}+\frac{1}{2}\Delta t L(U^{(0)})+\frac{1}{2}\Delta t L(U^{(1)}),
\end{aligned}
\end{equation}
while the third-order method (RK3) is given by:
\begin{equation}
\begin{aligned}
U^{(1)}=U^{(0)}+\Delta t L(U^{(0)})\\
U^{(2)}=U^{(0)}+\frac{1}{4}\Delta t L(U^{(0)})+\frac{1}{4}\Delta t L(U^{(1)})\\
U^{(3)}=U^{(0)}+\frac{1}{6}\Delta t L(U^{(0)})+\frac{1}{6}\Delta t L(U^{(1)})+\frac{2}{3}\Delta t L(U^{(2)}),
\end{aligned}
\end{equation}
where $L(U^{(n)})$ is the upwind differencing operator:

\begin{equation}
\begin{aligned}
    L(U^{(n)})=-\frac{1}{\Delta x}\left(\hat{\boldsymbol{F}}^x_{i+1/2,j,k}-\hat{\boldsymbol{F}}^x_{i-1/2,j,k}\right)\\
               -\frac{1}{\Delta y}\left(\hat{\boldsymbol{F}}^y_{i,j+1/2,k}-\hat{\boldsymbol{F}}^y_{i,j-1/2,k}\right)\\
               -\frac{1}{\Delta z}\left(\hat{\boldsymbol{F}}^z_{i,j,k+1/2}-\hat{\boldsymbol{F}}^z_{i,j,k-1/2}\right).
\end{aligned}               
\end{equation}

\subsubsection{Divergence-free constraint}

In general, HRSC methods do not preserve the divergence-free condition given by Eq. \ref{solenoidal}. This can produce numerical artificial forces that can lead to wrong solutions and the eventual failure of the code. Thus, there exist different approaches to control the solenoidal condition in RMHD codes, the most popular one being \citep[see e.g.,][and references therein]{toth00,marti15}: the eight-wave method \citep{powell94}, the hyperbolic-parabolic divergence cleaning \citep{dedner02}, the projection scheme \citep{brackbill80} and the Constrained Transport method \citep[CT]{evans88,Ryu98,balsara99}. The latter has been probed to be robust and accurate since it guarantees that the divergence-free constraint is fulfilled up to machine round-off errors by definition, preserving the divergence of the initial setup during the simulation. Thus, it is the only scheme we have implemented in \JLM{Lóstrego v1.0}. The method was originally proposed by \cite{evans88} and adapted to Godunov-type Riemann solvers by \cite{Ryu98,balsara99}. The main characteristic of this approach is that it requires the introduction of a staggered representation of the magnetic field at cell interfaces, which is evolved according to a semi-discrete version of the induction equation: \\
\begin{equation}
    \begin{aligned}
        \frac{dB^x_{i+1/2,j,k}}{dt}=\frac{\hat{\Omega}^z_{i+1/2,j+1/2,k}-\hat{\Omega}^z_{i+1/2,j-1/2,k}}{\Delta y}-\\
        \frac{\hat{\Omega}^y_{i+1/2,j,k+1/2}-\hat{\Omega}^y_{i+1/2,j,k-1/2}}{\Delta z},        
    \end{aligned}
\end{equation}
\begin{equation}
    \begin{aligned}
        \frac{dB^y_{i,j+1/2,k}}{dt}=\frac{\hat{\Omega}^x_{i,j+1/2,k+1/2}-\hat{\Omega}^x_{i,j+1/2,k-1/2}}{\Delta z}-\\
        \frac{\hat{\Omega}^z_{i+1/2,j+1/2,k}-\hat{\Omega}^z_{i-1/2,j+1/2,k}}{\Delta x},        
    \end{aligned}
\end{equation}
\begin{equation}
    \begin{aligned}
        \frac{dB^z_{i,j,k+1/2}}{dt}=\frac{\hat{\Omega}^y_{i+1/2,j,k+1/2}-\hat{\Omega}^y_{i-1/2,j,k+1/2}}{\Delta x}-\\
        \frac{\hat{\Omega}^x_{i,j+1/2,k+1/2}-\hat{\Omega}^x_{i,j-1/2,k+1/2}}{\Delta y},        
    \end{aligned}
\end{equation}
where $B^x_{i+1/2,j,k},B^y_{i,j+1/2,k},B^z_{i,j,k+1/2}$ are the staggered field components which represent an average of the magnetic field at cell surface:
\begin{equation}
    B^x_{i+1/2,j,k}=\frac{1}{\Delta S_{i+1/2,j,k}}\int_{S_x}\boldsymbol{B}\cdot d\boldsymbol{S}_x,
\end{equation}
\begin{equation}
    B^y_{i,j+1/2,k}=\frac{1}{\Delta S_{i,j+1/2,k}}\int_{S_y}\boldsymbol{B}\cdot d\boldsymbol{S}_y,
\end{equation}
\begin{equation}
    B^z_{i,j,k+1/2}=\frac{1}{\Delta S_{i,j,k+1/2}}\int_{S_z}\boldsymbol{B}\cdot d\boldsymbol{S}_z.
\end{equation}

The electromotive forces (emf) $\Omega^i$ are discrete representations of the components of $\boldsymbol{\Omega}=\boldsymbol{v}\times\boldsymbol{B}$ defined at cell corners. To compute them, we follow the flux-CT formalism \citep{balsara99,giacomazzo07}, taking the simple average of the neighboring upwind numerical fluxes:
\begin{equation}
\begin{aligned}
      \bar{\Omega}^x_{i,j+1/2,k+1/2}=\frac{1}{4}(\hat{F}^{yz}_{i,j+1/2,k}+\hat{F}^{yz}_{i,j+1/2,k+1} \\
       -\hat{F}^{zy}_{i,j,k+1/2}-\hat{F}^{zy}_{i,j+1,k+1/2}),     
\end{aligned}
\label{omegax}
\end{equation}
\begin{equation}
\begin{aligned}
      \bar{\Omega}^y_{i+1/2,j,k+1/2}=\frac{1}{4}(\hat{F}^{zx}_{i,j,k+1/2}+\hat{F}^{zx}_{i+1,j,k+1/2} \\
       -\hat{F}^{xz}_{i+1/2,j,k+1/2}-\hat{F}^{xz}_{i+1/2,j,k+1}) ,    
\end{aligned}
\label{omegay}
\end{equation}
\begin{equation}
\begin{aligned}
      \bar{\Omega}^z_{i+1/2,j+1/2,k}=\frac{1}{4}(\hat{F}^{xy}_{i+1/2,j,k}+\hat{F}^{xy}_{i+1/2,j+1,k} \\
       -\hat{F}^{yx}_{i,j+1/2,k}-\hat{F}^{yx}_{i+1,j+1/2,k}),     
\end{aligned}
\label{omegaz}
\end{equation}
where $\hat{F}^{\mu\nu}$ is the $\mu$-component of the flux along the $\nu$-direction.

Arithmetic averaging is the simplest and most often suggested CT scheme, but this approach may suffer from insufficient dissipation and it does not reduce to the plane-parallel algorithm for grid-align flows \citep{gardiner05,mignone21}. Thus, we modified the definitions of Eqs. \ref{omegax} to \ref{omegaz} following the CT-contact scheme of \cite{gardiner05}, where the CT algorithm is presented as a spatial integration procedure. This scheme has been probed to be robust and accurate in practical applications and it does reduce to the base upwind method for grid-aligned planar flows. In the CT-contact approximation, the average electromotive forces are modified according to:
\begin{equation}
\begin{aligned}
\hat{\Omega}^z_{i+1/2,j+1/2,k}=  \bar{\Omega}^z_{i+1/2,j+1/2,k}\\
      +\frac{\delta y}{8}\left(\left(\frac{\partial\Omega^z}{\partial y}\right )_{i+1/2,j+1/4}-\left(\frac{\partial\Omega^z}{\partial y}\right )_{i+1/2,j+3/4}\right) \\
            +\frac{\delta x}{8}\left(\left(\frac{\partial\Omega^z}{\partial x}\right )_{i+1/4,j+1/2}-\left(\frac{\partial\Omega^z}{\partial x}\right )_{i+3/4,j+1/2}\right),
\end{aligned}
\label{omegazct}
\end{equation}
where similar expressions hold for $\Omega^x,\Omega^y$ by cyclic permutation of the spatial indices. \cite{gardiner05} proposed several ways to compute the derivatives of Eq. \ref{omegazct}. Here, we follow the $\epsilon_z^c$-CT algorithm, where derivatives are selected in an upwind way according to the sign of the velocity of the contact mode:
\begin{equation}
\begin{aligned}
\left(\frac{\partial\Omega^z}{\partial y}\right)_{i+1/2,j+1/4}=\left\{\begin{matrix}
(\partial\Omega^z/\partial y)_{i,j+1/4}, & \text{if}\hspace{0.3cm} v_{x,i+1/2,j}>0, \\ 
&\\
(\partial\Omega^z/\partial y)_{i+1,j+1/4}, & \text{if}\hspace{0.3cm} v_{x,i+1/2,j}<0, \\
&\\
\frac{1}{2}((\partial\Omega^z/\partial y)_{i,j+1/4}+\\
(\partial\Omega^z/\partial y)_{i+1,j+1/4}), & \text{otherwise}.
\end{matrix}\right.
\end{aligned}
\label{bigif}
\end{equation}

To obtain the derivatives that appear in Eq. \ref{bigif}, \cite{gardiner05} propose to subtract the face centered emf $\Omega^z_{i,j+1/2}$ computed with the upwind fluxes and the emf $\Omega^z_{i,j}$ evaluated at the cell center:
\begin{equation}
\left(\frac{\partial\Omega^z}{\partial y}\right)_{i,j+1/4}=\frac{2}{\delta y}\left(\Omega^z_{i,j+1/2}-\Omega^z_{i,j}\right).
\end{equation}

The cell-center magnetic field can be recovered from the staggered representation using a simple linear interpolation at the end of the CT step:
\begin{equation}
    B^x_{i,j,k}=\frac{1}{2}(B^x_{i-1/2,j,k}+B^x_{i+1/2,j,k}),
\end{equation}
\begin{equation}
    B^y_{i,j,k}=\frac{1}{2}(B^y_{i,j-1/2,k}+B^y_{i,j+1/2,k}),
\end{equation}
\begin{equation}
    B^z_{i,j,k}=\frac{1}{2}(B^z_{i,j,k-1/2}+B^z_{i,j,k+1/2}).
\end{equation}

The interpolations performed to obtain the required fluxes at cell edges in the CT-contact formalism, and to recover the cell centered magnetic field from the staggered solution, limit the accuracy of the algorithm to second order. Other strategies like the UCT-HLL(D) method of \cite{mignone21} shall be followed to extend the CT method to higher than second order.\\


\subsubsection{Recovery of primitive variables}

Finally, in order to start a new time iteration, we must recover the primitive variables from the already evolved conserved solution. In RHD and RMHD codes, this task requires to solve a highly nonlinear algebraic system of equations, constituting one of the most challenging tasks from the point of view of computational efficiency. In fact, the recovery of primitive variables is usually considered as one of the bottlenecks for the speed-up of the code \citep{wright19}. Among the different strategies that exist in the literature \citep[see e.g,][ and references therein]{marti15}, we choose to recover the set of primitive variables following the inversion scheme of \cite{mignone07}. This algorithm can be extended to general EoS and it has been probed to avoid numerical problems due to loss of precision in the non-relativistic and ultra-relativistic limit. For example, this might be relevant for microquasar jet simulations, where we probed that other inversion schemes like the one used by \cite{leismann05,marti153} failed to recover the low pressure of the non-relativistic stellar wind with acceptable accuracy. The relevant equations of this inversion scheme are:
\begin{equation}
    E'=Z'-p+\frac{|\boldsymbol{B}|^2}{2}+\frac{|\boldsymbol{B}|^2|\boldsymbol{S}|^2-S_B^2}{2(|\boldsymbol{B}|^2+Z'+D)^2},
    \label{mignone1}
\end{equation}
\begin{equation}
    |\boldsymbol{S}|^2=(Z+|\boldsymbol{B}|^2)^2\frac{|\boldsymbol{u}|^2}{1+|\boldsymbol{u}|^2}-\frac{S_B^2}{Z^2}(2Z+|\boldsymbol{B}|^2),
    \label{mignone2}
\end{equation}
\noindent
where $E'=E-D$, $Z'=Z-D$, $Z=DhW$, $S_B=\boldsymbol{S}\cdot\boldsymbol{B}$ and $|\boldsymbol{u}|^2=W^2|\boldsymbol{v}|^2$. Equation~\ref{mignone1} can be solved for $Z'$ by means of a one-dimensional Newton-Raphson iterative method using Eq.~\ref{mignone2} to express $|\boldsymbol{u}|^2$ as a function of $Z'$. The method involves the computation of the derivative:
\begin{equation}
    \frac{dE}{dZ'}=1-\frac{dp}{dZ'}-\frac{|\boldsymbol{B}|^2|\boldsymbol{S}|^2-S_B^2}{(|\boldsymbol{B}|^2+Z'+D)^3},
\end{equation}
where $dp/dZ'$ depends on the EoS of the code. To avoid catastrophic losses of accuracy in the non-relativistic and ultrarelativistic limits, \cite{mignone07} choose $p = p(\chi, \rho)$ with $\chi=\rho\epsilon+p$ to get:
\begin{equation}
    \frac{dp}{dZ'}=\left.\frac{\partial p}{\partial \chi}\right|_\rho \frac{d\chi}{dZ'} + \left.\frac{\partial p}{\partial\rho}\right|_\chi \frac{d\rho}{dZ'}.
\end{equation}

For ideal gases, $\partial p/\partial\chi$ and $\partial p / \partial \rho$ become:
\begin{equation}
    \frac{\partial p}{\partial \chi}=\frac{\Gamma-1}{\Gamma}\chi; \hspace{0.5cm} \frac{\partial p}{\partial\rho}=0.
\end{equation}

Moreover, the variable $\chi$ can be written as a function of $Z'$ as:
\begin{equation}
    \chi=\frac{Z'}{W^2}-\frac{D|\boldsymbol{u}|^2}{(1+W)W^2},
\end{equation}
and its derivative with respect to $Z'$ as:
\begin{equation}
\frac{d\chi}{dZ'}=\frac{1}{W^2}-\frac{W}{2}(D+2W\chi)\frac{d|\boldsymbol{v}|^2}{dZ'},
    \label{chi}
\end{equation}
where:
\begin{equation}
    \frac{d|\boldsymbol{v}|^2}{dZ'}=\frac{-2}{Z^3}\frac{S_B^2[3Z(Z+|B|^2)+|B|^4]+|S|^2Z^3}{(Z+|B|^2)^3}.
    \label{dvdz}
\end{equation}
On the other hand:
\begin{equation}
    \frac{d\rho}{dZ'}=-\frac{DW}{2}\frac{d|\boldsymbol{v}|^2}{dZ'}.
\end{equation}

Once $Z'$ and $p$ have been found with some desirable degree of accuracy, the inversion scheme is completed by recovering the three-velocity and the density of the gas through:
\begin{equation}
    v^j =\frac{1}{Z+|B|^2} \left(S^j+\frac{S_B}{W}B^j\right),
\end{equation}
\begin{equation}
    \rho=\frac{D}{W}.
\end{equation}


We refer the reader to \cite{mignone07} for further details on the full algorithm. If the inversion step fails to recover a proper physical solutions (i.e., $p<0$), we
follow the same strategy than in the PLUTO code, fixing pressure to a positive threshold before solving for $|v|^2$ using a bisections root finder with the following function:
\begin{equation}
    |v|^2-\frac{S_B^2(2Z+|B|^2)+|S|^2Z^2}{(Z+|B|^2)^2Z^2}=0.
\end{equation}
Using this approach, $Z$ has to be recomputed after each iteration using Eqs. \ref{mignone1} to \ref{dvdz} .Then, total energy needs to be corrected according to the new solution.\\


\subsubsection{Correction of conserved variables}

RMHD codes based on constrained transport algorithms are subject to several pathologies that may lead to non-physical solutions, specially for highly magnetized flows under the influence of strong shocks. The decoupled evolution of the staggered and cell-centered magnetic field makes the algorithm inconsistent, since the conserved variables have been evolved according to the fluxes computed from the cell-centered field, while the staggered solution has been computed with the CT method. This inconsistency relies on the basis to develop correction algorithms of the conserved variables, which may be specially important when the magnetic pressure dominates over the gas pressure by more than two orders of magnitude \citep{marti152}. \JLM{Lóstrego v1.0} admits both the non-relativistic energy correction of \cite{Mignone06}:
\begin{equation}
    E_{\rm stag}=E_{\rm cell}-\frac{B_{\rm cell}^2-B_{\rm stag}^2}{2},
    \label{ca1}
\end{equation}
and the full-relativistic correction of both energy and momentum of \cite{marti152}. 
In this last approach, the non-relativistic correction of the energy, now $E^{(1)}_{\rm stag}$, is used to obtain a first approximation of the primitive variables and then use the new flow velocity ${\bf v}^{(1)}$ to complete the relativistic correction of the momentum and energy:
\begin{equation}
S^{i\,(2)}_{\rm stag} = S^i - ({B}^2_{\rm cell} - {B}^2_{\rm
  stag}) v^{i\,(1)} + {\bf v}^{(1)} \cdot (B^i_{\rm cell} {\bf B}_{\rm cell} - B^i_{\rm stag} {\bf B}_{\rm stag}),
  \label{ca2m}
\end{equation}
\begin{equation}
E^{(2)}_{\rm stag} = E^{(1)}_{\rm stag} - \frac{({v}^{(1)})^2}{2} ({B}^2_{\rm cell} - {B}^2_{\rm
  stag}) + \frac{({\bf v}^{(1)} \cdot {\bf B}_{\rm
    cell})^2 - ({\bf v}^{(1)} \cdot {\bf B}_{\rm
    stag})^2}{2}.
    \label{ca2e}
\end{equation}
Although these algorithms can be applied before or after recovering the primitive variables, in \JLM{Lóstrego} we decided to do the correction before. Following the notation of \cite{marti152}, these algorithms are called CA1 (Eq. \ref{ca1}) and CA2 (Eq. \ref{ca2m} and Eq. \ref{ca2e}), respectively.

\subsection{Parallelization}

\JLM{Lóstrego v1.0} is parallelized with an hybrid scheme using OpenMP (OMP) and MPI library instructions in order to exploit the computational power of both distributed and shared memory. This parallelization configuration is based on the parallel architecture of the code Ratpenat \citep{perucho101}. The hybrid scheme is only available in the code for full three dimensional simulations, while two-dimensional computations like the tests described in the next section might benefit from shared memory parallelization with OMP directives. For those large 3D simulations where the hybrid scheme is available, the computational box is initially decomposed in several sub-domains such that each one is assigned to a MPI node that will perform all the calculations independently using shared-memory OMP threads. At the beginning of each time step, all MPI blocks that share internal boundaries must interchange a collection of \textit{ghost} cells with the neighbours in the three spatial directions. These cells will be used as boundary conditions for those blocks that do not limit with the physical boundaries of the grid. The latter are imposed according to the physical conditions on the box boundaries, which are particular for each numerical simulation. Moreover, the implementation of the CT-contact algorithm (see Sec.~\ref{methods}) requires one more MPI communication step before the time evolution of the staggered components of the magnetic field. This CT algorithm requires not only the direction-wise boundaries in the three spatial directions, but also the corners of each sub-box. Nevertheless, this information can be also transmitted direction-wise, since the corners are already available in the neighbouring boundaries after the first MPI comunnication.

\subsection{Testing benchmark}


We provide an extensive collection of one-dimensional and multi-dimensional numerical tests to probe the performance of the new \JLM{Lóstrego v1.0} code. Unless otherwise stated, all tests shown in this section were performed with the HLLD Riemann solver and the second-order piecewise linear method (PLM) with the VAN LEER slope limiter for cell reconstruction. A weak form
of flattening is introduced degrading the slope limiter to MINMOD
whenever a strong shock is detected \citep{Mignone06}, but no degradation of the HLLD Riemann Solver is applied in the tests. We used a third order TVD-preserving Runge-Kutta algorithm for time integration with CFL=0.3. The magnetic field divergence-free constraint is preserved with the Constrained Transport (CT) method, where electromotive forces were averaged according to the CT-contact formalism. The relativistic correction scheme CA2 has been used to correct the conserved variables after each time integration.

\subsubsection{One-dimension}


\paragraph{Sinusoidal perturbation}

\begin{figure}
  \centerline{\includegraphics[width=\columnwidth]{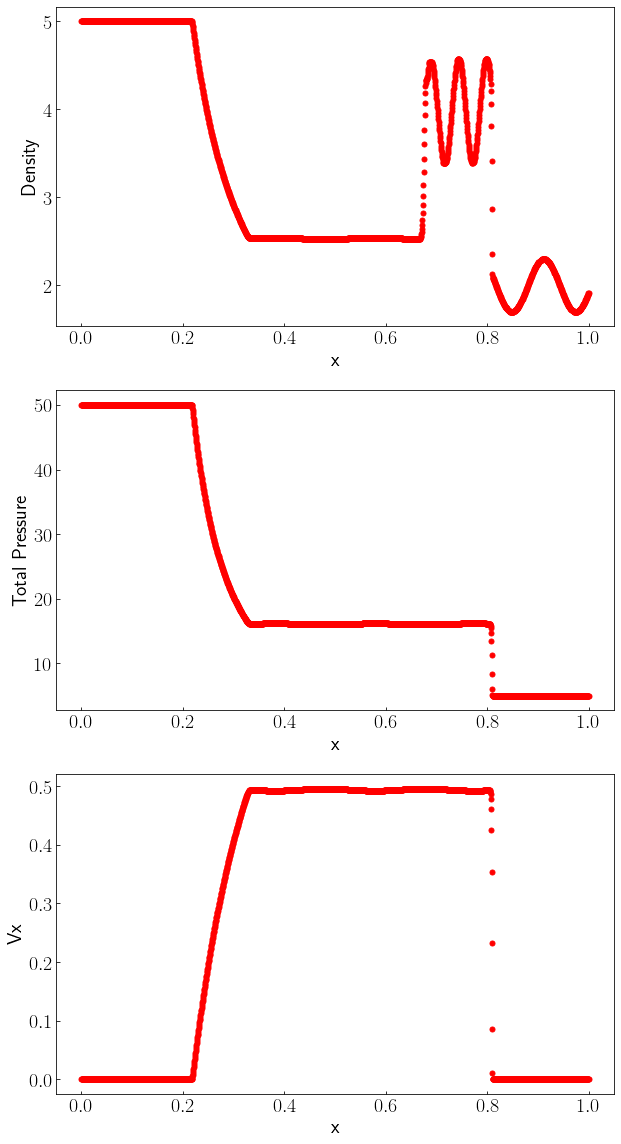}}
  \caption{Gas density (top), pressure (middle) and $x$-velocity (bottom) at $t=0.35$ for the hydrodynamical sinusoidal perturbation test. The shock tube contains 2000 grid points in the $x$-direction.}
\label{DZBU}
\end{figure}

As a first 1-D application we show a pure hydrodynamical problem initially formulated by \cite{Dolezal95} with a nuclear EoS and adapted to ideal gases in \cite{Zanna02}. A one-dimensional unitary shock tube with a resolution of 2000 computational zones is perturbed in the right-hand zone $(0.5<x<1.0)$ with a density sinusoidal profile such that $\boldsymbol{U}_L=(\rho,v,p)=(5,0,50)$ and $\boldsymbol{U}_R=(\rho,v,p)=(2+0.3\sin{(50x)},0,5)$. Figure \ref{DZBU} shows the solution of the problem at $t=0.35$. The original jump in pressure generates a blast wave that interacts with the density perturbation. \JLM{Qualitatively}, our results are similar to those presented in \cite{Zanna02}.

\paragraph{RMHD shock tubes}

\begin{figure*}
  \centerline{\includegraphics[width=\textwidth]{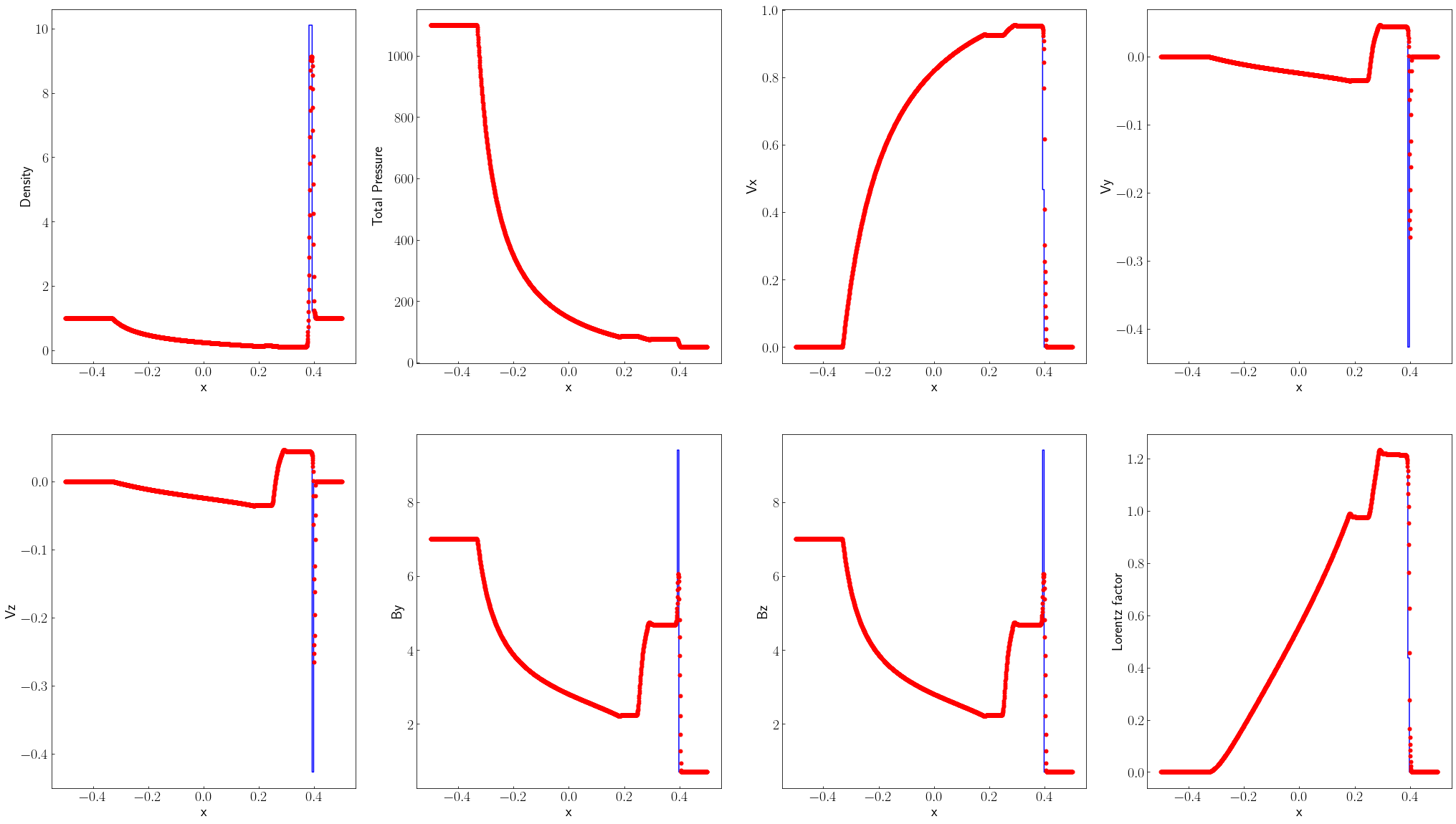}}
  \caption{From left to right (top): Density, total pressure, $x$-velocity and $y$-velocity. From left to right (bottom):$z$-velocity, $B_y$, $B_z$ and logarithmic Lorentz factor at $t=0.4$ for the test BA3 (red dots). The analytic solution of the Riemann problem is over-plotted (blue solid line). The shock tube contains 1600 grid points in the $x$-direction.}
\label{Ba3}
\end{figure*}

\begin{figure*}
  \centerline{\includegraphics[width=\textwidth]{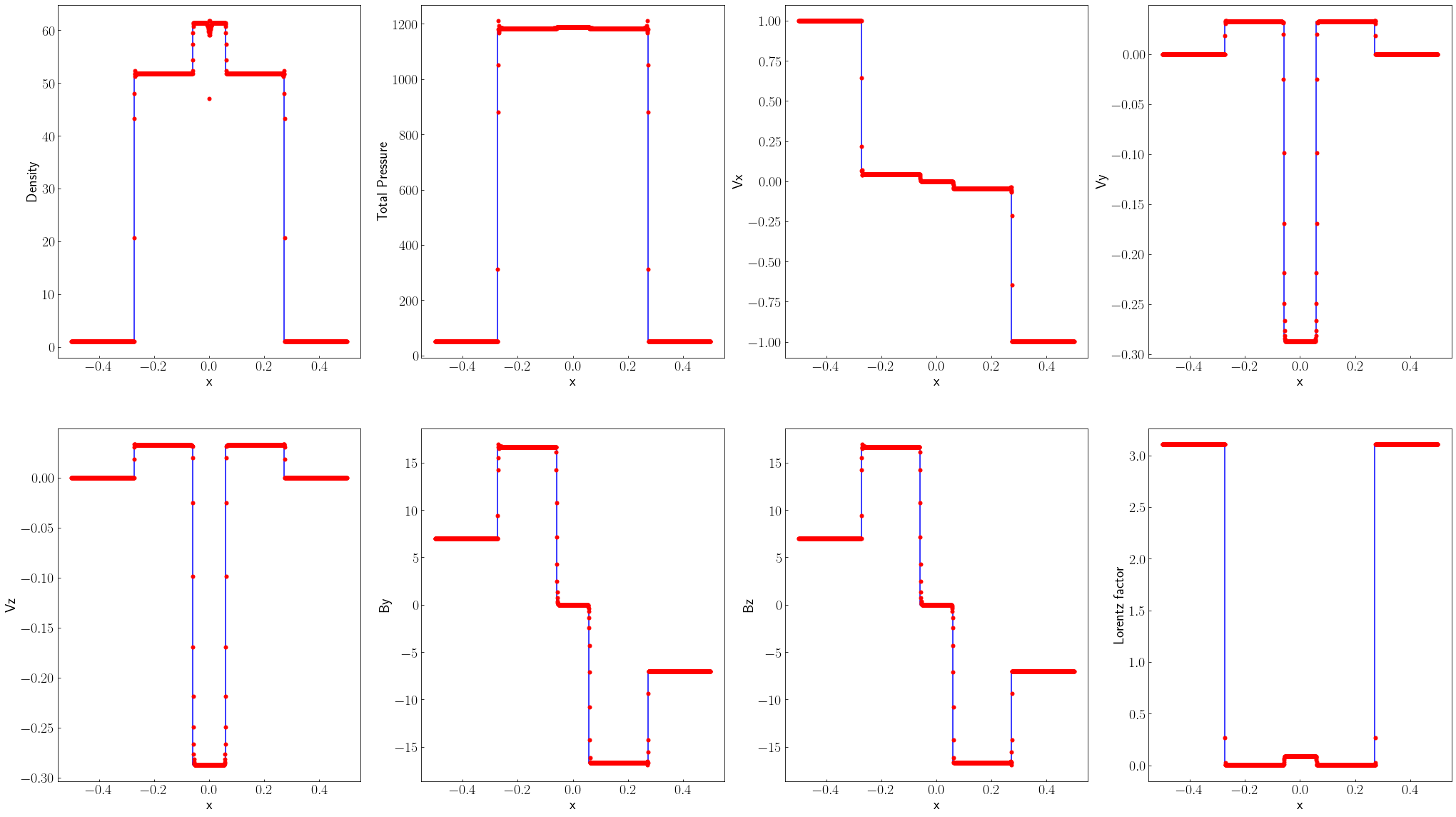}}
  \caption{From left to right (top): Density, total pressure, $x$-velocity and $y$-velocity. From left to right (bottom): $z$-velocity, $B_y$, $B_z$ and logarithmic Lorentz factor at $t=0.4$ for the test BA4 (red dots). The analytic solution of the Riemann problem is over-plotted (blue solid line). The shock tube contains 1600 grid points in the $x$-direction.}
\label{Ba4}
\end{figure*}

A collection of 1D RMHD test problems were initially proposed by \cite{dubal91,putten93} and extended in \cite{komissarov99,balsara01}. These tests constitute a set of Riemann problems that have been established as a benchmark to test the robustness, accuracy, stability and diffusion of any RMHD code. We consider the five shock tube problems of \cite{balsara01} (hereinafter, BA1-5), \JLM{although we only show the solution of two selected tests}. The initial conditions of these \JLM{five} tests can be found in the aforementioned reference, so we do not reproduce them here. For all of these tests, we considered a one-dimensional unit grid with a resolution of 1600 cells. The adiabatic coefficient is set to $\Gamma=5/3$ for all tests except for BA1, where $\Gamma=2.0$.  \JLM{The BA1 problem}, which was originally proposed by \cite{brio88} and extended to relativistic MHD by \cite{putten93}, describes the formation of a left-going fast rarefaction wave, a left-going compound wave (that only appears in the numerical solution), a contact discontinuity, a right-going slow shock and a right-going fast rarefaction wave in a moderate relativistic flow. The right part of the shock tube is magnetically dominated. \JLM{The solution of this test is not shown in the paper}. BA2 and BA3 tests are both blast waves problems, although the initial pressure jump is moderate for BA2 and strong for BA3.  The \JLM{BA2 problem, which is not shown in this paper neither,} is well-resolved \JLM{with our code} and we clearly identify a left-going fast rarefaction wave, a left-going slow rarefaction, a contact discontinuity, a right-going slow shock and a right-going fast shock. Figure \ref{Ba3} shows the solution of the BA3 test at $t=0.4$ (red dots) and the analytical solution provided by \cite{giacomazzo06} (blue line). The collection of waves that develops as a consequence of the initial pressure discontinuity is similar to those obtained in BA2, but due to the jump in pressure of several orders of magnitude, it develops a strong relativistic flow. As a direct consequence of the large Lorentz factors achieved, the contact discontinuity and the right-going shocks are under-resolved, even with the less diffusive HLLD Riemann solver. This was not the case of BA2, where a weaker blast wave leads to lower Lorentz factors and thus, mildly relativistic flows. The solution of the BA4 test at $t=0.4$ is shown in Fig. \ref{Ba4} (red dots), over-plotted with the analytical solution provided by \cite{giacomazzo06} (blue line) for this test. This problem was originally proposed in \cite{noh87} and it describes the interaction of two streams moving in opposite direction with high Lorentz factor, so the problem is strongly relativistic. As a byproduct of this collision, four shocks are generated: two strong external fast shocks and two internal slow shocks, one of each left-going and right-going, respectively.  \JLM{Finally, BA5} is the only generic RMHD Riemann problem that allows all seven waves to appear after the decay of the initial discontinuity: a left-going fast shock, a left-going Alfvén wave, a left-going slow rarefaction, a contact discontinuity, a right-going slow shock, a right-going Alfven wave and a right-going fast shock. The initial setup includes both transverse components of velocity and magnetic field vectors. \JLM{The results of this test, which are not shown in this section, show that the code is able to capture the whole set of waves predicted by the analytical solution}.



\subsubsection{Two-dimensions}
\label{2D}


\paragraph{Cylindrical Magnetized Blast Wave}

\begin{figure*}
  \includegraphics[width=\columnwidth]{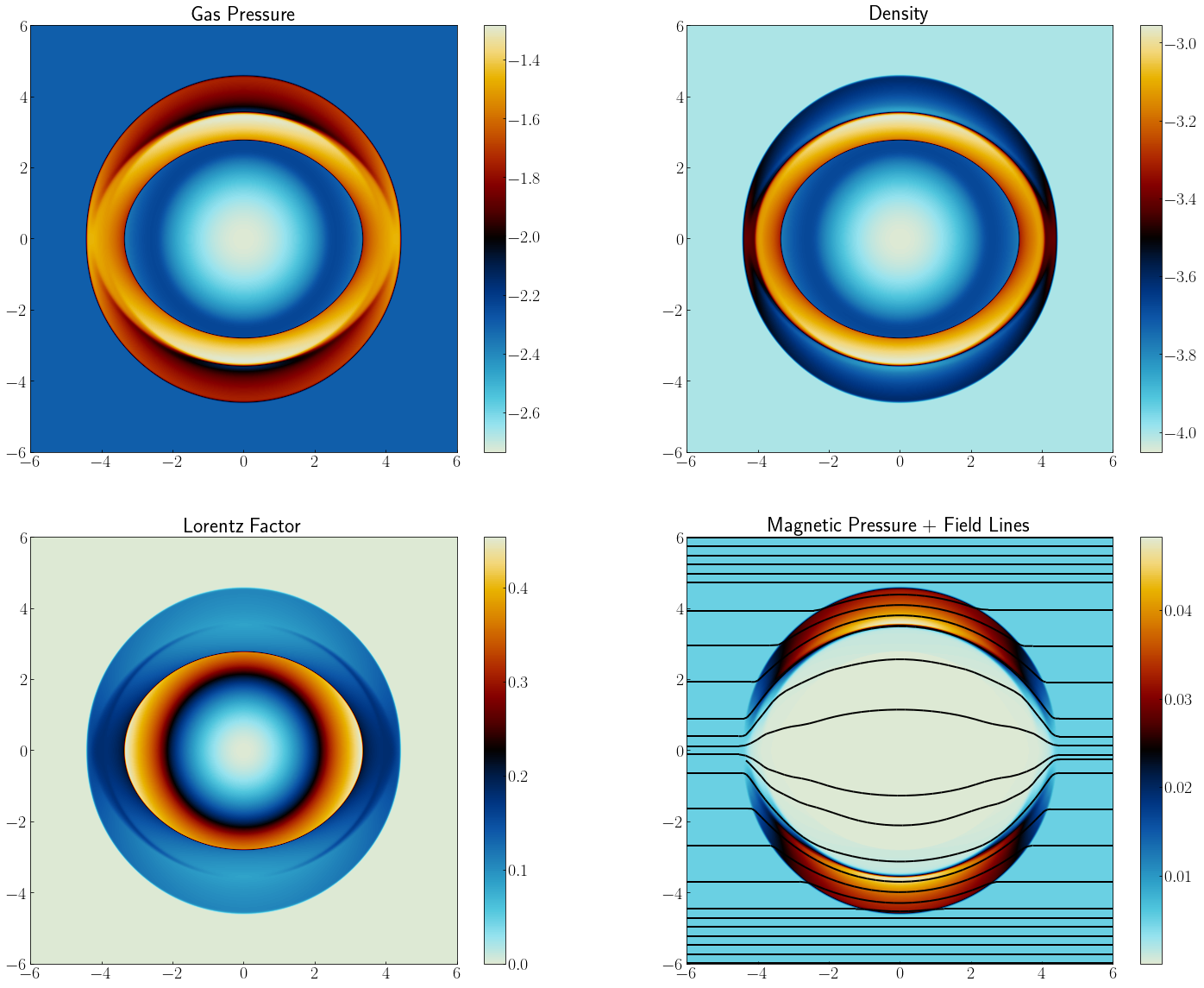}
  \includegraphics[width=\columnwidth]{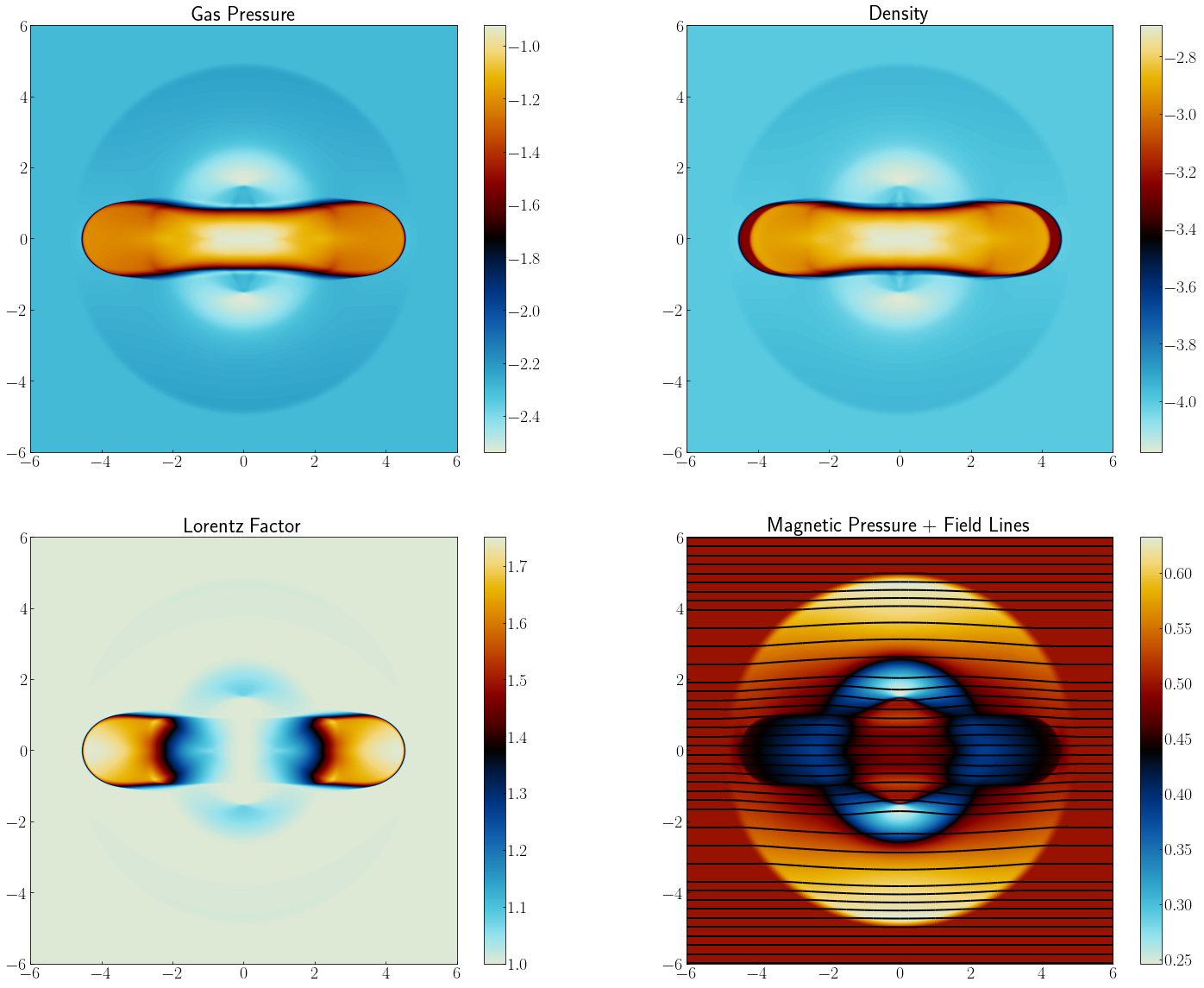}
  \caption{Logarithmic gas pressure (top left), logarithmic density (top right), logarithmic Lorentz factor (bottom left) and logarithmic magnetic pressure (bottom right) at $t=0.4$ for the two-dimensional cylindrical magnetized blast wave with moderate (two left hand columns) and strong (two right hand columns) magnetization. Magnetic field lines are superposed to the magnetic pressure. We consider the Cartesian grid $[-6,6]\times[-6,6]$ with 1024 cells per spatial dimension.}
\label{CMBW}
\end{figure*}

The cylindrical magnetized blast wave is a classical problem in RMHD to test the performance of the code handling MHD wave degeneracies parallel and perpendicular to the field orientation \citep[see e.g.,][and references therein]{marti15}. Our initial setup follows the version proposed by \cite{beckwith11} and adapted from \cite{leismann05}. We consider the domain $[-6.0,6.0]^2$ with $1024^2$ cells and free-flow boundaries everywhere. An over-pressured ($p_{\mathrm{s}}=1$) and over-dense ($\rho_{\mathrm{s}}=10^{-2}$) cylinder of radius $r=0.8$ is placed at the center of the grid, filled with an homogeneous ambient medium of density $\rho_{\mathrm{a}}=10^{-4}$ and pressure $p_{\mathrm{a}}=5\times 10^{-3}$. A transition layer between $r=0.8$ and $r=1$ is established to smooth the initial jump and avoid numerical problems when the flow starts to propagate outwards. All velocities are initially set to zero and the magnetic field is aligned with the x-axis in the whole grid. For this test, we consider two degrees of magnetization: moderate ($B=0.1$) and strong ($B=1.0$). The adiabatic coefficient is set to the relativistic value, $\Gamma=4/3$. The solution of the problem at $t=4.0$ is shown in Fig. \ref{CMBW}. The difference in pressure between the cylindrical region and the ambient medium produces the expansion of the central region which is delimited by a strong forward shock propagating radially near the speed of light. When we increase the magnetization of the medium ($B=1.0$), the strong sideways magnetic confinement produces an elongated flow structure. The test preserves the symmetry with good accuracy and no numerical artifacts nor instabilities appear in our simulation.

\paragraph{Rotor}

\begin{figure}
  \centerline{\includegraphics[width=\columnwidth]{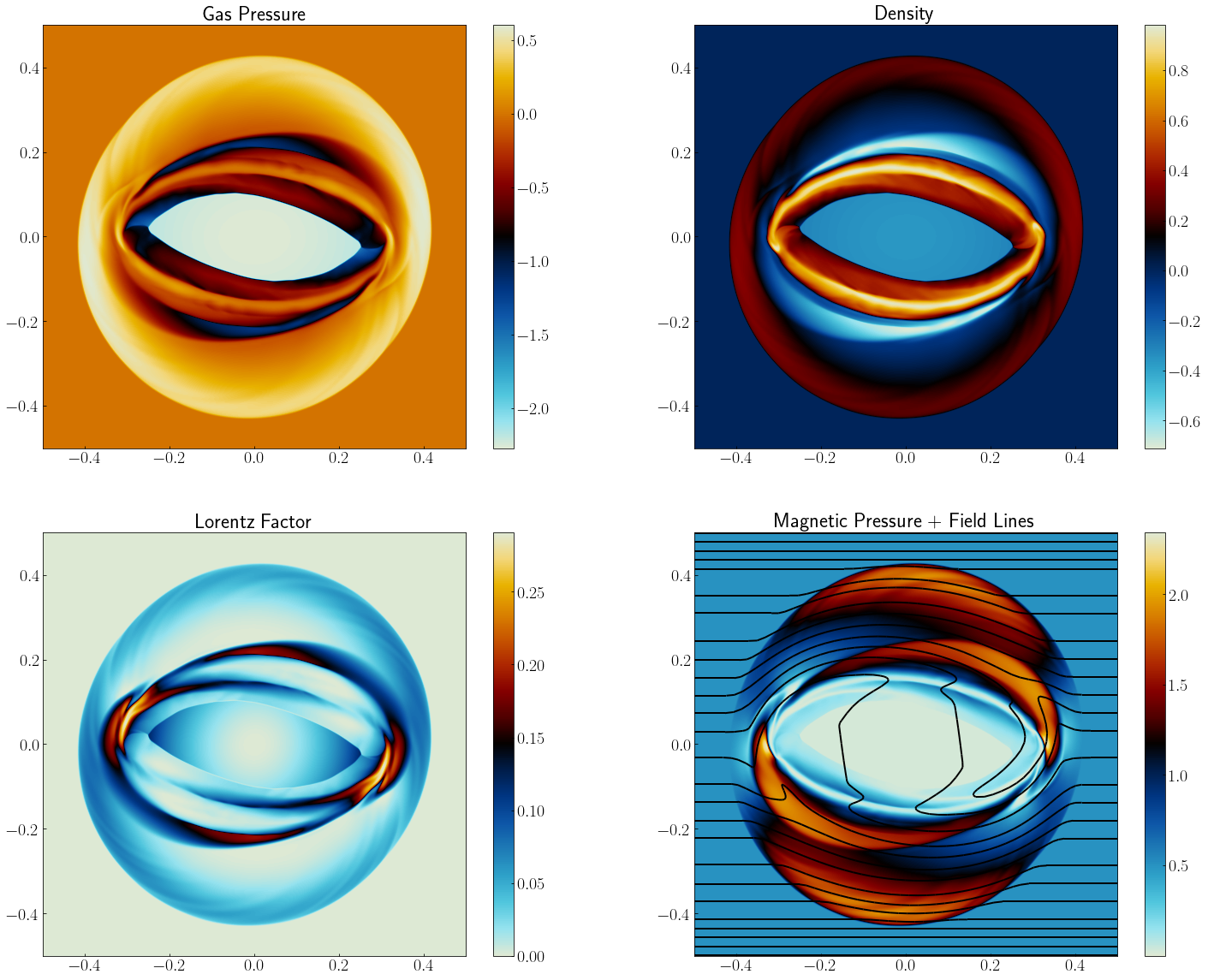}}
  \caption{Logarithmic gas pressure (top left), logarithmic density (top right), logarithmic Lorentz factor (bottom left) and logarithmic magnetic pressure (bottom right) at $t=0.4$ for the two-dimensional relativistic rotor problem. Magnetic field lines are superposed to the magnetic pressure.We consider the unit square $[-0.5,0.5]\times[-0.5,0.5]$ with 1024 cells per spatial dimension.}
\label{RROT}
\end{figure}

We consider the relativistic rotor problem of \cite{zanna03} in a unitary Cartesian two-dimensional grid with a resolution of $1024^2$ cells and free-flow boundary conditions in all directions. The initial setup consists of a disk with radius $r=0.1$ rotating with an angular relativistic velocity $\omega=0.95$. The disk is ten times denser than the ambient medium ($\rho_{\mathrm{r}}=10$, $\rho_{\mathrm{a}}=1$) and all the system is in pressure equilibrium with $p=1$. The background is initially at rest ($\boldsymbol{v}=0$) and the magnetic field is aligned with the $x$ direction in the whole grid, $B_x=1.0,B_y=0,B_z=0$. An adiabatic coefficient of $\Gamma=5/3$ was used for this test. The solution of the problem at $t=0.4$ is shown in Fig. $\ref{RROT}$. The results show complex wave patterns and torsional Alfvén waves that are generated due to the rotation of the disk. At the end of the simulation, the magnetic field lines inside the rotor are almost perpendicular with respect to the background field. All the system is embedded in a fast rarefaction.  The initial over-density was swept away by the torsional waves and distributed in a thin oblique shell.

\paragraph{Shock-Cloud interaction}

\begin{figure}
  \centerline{\includegraphics[width=\columnwidth]{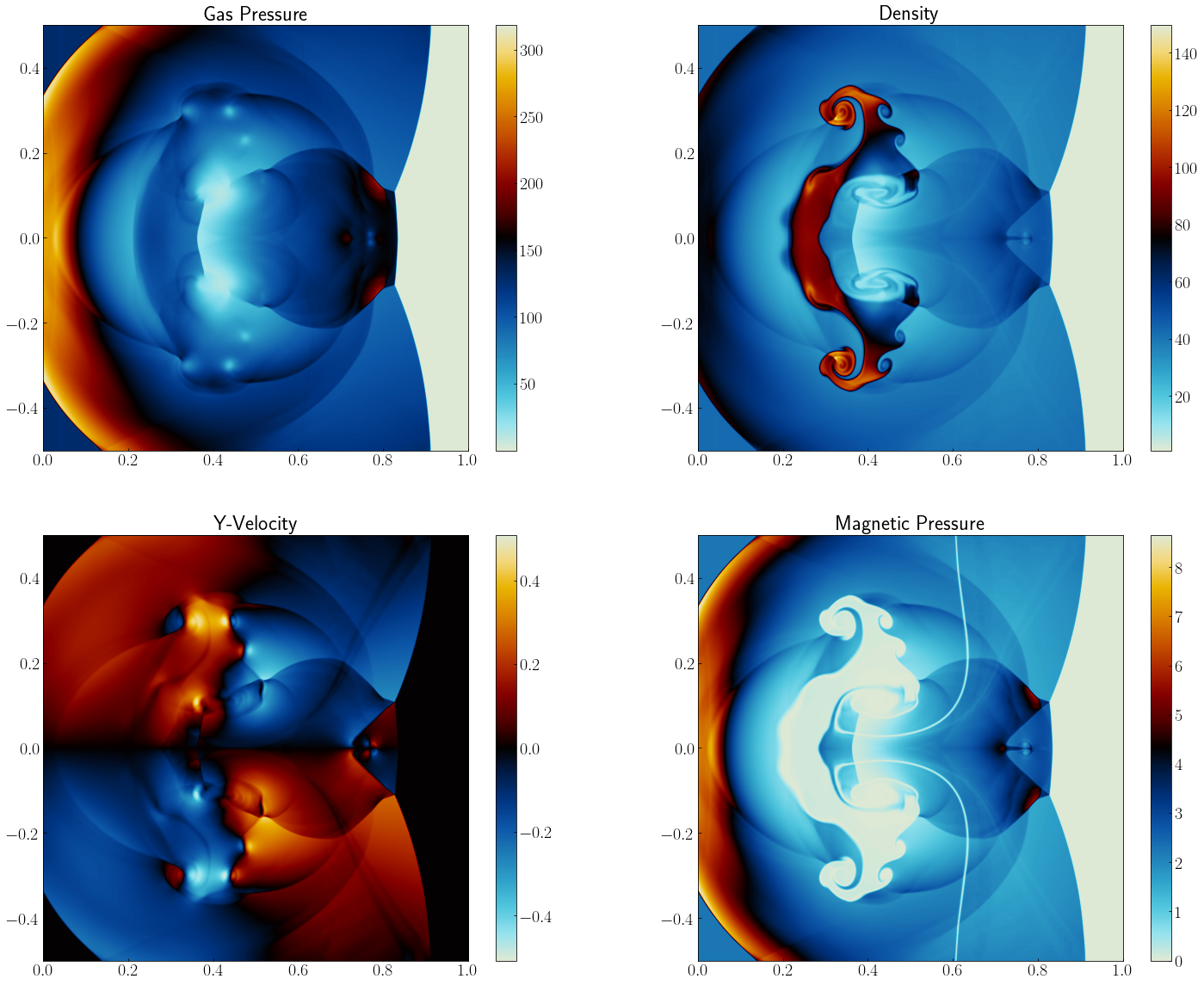}}
  \caption{Gas pressure (top left), density (top right), $y$-velocity (bottom left) and magnetic pressure (bottom right) at $t=1.0$ for the two-dimensional shock-cloud interaction test. We consider the unit square $[0,1]\times[-0.5,0.5]$ with 1024 cells per spatial dimension.}
\label{IMPL}
\end{figure}

A multi-dimensional test with relevance in astrophysical applications is the interaction of a strong shock wave with a density clump. Here, we consider the relativistic version of the problem proposed by \cite{Mignone06}, where the magnetic field is orthogonal to the plane and carries a rotational discontinuity. We set a Cartesian two-dimensional grid with dimensions $[0,1]\times[-0.5,0.5]$ and a resolution of $1024^2$ cells, with outflow boundaries in all directions. The shock wave is initially located at $x=0.6$ such that the pre-shocked values (i.e., $x>0.6$) are $(\rho, W_x, p_{\mathrm{g}}, B_z)=(1.0,10,10^{-3},0.5)$ (with the flow propagating to the left) and the shocked gas at $x<0.6$ is given by $(\rho,W_x, p_{\mathrm{g}}, B_z )=(42.5942,1.0,127.9483,-2.12917)$, where $\rho$ is the density, $W_x$ is the Lorentz factor, $p_{\mathrm{g}}$ is the gas pressure and $B_z$ is the z-component of the magnetic field. The transverse components of the velocity $v_y$ and $v_z$ and the components of the magnetic field $B_x$ and $B_y$ are initially equal to zero in the whole domain. The cloud is characterized as a cylinder with radius $r=0.15$ and density $\rho=10.0$, centered at $x=0.8$ in pressure equilibrium with the pre-shocked material. An adiabatic coefficient of $\Gamma=4/3$ was used for this test. The solution of the problem at $t=1.0$ is shown in Fig. $\ref{IMPL}$. Immediately after the impact between the cloud and the shock wave, the sphere experiments a strong compression that increases significantly the density  of  the  clump.  As a byproduct  of  this  collision,  a bow shock propagates to the left in the shocked material and a reverse shock is transmitted to the right, penetrating into the cloud and producing a mushroom-shaped structure. The overall evolution of the cloud after the impact of the shocked wave and the development of a mushroom-shaped shell is in good agreement with \cite{Mignone06}. However, the higher resolution employed in our work and the use of the HLLD Riemann solver produce a less diffusive result where several complex wave patterns and eddies are noticeable in the arms of the shell.

\paragraph{Relativistic Orszag-Tang vortex}

The Orszag-Tang vortex problem was originally proposed in \cite{Orszag79} and it has become a classical test for newtonian MHD applications \citep{Ryu98,Londrillo00}. We adapted the relativistic version of the test proposed by \cite{castro17} into a two dimensional \JLM{Cartesian} grid with dimensions $[0,6]\times[0,6]$ and a resolution of $1024^2$ cells, with periodic boundary conditions in all directions. Initially, density ($\rho$) and pressure ($p$) are set to $1.0$ with an adiabatic index of $\Gamma=4/3$. The velocity field in the laboratory frame is given by:

\begin{equation}
    \boldsymbol{v}=(-\frac{0.75}{\sqrt{2}}\sin{y},\frac{0.75}{\sqrt{2}}\sin{x},0),
\end{equation}

\noindent
while the proposed magnetic field configuration is:

\begin{equation}
    \boldsymbol{B_0}=(-\sin{y},\sin{2x},0).
\end{equation}

\begin{figure}
  \centerline{\includegraphics[width=\columnwidth]{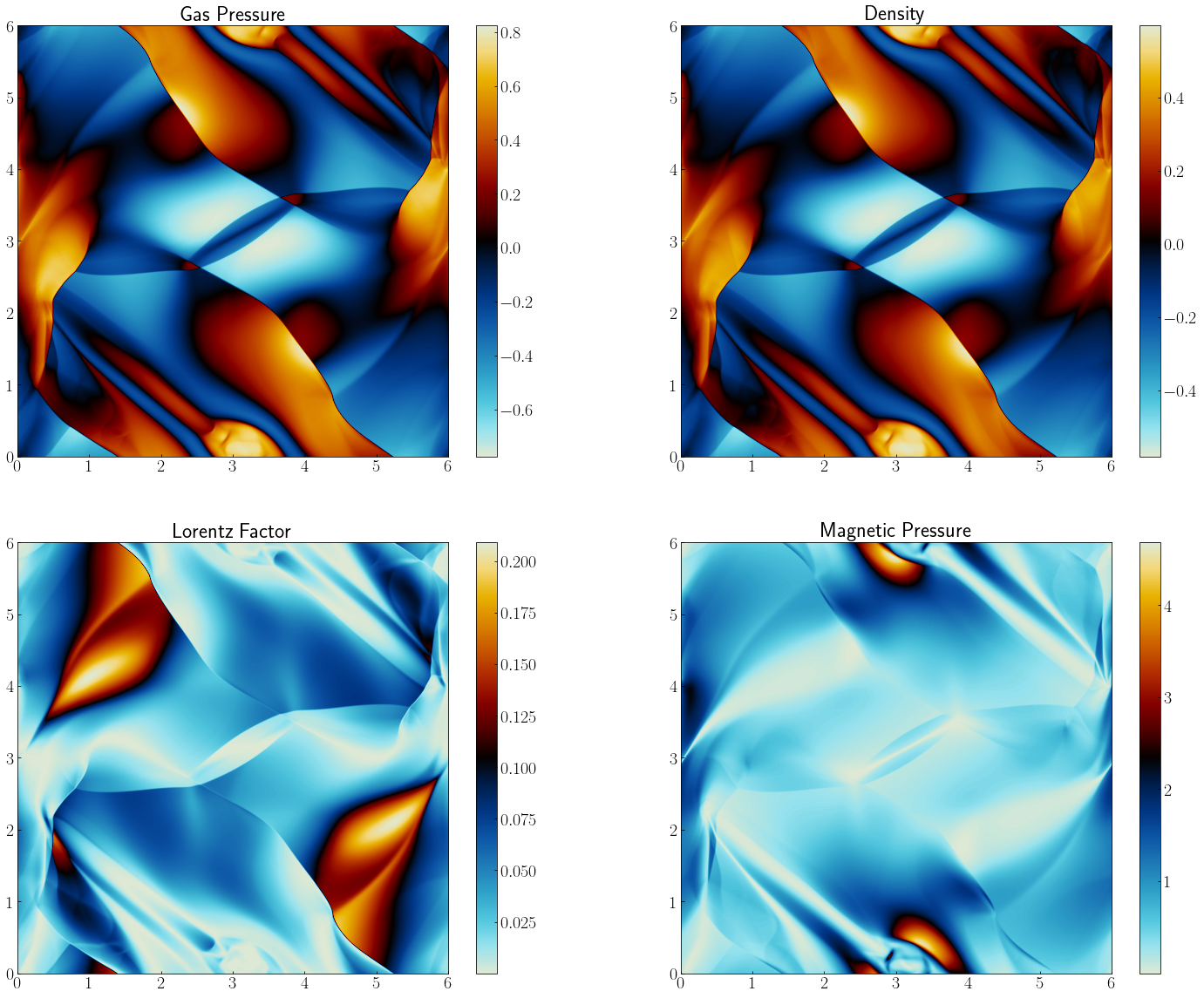}}
  \caption{Logarithmic pressure (top left), logarithmic density (top right), logarithmic Lorentz factor (bottom left) and magnetic pressure (bottom right) at t=4.0 for the relativistic Orszag-Tang vortex test. We consider the two-dimensional \JLM{Cartesian} grid $[0.0,6.0]\times[0.0,6.0]$ with 1024 cells per spatial dimension.}
\label{ORZ}
\end{figure}

Figure $\ref{ORZ}$ shows the solution to the problem at $t=4.0$. Our results are in good agreement with those obtained in \cite{castro17}. This test validates the performance of the code describing the transition to supersonic MHD turbulence and its ability to handle the formation of shocks and shock-shock interactions in the relativistic domain. 

\paragraph{Relativistic Kelvin-Helmholtz Instability}

\begin{figure}
  \centerline{\includegraphics[width=\columnwidth]{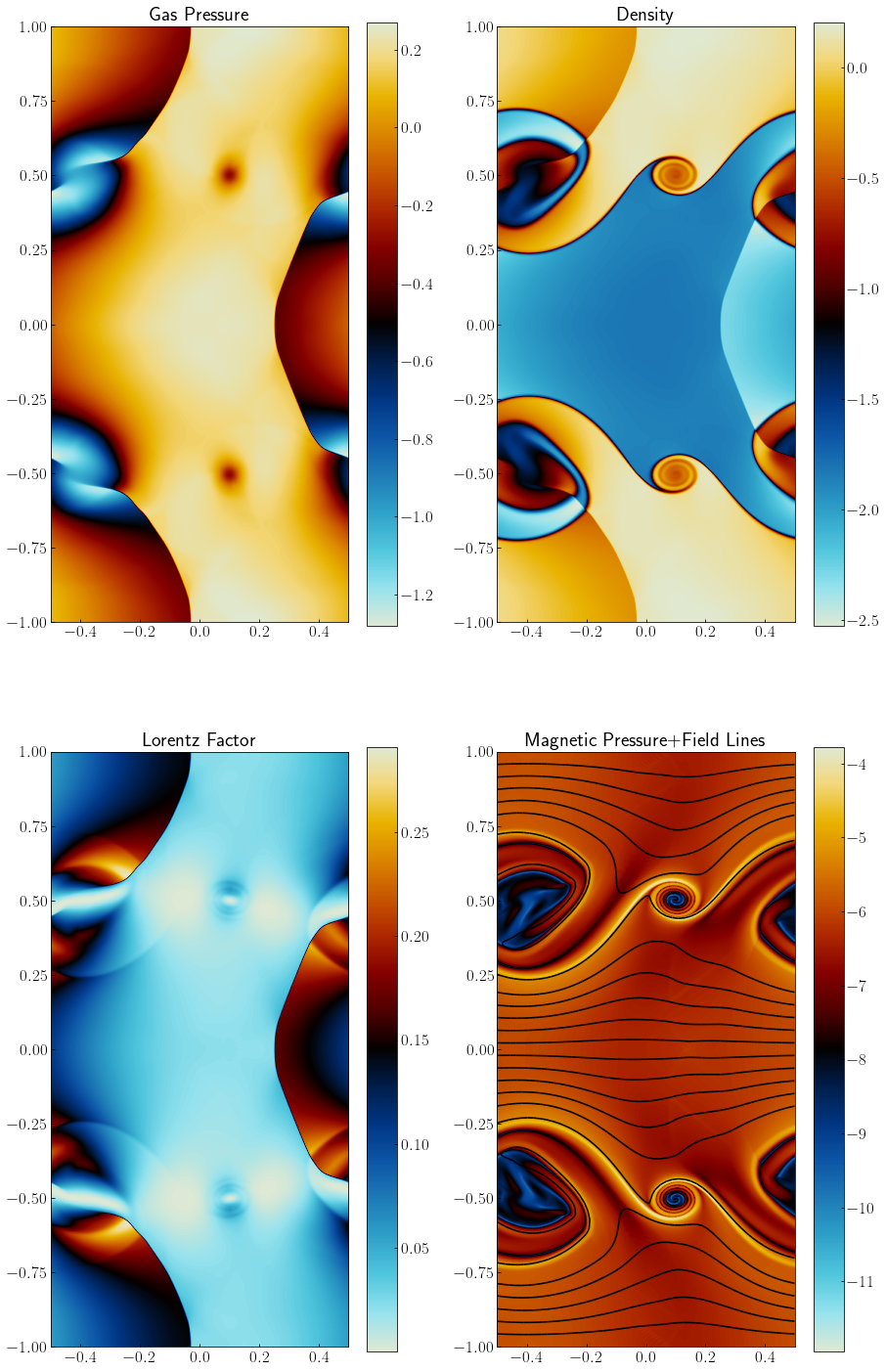}}
  \caption{Logarithmic pressure (top left), logarithmic density (top right), logarithmic Lorentz factor (bottom left) and magnetic pressure (bottom right) at t=3.0 for the relativistic Kelvin-Helmholtz test. Magnetic field lines are superposed to the magnetic pressure. We consider the two-dimensional \JLM{Cartesian} grid $[-0.5,0.5]\times[-1.0,1.0]$ with $512\times 1024$ cells.}
\label{KH2D}
\end{figure}

The last two-dimensional test that we propose is the linear growth phase of the relativistic Kelvin-Helmholtz instability (KHI). As the shock-cloud interaction problem, this test is also relevant for astrophysical application since KHIs are commonly present in nature. In particular, this test may be useful to probe the performance of the code describing turbulence zones and the description of the growth of small scale perturbations. KHIs develops when there is a velocity shear in a continuous flow or when there is a velocity difference across two separated flow states. Thus, it is an useful setup to test the response of the code when unstable initial conditions are provided. We consider a two dimensional Cartesian grid with dimensions $[-0.5,0.5]\times [-1.0,1.0]$ and resolution $512\times 1024$ cells, with periodic boundaries. We follow the initial configuration given by \cite{beckwith11}, which in turn is adapted from \cite{Mignone09} and exploited by other authors \citep{castro17}. The shear velocity and density profiles are given by:

\begin{equation}
V^x=\left\{\begin{matrix}
V_{\mathrm{shear}}\tanh{\displaystyle{\left(\frac{y-0.5}{a}\right)}} & \text{if}\hspace{0.3cm} y>0 \\ 
\\
-V_{\mathrm{shear}}\tanh{\displaystyle{\left(\frac{y+0.5}{a}\right)}} & \text{if}\hspace{0.3cm} y<0
\end{matrix}\right.,   
\end{equation}

\begin{equation}
\rho=\left\{\begin{matrix}
\rho_0+\rho_1\tanh{\displaystyle{\left(\frac{y-0.5}{a}\right)}} & \text{if}\hspace{0.3cm} y>0 \\ 
\\
\rho_0-\rho_1\tanh{\displaystyle{\left(\frac{y+0.5}{a}\right)}} & \text{if}\hspace{0.3cm} y<0
\end{matrix}\right.,   
\end{equation}

\noindent
where $a=0.01$ is the characteristic thickness of the shear layer and $V_{\mathrm{shear}}=0.5$ determines the velocity profile. The characteristic densities are $\rho_0=0.505$,  and $\rho_1=0.495$. The instability is triggered by a perturbation in the transverse velocity $V_y$ of the form:

\begin{equation}
V^y=\left\{\begin{matrix}
A_0V_{\mathrm{shear}}\sin{(2\pi x)}\exp{\left[-((y-0.5)/\sigma)^2\right]} & \text{if}\hspace{0.3cm} y>0 \\ 
-A_0V_{\mathrm{shear}}\sin{(2\pi x)}\exp{\left[-((y+0.5)/\sigma)^2\right]} & \text{if}\hspace{0.3cm} y<0 
\end{matrix}\right.,   
\end{equation}

\noindent
where $A_0=0.1$ is the amplitude of the perturbation and $\sigma=0.1$ the characteristic length scale. The gas pressure is constant everywhere, $p=1$, and we consider an ideal EoS with $\Gamma=4/3$. The magnetic field is aligned with the $x$-axis, $\boldsymbol{B}=(10^{-3},0,0)$. Figure \ref{KH2D} shows the solution of the problem at $t=3.0$ , which is qualitative similar to the one present in the aforementioned papers. In our simulation, we have also resolved the secondary vortex near $x=0.1$, although \cite{beckwith11} shows that it can be annihilated by using more diffusive Riemann solvers like HLL. As pointed by \cite{castro17}, these secondary instabilities may be non-physical and their appearance depends on the Riemann solver and the resolution employed in the simulation.

\subsubsection{Three-dimensions}


\paragraph{Spherical Magnetized Blast Wave} 

\begin{figure*}
  \centerline{\includegraphics[width=\textwidth]{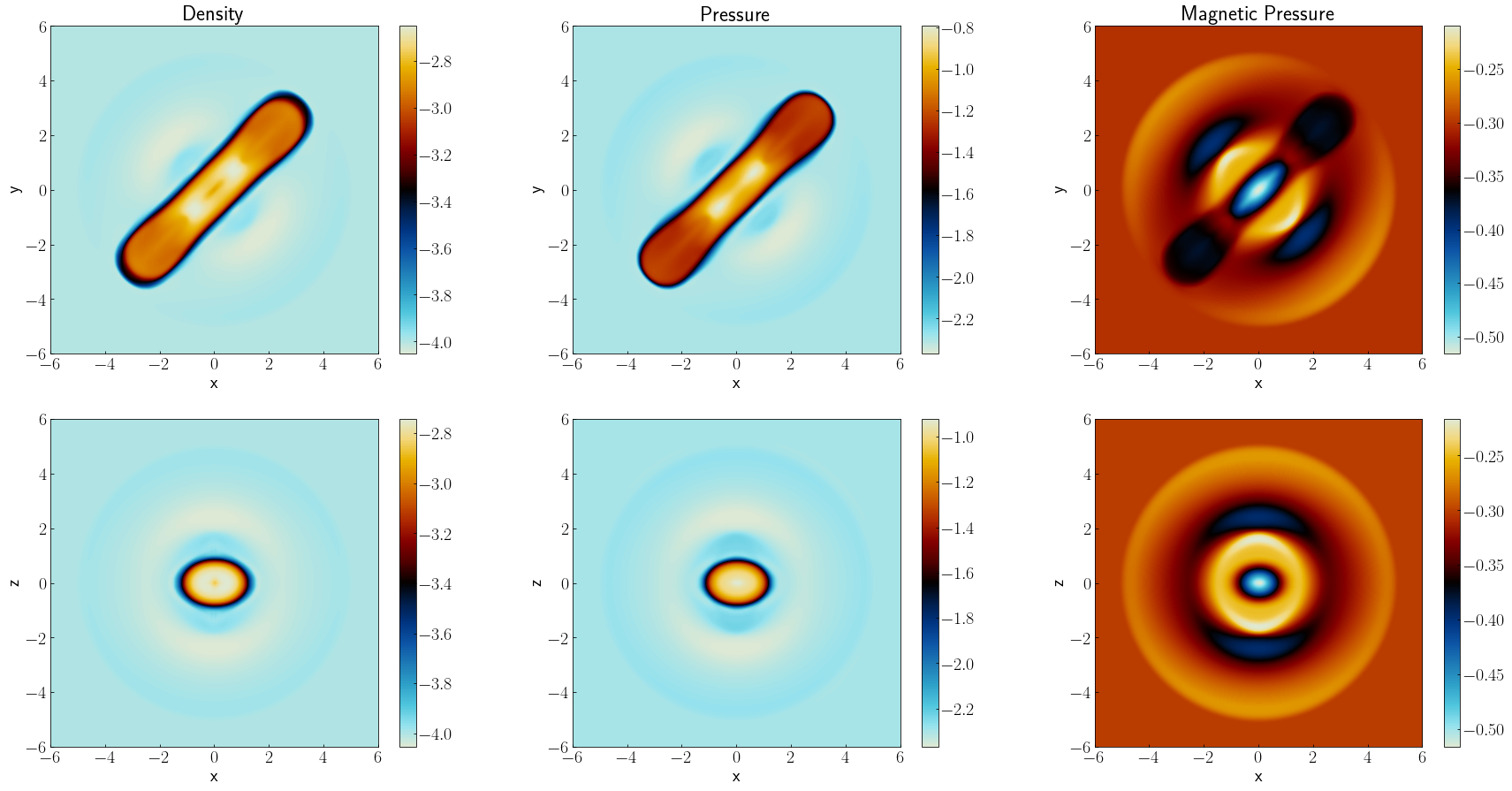}}
  \caption{Logarithmic density (left), logarithmic pressure (middle) and logarithmic magnetic pressure (right) at $t=4.0$ for the spherical magnetized blast wave in the $xy$ plane at $z=0$ (top) and $xz$ plane at $y=0$ (bottom). We consider the three-dimensional Cartesian grid $[-6.0,6.0]\times[-6.0,6.0]\times[-6.0,6.0]$ with 256 cells per spatial dimension.}
\label{SMBW}
\end{figure*}

The spherical magnetized blast wave allows to test the ability of the code to handle parallel and perpendicular strong three dimensional shocks in magnetized plasma. The initial setup is similar to the one described in the cylindrical blast wave problem in 2D (see Sec. \ref{2D}). We consider the domain $[-6.0,6.0]^3$ with $256^3$ cells and free boundaries everywhere. An over-pressured and over-dense sphere of radius $r=0.8$, density  $\rho_{\mathrm{s}}=10^{-2}$, and pressure $p_{\mathrm{s}}=1$ is placed in the center of the grid, where  $\rho_{\mathrm{a}}=10^{-4}$, $p_{\mathrm{a}}=5\times 10^{-3}$ are the density and pressure of the ambient medium, respectively. A transition layer between $r=0.8$ and $r=1$ is established to smooth the initial jump and to avoid numerical problems when the flow starts to propagate. All velocities are initially set to zero. The adiabatic coefficient is set to the relativistic value, $\Gamma=4/3$. In this test, we choose a magnetic field oblique to the grid as in \cite{castro17}:

\begin{equation}
    \boldsymbol{B}=B_0(\sin{\theta}\cos{\phi},\sin{\theta}\sin{\phi},\cos{\theta}),
\end{equation}

\noindent
where $B_0=1$ (high magnetization). The solution of the problem at $t=4.0$ is shown in Fig. \ref{SMBW}. Due to the high magnetization of the test, the blast wave propagates outwards following the oblique field lines. The test preserves the symmetry with good accuracy and no numerical artifacts nor undesired instabilities appear in our simulation.

\paragraph{3D Rotor problem} 

\begin{figure*}
  \centerline{\includegraphics[width=\textwidth]{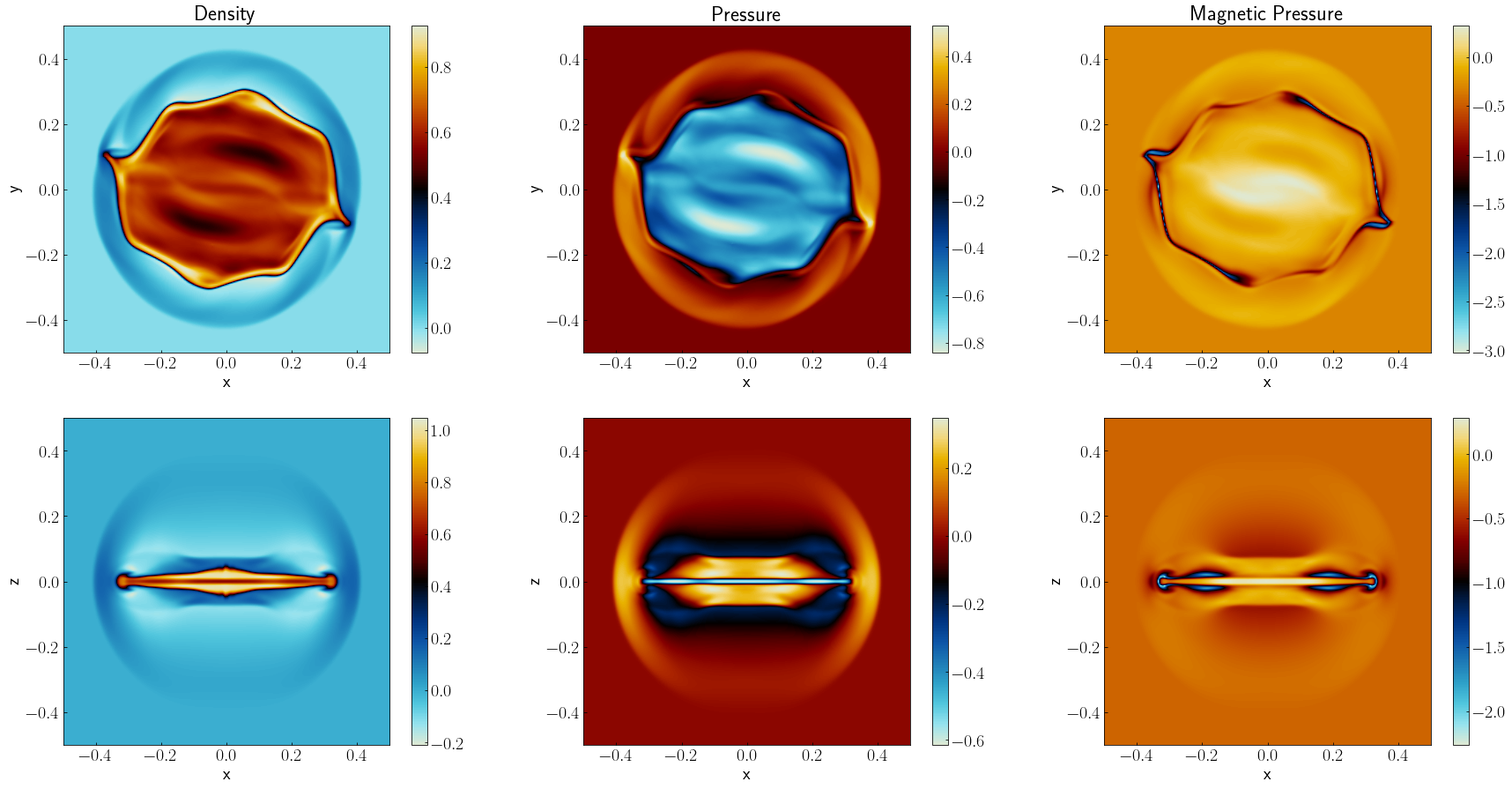}}
  \caption{Logarithmic density (left), logarithmic pressure (middle) and logarithmic magnetic pressure (right) at $t=0.4$ for the three dimensional relativistic rotor problem in the $xy$ plane at $z=0$ (top) and $xz$ plane at $y=0$ (bottom). We consider the three-dimensional Cartesian grid $[-0.5,0.5]\times[-0.5,0.5]\times[-0.5,0.5]$ with 256 cells per spatial dimension.}
\label{3ROT}
\end{figure*}

A three-dimensional version of the relativistic rotor problem is considered \citep{Mignone09}. The computational box covers the domain $[-0.5,0.5]^3$, with $256^3$ cells and free boundary conditions everywhere. The initial configuration consists of a sphere of density $\rho_{\mathrm{s}}=10$ and radius $r_{\mathrm{s}}=0.1$ which is rotating around the $z$-axis with relativistic velocity components $(v_x,v_y,v_z)=\omega~(-y,x,0)$, where $\omega=9.95$ is the angular velocity. The sphere is in pressure equilibrium with the ambient medium, whose pressure and density are $p_{\mathrm{a}}=1$ and $\rho_{\mathrm{a}}=1$, respectively. The magnetic field is aligned with the $x$-direction in the whole domain, $\boldsymbol{B}=(1,0,0)$ and we use the non-relativistic adiabatic index, $\Gamma=5/3$. The solution of the problem at $t=0.4$ is shown in Fig. \ref{3ROT}. Once the sphere starts spinning, complex torsional waves and shocks propagate outwards carrying angular momentum from the sphere to the medium, producing an octogonal-like disk in the $xy$ plane ($z=0$) surrounded by a shell of higher density and lower magnetic pressure, all embedded in a spherical fast rarefaction. The overall shape and the internal distribution of the flow is in good agreement with the results presented in \cite{Mignone09} using the HLLD Riemann solver. The test also shows the same flow distortions in the $xy$ plane along the $x$-axis, which are 3D numerical pathologies that do not manifest in the 2D version of the problem. This demonstrates that our version of HLLD works similar to the version in PLUTO for three dimensional applications.

\paragraph{3D Shock-Cloud interaction}

\begin{figure*}
  \centerline{\includegraphics[width=\textwidth]{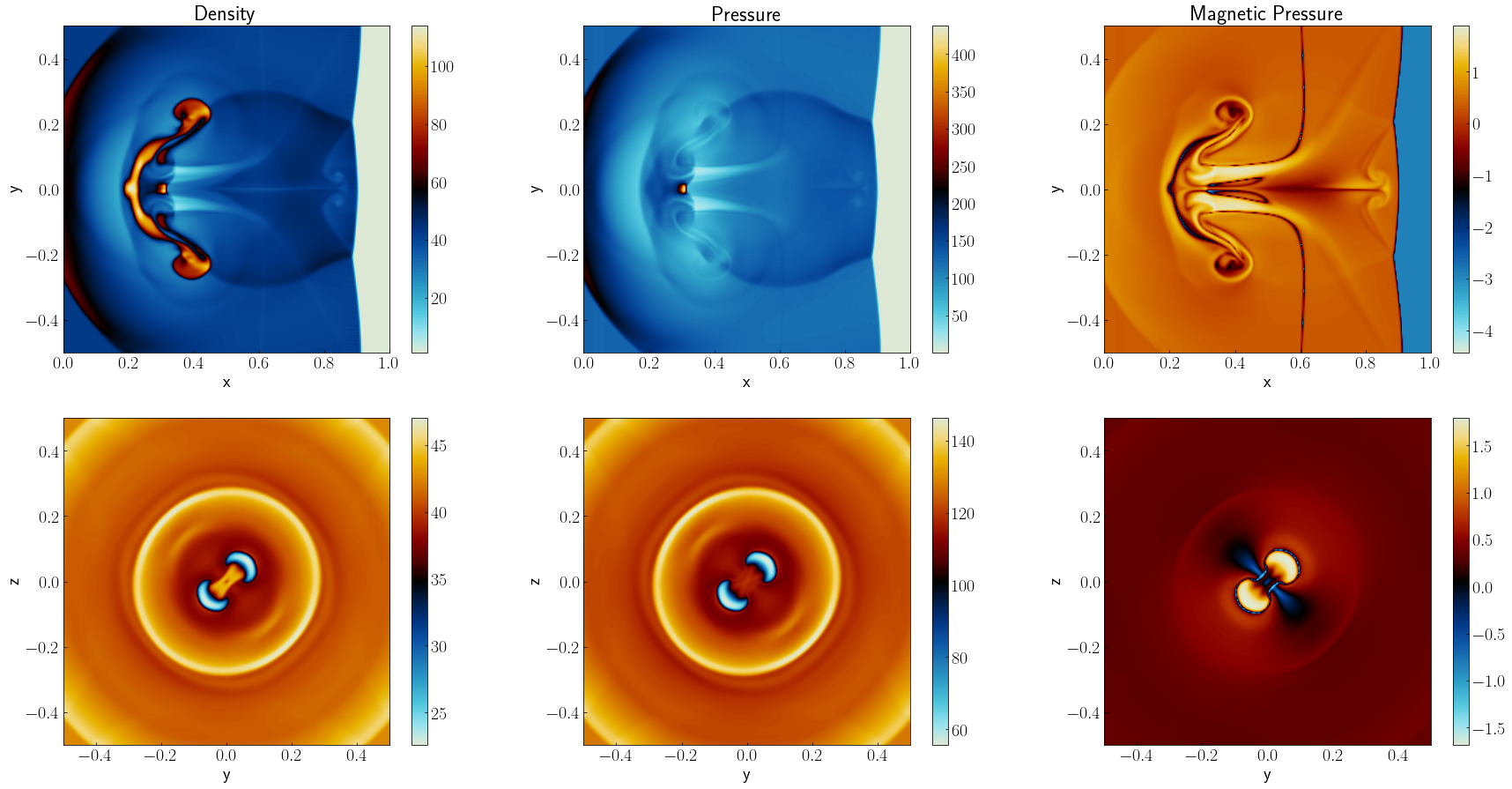}}
  \caption{Density (left), pressure (middle) and logarithmic magnetic pressure (right) at $t=1.0$ for the three dimensional relativistic shock-cloud interaction in the $xy$ plane at $z=0$ (top) and $xz$ plane at $y=0$ (bottom). We consider the three-dimensional Cartesian grid $[0,1.0]\times[-0.5,0.5]\times[-0.5,0.5]$ with 256 cells per spatial dimension.}
\label{SC3D}
\end{figure*}

Finally, we consider a three dimensional version of the shock-cloud collision problem of \cite{Mignone06}, following the methodology described in \cite{mignone12}. We perform the simulation in the computational box $[0,1]\times[-0.5,0.5]\times[-0.5,0.5]$, with $256^3$ cells and free boundary conditions in the whole domain. A shock-wave is initially located at $x=0.6$. Upstream ($x>0.6$), the pre-shocked material is defined by $\rho=1$, $p_{\mathrm{g}}=10^{-3}$, $W_x=10$ ($v_x = -0.995$), $B_z=0.5$, while downstream the medium is given by  $\rho=42.5942$, $p_{\mathrm{g}}=127.9483$, $W_x=1$, $B_z=-2.12971$. In this version, we reduced the magnitude of the $z$-component of the magnetic field with respect to the two dimensional test by $B_z=B_z^{2D}/\sqrt{2}$ and we set $B_y=B_z$. Transversal velocities $v_y,v_z$ and the component of the magnetic field parallel to the x-axis, $B_x$, are initially set to zero everywhere. At $x=0.8$, an over-dense sphere of radius $r=0.15$ and density $\rho=10$ is set in pressure equilibrium with the pre-shocked material and advected with the flow. For the ideal EoS, we consider the relativistic adiabatic factor, $\Gamma=4/3$. The solution of the problem at $t=1.0$ is shown in Fig. \ref{SC3D}. As in the 2D simulation, immediately after the impact between the cloud and the shock wave, the sphere experiments a strong compression that increases significantly the density of the clump. As a byproduct of this collision, a bow shock propagates to the left in the shocked material and a reverse shock is transmitted to the right, penetrating into the cloud and producing a mushroom-shaped structure like in the 2D shock-cloud interaction test. However, due to the lower resolution employed in this case, the pattern of waves that appear at the end of the test is more diffusive. 

\end{appendix}

\end{document}